\documentclass[prb,showpacs,twocolumn]{revtex4-1}
\usepackage{amsmath,graphicx}

\newcount\minute
\newcount\hour
\newcount\hourMins
\def\now
{
  \minute=\time
  \hour=\time \divide \hour by 60
  \hourMins=\hour \multiply\hourMins by 60
  \advance\minute by -\hourMins
  \zeroPadTwo{\the\hour}:\zeroPadTwo{\the\minute}
}
\def\timestamp
{
  \today\ \now
}
\def\today
{
  \zeroPadTwo{\the\day}-\zeroPadTwo{\the\month}-\the\year%
}
\def\zeroPadTwo#1%
{
  \ifnum #1<10 0\fi
  #1%
}
\def \dif {\mathrm{d}}
\def \Dif {\mathcal{D}}

\newcommand{\nn}{\nonumber}
\newcommand {\eq}[1]{(\ref{#1})}

\newcommand{\be}{\begin{equation}}
\newcommand{\ee}{\end{equation}}
\newcommand{\beq}{\begin{equation}}
\newcommand{\eeq}{\end{equation}}
\newcommand{\bea}{\begin{eqnarray}}
\newcommand{\eea}{\end{eqnarray}}

\newcommand{\rme}{{\mathrm{e}}}
\newcommand{\rmd}{{\mathrm{d}}}

\newcommand{\E}{\epsilon}
\newlength{\bilderlength}
\newcommand{\bilderscale}{0.35}

\newcommand{\bilderskip}{\hspace*{0.8ex}}

\newcommand{\diagram}[1]{%
\settowidth{\bilderlength}{\bilderskip%
\includegraphics[scale=\bilderscale]{#1}\bilderskip}%
\parbox{\bilderlength}{\bilderskip%
\includegraphics[scale=\bilderscale]{#1}\bilderskip}}

\def\be{\begin{equation}}
\def\ee{\end{equation}}

\date{\timestamp}

\begin{document}

\title{Super-rough phase of the random-phase sine-Gordon model: Two-loop results}

\author{Zoran Ristivojevic, Pierre Le Doussal, and Kay J\"{o}rg Wiese}

\affiliation{Laboratoire de Physique Th\'{e}orique--CNRS, Ecole Normale Sup\'{e}rieure, 24 rue Lhomond, 75005 Paris, France}

\begin{abstract}
We consider the two-dimensional random-phase sine-Gordon and study the vicinity of its glass transition temperature $T_c$, in an expansion in small $\tau=(T_c-T)/T_c$, where $T$ denotes the temperature. We derive renormalization group equations in cubic order in the anharmonicity, and show that they contain two universal invariants. Using them we obtain that the correlation function in the super-rough phase for temperature $T<T_c$ behaves  at large distances as $\overline{\langle[\theta(x)-\theta(0)]^2\rangle} = \mathcal{A}\ln^2(|x|/a) + \mathcal{O}[\ln(|x|/a)]$, where the amplitude $\mathcal{A}$ is a universal function of temperature $\mathcal{A}=2\tau^2-2\tau^3+\mathcal{O}(\tau^4)$. This result differs at two-loop order, i.e., $\mathcal{O}(\tau^3)$,
from the prediction based on results from the ``nearly conformal'' field theory of a related fermion model. We also obtain the correction-to-scaling exponent.
 \end{abstract}

\maketitle

\section{Introduction}

Although in two-dimensional (2D) systems with continuous symmetry and short-range interactions thermal fluctuations prevent the existence of long-range order \cite{Mermin+66}, they do not prevent phase transitions. The 2D XY model, much studied in that context, describes a large class of physical systems with continuous symmetry, which includes superfluid and superconducting films, magnetic systems and one-dimensional quantum liquids. It exhibits a topological phase transition between the low-temperature phase with quasi-long-range order and a disordered phase at high temperatures. This Berezinskii-Kosterlitz-Thouless transition is driven by unbinding of vortices due to an increasing amount of thermal fluctuations \cite{Berezinskii+72,Kosterlitz+73}.
From the technical side, the XY model is conveniently studied within the equivalent dual 2D sine-Gordon (SG) model, which is amenable to powerful field-theoretical treatments \cite{Amit+80}. 
It exhibits a high temperature quasi-long-range ordered phase and a low temperature massive phase. When additional terms are added to the SG model new universality classes can emerge \cite{Korshunov06}. For a pure system, the simplest example is an additional field gradient in one direction \cite{Pokrovsky+79}, which describes the commensurate-incommensurate transition in 2D and realizes, for example, when an atomic layer of noble gases is deposited on the periodic substrate of graphite \cite{Chaikin+}. Both
models, with and without the tilt, are exactly solvable \cite{Haldane+83,Korepin+} and are by now well understood.

\begin{figure}
\includegraphics[width=0.7\columnwidth]{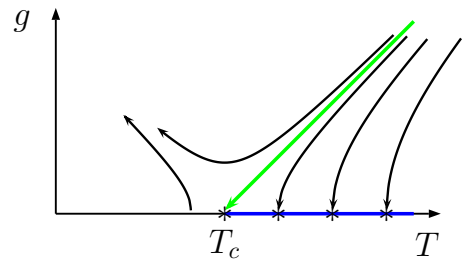}
\caption{The renormalization group flow diagrams of the sine-Gordon model. At small temperatures, $T<T_c$ the tentative fixed point is disconnected from the line of Gaussian fixed points that exist for $T>T_c$ when the strength of the anharmonic term $g$ of the SG model flows to zero. In the context of one-dimensional fermions with disorder the same flow diagram apply provided the temperature is replaced by the strength of interaction and that $g$ denotes the disorder strength \cite{Giamarchi}.}\label{Fig:flowsg}
\end{figure}

The random versions of the 2D SG model allow for more
scenarios and much less exact results are known. Via
bosonization they are related to fermions with disorder
and have also been much studied in that context  \cite{Giamarchi}. A well known example is the 2D SG model with a quenched random phase that depends on
only one coordinate in the cosine term. Such a model describes a classical 2D model with correlated disorder, and also a 1D quantum system with point disorder (the second direction being imaginary time). In the latter case it is related to 1D disordered Luttinger liquids and belongs to the Berezinskii-Kosterlitz-Thouless class \cite{Giamarchi}. Seen as a 2D classical model it exhibits quasi-long-range order (i.e.,~ an infinite correlation length) in its high-temperature phase, which is described by a line of Gaussian fixed points of the renormalization group (RG) where the cosine term is irrelevant. Its low-temperature phase is glassy and described by RG fixed points at large disorder strengths, see Fig.~\ref{Fig:flowsg}. This scenario first found in one-loop order \cite{Giamarchi+88} is not changed at two-loop order \cite{Ristivojevic+12}.

\begin{figure}
\includegraphics[width=0.7\columnwidth]{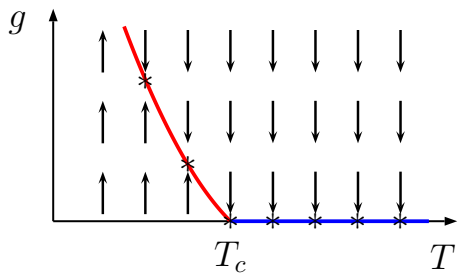}
\caption{The renormalization group flow diagram of the random-phase sine-Gordon model (\ref{H}). The line of fixed points (red curve) in the low-temperature phase $T<T_c$ occurs at finite disorder strength and it is continuously connected with the line (blue curve) of Gaussian fixed points at $T>T_c$. While the line of nonzero fixed points is a linear function of temperature to the lowest one loop order, it gets quadratic correction beyond that. The correlation function at $T<T_c$ has a super-rough logarithm-squared form \cite{Toner+90}.}\label{Fig:flow-co}
\end{figure}

In this paper we study the 2D random-phase sine-Gordon (RPSG) model  \cite{Toner+90,Hwa+94,Cardy+82} where the quenched random phase depends on both coordinates. This model can also be understood as the random field XY model provided one excludes vortices by hand \cite{vortex}. The RPSG model describes 2D periodic disordered elastic systems, such as a randomly pinned planar array of vortex lines \cite{Hwa+94} or surfaces of crystals with quenched disorder \cite{Toner+90}. It also exhibits a phase transition at a critical temperature $T_c$ below which the random cosine term becomes relevant. This transition was first studied in an expansion in $\tau= (T_c-T)/T$ using a one-loop RG approach in Ref.~\cite{Cardy+82}. The physics of the RPSG model is however quite different from the previously mentioned disordered model. While the high-temperature phase is described by a line of Gaussian fixed points, similarly to the SG model, the low-temperature phase is glassy and described by a line of non-Gaussian fixed points where the renormalized disorder gradually increases from zero when decreasing the temperature below $T_c$, see Fig.~\ref{Fig:flow-co}. The glass phase for $T<T_c$ is super-rough, i.e.,~the variance of the fluctuations of the displacement field grows as logarithm-squared of the distance in contrast to the standard rough logarithmic form at high temperatures \cite{Toner+90}. Within the RG approach this is due to an unbounded growth of the off-diagonal disorder (in replica space) at low temperatures that determines the correlation function. Further RG studies predicted that the amplitude $\mathcal{A}$ of the logarithm-squared correlations is a universal function of temperature\cite{Hwa+94}, with $\mathcal{A} = 2 \tau^2 + \mathcal{O}(\tau^3)$ to one loop order accuracy \cite{Carpentier+97,SchehrLeDoussal2003}.

The existence of the super-rough phase has been confirmed in several numerical studies at zero temperature \cite{Zeng+96,Rieger+97}
and for all temperatures $0<T<T_c$, see Refs.~\cite{ZengLeathHwa1999,PerretSchehr2011}. These studies consider discrete random height models, or discrete-line models, believed to be in the same universality class as the RPSG model. These are further mapped onto the dimer covering problem with random weights \cite{henley,Kenyon,Bogner2004}. Powerful polynomial algorithms then allow to generate all possible coverings of the lattice by dimers \cite{Propp}. Using such algorithms $\mathcal{A}(\tau)$ was estimated in \cite{ZengLeathHwa1999} where the quadratic behavior $\mathcal{A}(\tau) \propto \tau^2$ at small $\tau$ was confirmed. Very recently more accurate data have been obtained in \cite{Perret+12,PerretSchehr2011} (see below).

Some recent studies opened the hope that $\mathcal{A}$ could be obtained non-perturbatively. Considering a model of disordered noninteracting fermions in 2D, Guruswamy, LeClair, and Ludwig \cite{Guruswamy+00} used methods of ``nearly conformal'' field theory to predict the exact form of the correlation functions as well as the scaling equations for their fermionic model. Upon bosonization, these results where interpreted as corresponding to the RPSG model exactly on its fixed points at finite disorder (see Fig.~\ref{Fig:flow-co}). In Ref.~\cite{LeDoussal+07} this correspondence and the translation to the parameters of the RPSG model was performed in details, with the conclusion that if the exact beta function of \cite{Guruswamy+00} is correct then one should have $\mathcal{A}=\mathcal{A}_{NCFT} = 2\tau^2(1-\tau)^2$ exactly in the whole super-rough phase $0<\tau<1$. As discussed in \cite{LeDoussal+07} this however raises some puzzle: numerics exclude the amplitude vanishing at $T=0$, and the non-monotonous behavior of $\mathcal{A}_{NCFT}$ with temperature is surprising. Hence the formula, correct to one loop accuracy, can hold exactly at best in a vicinity of $T_c$, i.e.,~for $\tau < \tau^*$ with some unknown $\tau^*$. Since the numerical values are larger by a factor $\approx 4$ than the maximum $\mathcal{A}_{NCFT}(1/2)$ the true amplitude should be {\it larger} than the predicted one. Possible scenarios are discussed in Ref.~\cite{LeDoussal+07} such as the mapping between free fermion models and the RPSG model failing below some temperature, or some new operators becoming relevant at $\tau^*$. In addition a functional RG study performed in Ref.~\cite{LeDoussal+07} leads to a non-vanishing amplitude at $T=0$ as a result of including higher harmonics of the disorder that are relevant there.

In the present paper we revisit the  model using perturbative renormalization group methods to the next two-loop order
and compute the amplitude of correlations in the super-rough phase including $\mathcal{O}(\tau^3)$ terms. A short summary of the present work has been presented in Ref.~\cite{Perret+12}. Here we give all the details. We perform a systematic calculation in terms of the strength of the anharmonic term $g$, see Eq.~(\ref{H1}). The problem is studied within a bosonic formulation using field theory methods. We use two complementary methods, which are explained in a pedagogical way. The first one is based on the calculation of the effective action and the second on the operator product expansion. We study both the theory regularized by a small distance cutoff $a$ and, for $T<T_c$, directly in the continuum limit, and obtain the precise dependence of the results on the cutoff functions. Our main findings are the scaling equations (\ref{RGgfinal}) and (\ref{RGsigmafinal}), the correlation function (\ref{correlationfunctionfinal}) and the correction-to-scaling exponent (\ref{scalingexponent}). The equations beyond lowest order contain non-universal coefficients that are connected by relation (\ref{inv}). For the amplitude of correlations in the glass phase $T<T_c$ we find $\mathcal{A}=2\tau^2-2\tau^3+\mathcal{O}(\tau^4)$. Hence it confirms the conclusion of \cite{LeDoussal+07} that the
translation of the results of Ref. \cite{Guruswamy+00} into an exact result to all orders for the RPSG model cannot be correct. Since the discrepancy arises already at two loop order (i.e., $\tau^*=0$) it also casts some doubts on the fermion calculation done in Ref.~\cite{Guruswamy+00} or on its consequences for the bosonic model as given in Ref.~\cite{Guruswamy+00} and in Ref.~\cite{LeDoussal+07}. Note that our result for the amplitude is indeed larger than $\mathcal{A}_{NCFT}$ (for small $\tau$) hence it goes in the right direction. Interestingly, it also appears to better fit the most recent numerical results of \cite{PerretSchehr2011,Perret+12} up to $\tau \approx 0.5$, although it is a perturbative calculation around $T_c$, i.e.,~around $\tau=0$.

To be complete let us mention that many other studies have addressed the RPSG model, its thermodynamics \cite{Emig+01,Bogner2004}, its stability to RSB and links to fermions \cite{rsb}, its dynamics near $T_c$
\cite{TsaiShapir1992,TsaiShapir1994a,TsaiShapir1994b,culeShapir1995}, its equilibrium dynamics at all $T$ \cite{SchehrLeDoussal2005,SchehrRieger2005} and its aging dynamics \cite{SchehrLeDoussal2003,SchehrLeDoussal2004,Schehr+05}. It would be interesting
to push such methods to two loop accuracy, as done here for the statics.

The outline of this paper is as follows. In section II we introduce the model. Using the replica method we derive the replicated Hamiltonian which is our starting point for the systematic field-theoretic renormalization group procedure. We also define several correlation functions of interest. In section III we calculate the effective action of the model order by order in the disorder strength and from it we derive the beta functions. Further we examine the universality of coefficients in the beta functions. In section IV we evaluate the coefficients in the beta functions using two different methods and find three universal coefficients, one of them from one-loop and two of them from two-loop. In section V we give the final form of the scaling equations, find the correction-to-scaling exponent and obtain the correlation function (that measures the fluctuations due to the disorder) in the super-rough low-temperature phase. In section VI we present a first principle derivation of the correlation function.
In section VII we use the operator product expansion method which is found in agreement with the results of the effective action method. Section VIII contains conclusions. Numerous technical details are relegated to appendices.

\section{Model, replicated Hamiltonian, and correlation functions}\label{s:model}

\subsection{Model}

We consider the 2D random-phase sine-Gordon model. In terms of a real displacement field $\theta(x) \in (-\infty,\infty)$ its Hamiltonian reads
\begin{align}\label{H}
H=\int\dif^2 x\left[\frac{\kappa}{2}\left(\nabla_x\theta\right)^2-h\cdot\nabla_x\theta- \frac{1}{a}\left(\xi\mathrm{e}^{i\theta}+\mathrm{h.c.}\right)\right],
\end{align}
where $\kappa$ is the elastic constant, $a$ is the short-length-scale cutoff, and $h(x)$ and $\xi(x)$ are quenched Gaussian random fields, the first one real and the other complex. Their nonzero correlations are given by
\begin{align}
&\overline{h^i(x)h^j(y)}= \Delta_h \delta^{ij}\delta(x-y),\\
&\overline{\xi(x)\xi^*(y)}=\Delta_\xi \delta(x-y),
\end{align}
where $i,j\in\{1,2\}$ denote the components of $h$. For future convenience we define the disorder strengths $\Delta_{h,\xi}$ in terms of the dimensionless parameters $\sigma$ and $g$ as follows:
\begin{align}
\Delta_h = T^2 \frac{\sigma}{2 \pi}, \quad
\Delta_\xi = T^2 \frac{g}{2 \pi},
\end{align}
where $T$ is the temperature. Note that the disorder $h(x)$ must be introduced as it is generated by the symmetry-breaking field under coarse graining. We denote disorder averages by $\overline{.\phantom{1}.\phantom{1}.}$. Depending on context $x$ and $y$ will be used either to denote 2D coordinates (as in previous equations) or as their norms, i.e., $x$ stands either for $\mathbf{x}$ or $|\mathbf{x}|$.

\subsection{Replicated Hamiltonian}

We use the replica method to treat the disorder \cite{Giamarchi}. The partition function for the model (\ref{H}) is given by $Z=\int\mathcal{D}\theta\mathrm{e}^{-H/T}$. In order to perform the disorder average, we use the replica trick, and the free energy of the system $F=-T\ln Z$ is written as
\begin{align}
F=\lim_{n\to0}(Z^n-1)/n.
\end{align}
The average with respect to disorder can now be done since one can write $\overline{Z^n}=\int\left(\prod_{\alpha=1}^n\mathcal{D} \theta_\alpha\right)\mathrm{e}^{-H^{rep}/T}$
where $\theta_\alpha, \alpha=1\ldots n$ are the replicated fields. In the following by greek indices $\alpha,\beta,\ldots$ we denote replicated fields and we do not write explicitly the boundaries in the sums.

The replicated Hamiltonian reads
\begin{align}
H^{rep}=H_0^{rep}+H_1^{rep},
\end{align}
where the harmonic part is
\begin{align}\label{H0}
\frac{H_0^{rep}}{T}=&\sum_{\alpha\beta}\int\dif^2x\Big\{ \frac{\kappa}{2T}\delta_{\alpha\beta}\left[ (\nabla_x\theta_\alpha)^2+m^2(\theta_\alpha)^2\right]\notag\\
&-\frac{\sigma}{4 \pi}\nabla_x\theta_\alpha \cdot\nabla_x\theta_\beta\Big\}.
\end{align}
The mass $m$ is introduced in the model as an infrared cutoff.
We will perform calculations with finite $m$ and study the limit $m\to 0$ at the end.
The system size is infinite throughout the paper. The anharmonic part reads
\begin{align}
\label{H1}
\frac{H_1^{rep}}{T}=-\frac{g}{2 \pi a^2}\sum_{\alpha\beta}\phantom{ }^{'} \int\dif^2x\cos(\theta_\alpha-\theta_\beta).
\end{align}
We introduced the symbol $\sum\phantom{ }^{'}$ which denotes a summation where all replica indices are different. While after replicating the model (\ref{H}) one formally obtains a sum over all unconstrained replica indices, for convenience we use (\ref{H1}).

\subsection{Correlation functions}

Our aim is to compute the two correlation functions:
\begin{align} \label{calG}
&{\cal G}(x) = \overline{\langle \theta(x) \theta(0) \rangle } - \overline{ \langle \theta(x) \rangle \langle \theta(0) \rangle}, \\
&{\cal G}_0(x) = \overline{\langle \theta(x) \rangle \langle \theta(0) \rangle },
\end{align}
where ${\cal G}(x)$ measures the (disorder averaged) thermal fluctuation while
${\cal G}_0(x)$ measures the fluctuations due to disorder of the (thermally averaged) displacement field.

These disorder averaged correlations can be obtained from correlation functions of replicated fields. For instance
\begin{align}
&{\cal G}(x) = \lim_{n \to 0} \frac{1}{n} \sum_{\alpha \beta} {\cal G}_{\alpha\beta}(x), \\
&{\cal G}_0(x) = \lim_{n \to 0}(1-\delta_{\alpha\beta}) {\cal G}_{\alpha \beta}(x),
\end{align}
where
\begin{align}
{\cal G}_{\alpha\beta}(x)=\langle \langle \theta_\alpha(x)\theta_\beta(0)\rangle\rangle.
\end{align}
It can also be expressed as
\begin{align} \label{def22}
{\cal G}_{\alpha\beta}(x)=\delta_{\alpha\beta}{\cal G}(x)+{\cal G}_0(x),
\end{align}
which contains both correlations defined in (\ref{calG}). ${\cal G}(x)$ is also called the connected part and
${\cal G}_0(x)$ the off-diagonal part. To this aim we will use the harmonic part $H_0^{rep}$ as the "free" theory and treat $H_1^{rep}$ in perturbation theory, i.e perform a perturbation theory in $g$. Here and below we denote by $\langle  \langle .. \rangle \rangle$ (exact) averages over the complete Hamiltonian $H^{rep}$ and by $\langle .. \rangle$
averages over the free part $H_0^{rep}$ (it also designates thermal averages in the unreplicated theory, as no ambiguity can arise).

We start by computing the correlation function for the harmonic part, i.e.,~for $g=0$. It is easily found in Fourier space:
\begin{align}\label{G-fourier}
G_{\alpha\beta}(q)=&\langle\theta_\alpha(q)\theta_\beta(-q)\rangle= \left[\frac{\kappa}{T}(q^2+m^2)\delta_{\alpha\beta}-\frac{\sigma}{2 \pi} q^2\right]^{-1}\notag\\
=&\frac{T}{\kappa}\frac{1}{q^2+m^2}\delta_{\alpha\beta}+\frac{\sigma}{2\pi} \frac{T^2}{\kappa^2}\frac{q^2}{(q^2+m^2)^2}+\mathcal{O}(n),
\end{align}
where in the last step we inverted a replica matrix keeping only non-vanishing terms in the replica limit $n\to 0$.  Going to real space one obtains
\begin{align}
G_{\alpha\beta}(x)=\langle \theta_\alpha(x)\theta_\beta(0)\rangle=\delta_{\alpha\beta}G(x)+G_0(x).
\end{align}
The connected part reads
\begin{align}\label{G}
G(x)=2(1-\tau)K_0(m\sqrt{x^2+a^2}),
\end{align}
and everywhere in the paper the parameter $\tau$ denotes
\begin{align}
\tau=1-T/T_c.
\end{align}
By $K_0$ we denote the modified Bessel function of the second kind \cite{Abramowitz}.
Our model (\ref{H}) has a phase transition at temperature $T_c=4\pi\kappa$ (see Fig.~\ref{Fig:flow-co}), i.e.,~for $\tau=0$. In the following we will repeatedly use the expression for $G(x)$, which has the following behavior at small distances $|x|\ll (cm)^{-1}$
\begin{align}\label{G-smallx}
G(x)=-(1-\tau)\ln \left[c^2 m^2(x^2+a^2)\right ],
\end{align}
with the constant $c=\mathrm{e}^{\gamma_E}/2$ and $\gamma_E$ is the Euler constant. In (\ref{G}) we have introduced the ultraviolet regularization by the parameter $a$. Such choice of regularization is preferable to some other choices in momentum space, since our RG procedure is most easily done in coordinate space.\footnote{One should notice that the Fourier transform of (\ref{G}) is positive for all values of $a$ and $m$, as it must be. See Eq.~(2.3) of Ref.~\cite{Amit+80}.}

The off-diagonal part of the correlation function for $g=0$ reads
\begin{align}
G_0(x)=\frac{\sigma T^2}{(2\pi)^2\kappa^2} \left[K_0(m|x|)-\frac{m|x|}{2}K_1(m|x|)\right]+\mathcal{O}(n),
\end{align}
which at small distances $|x|\ll (cm)^{-1}$ becomes
\begin{align}
G_0(x)=-2\sigma(1-\tau)^2 \left\{1+\ln\left[c^2m^2(x^2+a^2)\right]\right\}+\mathcal{O}(n). \label{G0x}
\end{align}

The model studied here possesses an important symmetry, the statistical tilt symmetry (STS), i.e., the non-linear part $H_1^{rep}$ is invariant under the change $\theta_\alpha(x) \to \theta_\alpha(x) + \phi(x)$ for an arbitrary function $\phi(x)$. As discussed in many works
\cite{Schulz+88,Hwa+94,Hwa+94,Carpentier+97,LeDoussal2008}, and recalled below, this implies two important properties:

(i) $G_0(x)$ does not appear to any order in perturbation theory in $g$ in the calculation of e.g.~the effective action (see the following section), and (ii) the disorder averaged thermal correlation is uncorrected to all orders:
\begin{align} \label{sts}
{\cal G}(x) = G(x),
\end{align}
i.e.,~independent of $g$. This implies that $\tau$ or $T/T_c$ can be {\it measured} from the amplitude of the logarithm in ${\cal G}(x) \sim 2 (1-\tau) \ln x$ at large $x$, hence they are uncorrected by disorder. \footnote{If one uses cutoff procedures which do not respect STS, the renormalized $\tau$, measured from ${\cal G}(x)$ may differ from the bare one from $G(x)$. Everywhere here $\tau$ means the renormalized one.}

Because of property (i) $G_0(x)$ only receives additive corrections, e.g.~corrections to $\sigma$ which, in the present model, change its logarithmic behavior (\ref{G0x}) into a squared-logarithm behavior for ${\cal G}_0(x)$, as discussed below.

The perturbation theory thus depends only on the function $G(x)$. While the precise form of the correlation function for the model (\ref{H0}) can be explicitly calculated and takes the form (\ref{G}), we will explicitly check below that the precise form of $G(x)$ is not important as long as it satisfies the two conditions: (a) the limiting behavior of the correlation function (\ref{G}) is logarithmic as given in (\ref{G-smallx}) and (b) the propagator tends (exponentially) to zero at large distances. The crossover length is given by the infrared cutoff, which is the inverse mass in our case. The freedom of the propagator that satisfies conditions (a) and (b) is manifested through the renormalization group equations which will contain several non-universal constants that, when appropriately combined, produce some universal numbers. These universal numbers determine in turn the amplitude of the correlation function in the super-rough phase, as well as the correction-to-scaling exponent.

\section{Effective action of the model}\label{sec:effectiveaction}

In this section we calculate the effective action functional $\Gamma$ for the model (\ref{H}). It will directly lead to the scaling equations of the model (\ref{H}) and critical properties of the system. It extends to the next order in perturbation theory the calculation of $\Gamma$ in  \cite{SchehrLeDoussal2003}. In the framework of diagrammatics, $\Gamma$ can be expressed as a sum over all one-particle irreducible graphs \cite{Zinn-Justin}. Here we will not calculate $\Gamma$ using a diagrammatic approach but via an equivalent algebraic method. The definition and the derivation of the final form of the effective action  (\ref{Gammafinal}), to the required order in perturbation theory, is presented in appendix \ref{appendix:effectiveaction}.
The difference with the standard Wilsonian procedure of Ref.~\cite{Kogut79} is that one  integrates out fields that live in the whole momentum space and not only in the high-momentum degrees of freedom. These fields that are integrated out are denoted by $\chi$ in (\ref{Gammafinal}). Translated to the replicated Hamiltonian, our aim is to evaluate the following expression
\begin{align}\label{Gamma-def}
\Gamma=\frac{H_0^{rep}}{T}+\Gamma_1+\Gamma_2+\Gamma_3+\mathcal{O}(g^4),
\end{align}
where $\Gamma_i$ is the corresponding term from (\ref{Gammafinal}) proportional to $g^i$.

\subsection{Derivation of $\Gamma$}

To lowest order in $g$ we have
\begin{align}
\label{Gamma1}
\Gamma_1&=\left \langle H_1^{rep}(\theta+\chi)/T\right\rangle^\chi\notag\\
&=-\frac{g}{2 \pi a^2} \mathrm{e}^{-G(0)}\sum_{\alpha\beta}\phantom{ }^{'}\int \dif^2x \cos(\theta_\alpha-\theta_\beta).
\end{align}
In the previous and all forthcoming terms of similar form, evaluation of averages of the type $\langle\cdots\rangle^\chi$ is quite simply done by making use of Wick's theorem when one gets contractions of $\chi$ fields with respect to the quadratic Hamiltonian (\ref{H0}). Due to ``charge neutrality'' of (\ref{H1}) only the diagonal part (\ref{G}) of the correlation function $G_{\alpha\beta}(x)$ survives in expressions of the type $\left\langle\left[ H_1^{rep}(\theta+\chi)/T\right]^p\right\rangle^\chi$, where $p$ is a positive integer.

In order to obtain $\Gamma_2$ we use the transformation of sum rule (\ref{sum4}) when evaluating the corresponding term from (\ref{Gammafinal}). The final result reads
\begin{widetext}
\begin{align}\label{Gamma2}
\Gamma_2=&-\frac{1}{2}\left(-\frac{g}{2 \pi a^2} \right)^2\mathrm{e}^{-2G(0)}\sum_{s=\pm 1}\Bigg\{\sum_{\alpha\beta} \phantom{}^{'} \int\dif^2x\dif^2y A(x-y,2s) \cos\left[\theta_\alpha(x)-s\theta_\alpha(y)- \theta_\beta(x)+s\theta_\beta(y)\right]\notag\\&
+\sum_{\alpha\beta\gamma} \phantom{}^{'} \int\dif^2x\dif^2y 2A(x-y,s) \cos\left[\theta_\alpha(x)-s\theta_\alpha(y)- \theta_\beta(x)+s\theta_\gamma(y)\right]\Bigg\},
\end{align}
where
\begin{align}\label{A}
&A(x,p)=\mathrm{e}^{pG(x)}-1-pG(x).
\end{align}
The second-order term $\Gamma_2$ consists of two- and three-replica contributions. It turns out that only terms with $s=+1$ give contributions to the renormalization. The term multiplied by $A(x-y,1)$ is responsible for renormalization of the coupling constant $g$, while the other term multiplied by $A(x-y,2)$ renormalizes the off-diagonal part of (\ref{H0}).

One should notice that the second term of Eq.~(\ref{Gammafinal}) proportional to $g^2$ that contains the integral basically makes the total contribution (proportional to $g^2$) of Eq.~(\ref{Gammafinal}) to be one-particle irreducible and produces terms $-aG(x)$ in Eq.~(\ref{A}).

In order to obtain the next-order contribution, similarly as we did for $\Gamma_2$, we use the sum transformation (\ref{sum6}) when we evaluate the term (\ref{Gammafinal}) proportional to $g^3$. The final result reads $\Gamma_3=\Gamma_3'+\Gamma_3''$ with
\begin{align}
\label{Gamma3'}
\Gamma_3'=&\frac{1}{6} \left(-\frac{g}{2 \pi a^2}\right)^3\mathrm{e}^{-3G(0)}\int\dif^2x\dif^2y\dif^2z \bigg\{\notag\\
&3B(x-y,y-z,z-x,-2,2,2)\sum_{\alpha\beta}\phantom{ }^{'}\cos[\theta_\alpha(x)+\theta_\alpha(y)-\theta_\alpha(z) -\theta_\beta(x)-\theta_\beta(y)+\theta_\beta(z)]\notag\\
&+12B(x-y,y-z,z-x,2,1,-1)\sum_{\alpha\beta\gamma}\phantom{ }^{'}\cos[\theta_\alpha(x)-\theta_\alpha(y)+\theta_\alpha(z) -\theta_\beta(x)+\theta_\beta(y)-\theta_\gamma(z)]\notag\\
&+2B(x-y,y-z,z-x,1,1,1)\sum_{\alpha\beta\gamma}\phantom{ }^{'}\cos[\theta_\alpha(x)-\theta_\alpha(y)-\theta_\beta(x) +\theta_\beta(z)+\theta_\gamma(y)-\theta_\gamma(z)]\notag\\
&+6B(x-y,y-z,z-x,1,1,0)\sum_{\alpha\beta\gamma\delta}\phantom{ }^{'}\cos[\theta_\alpha(x)-\theta_\alpha(y)+\theta_\beta(y) -\theta_\beta(z)+\theta_\gamma(z)-\theta_\delta(x)]\bigg\},
\end{align}
while the other part $\Gamma_3''$ is not important for critical properties of the model and is given in (\ref{Gamma3''}). In the previous equation the common term $B$ is defined as
\begin{align}
\label{Babc}
B(x,y,z,a,b,c)=&\mathrm{e}^{aG(x)+bG(y)+cG(z)}-\mathrm{e}^{aG(x)} -\mathrm{e}^{bG(y)}-\mathrm{e}^{cG(z)}+2-B_1(x,y,z,a,b,c),
\end{align}
where the part of $B$ that makes it one-particle irreducible reads
\begin{align}
B_1(x,y,z,a,b,c)=&\mathrm{e}^{aG(x)}[bG(y)+cG(z)] +\mathrm{e}^{bG(y)}[aG(x)+cG(z)]+\mathrm{e}^{cG(z)}[aG(x)+bG(y)]\notag\\
&-abG(x)G(y)-acG(x)G(z)-bcG(y)G(z)-2aG(x)-2bG(y)-2cG(z).
\end{align}
\end{widetext}
It arises from the last two terms in Eq.~(\ref{Gammafinal}).

The summands of $\Gamma_3$ can be distinguished by the sum $a+b+c$ of the corresponding $B$-functions. For purposes of renormalization of the model (\ref{H}) the only relevant terms are these for which $a+b+c$ equals either two or three. The former contribute to the renormalization of the coupling constant $g$, see Eq.~(\ref{H1}), while the latter renormalize the off-diagonal part of (\ref{H0}). All other summands produce nondivergent contributions to the effective action and hence can be neglected. For completeness of presentation they are given in Eq.~(\ref{Gamma3''}).

\subsection{Expansion of $\Gamma$}

Having obtained the effective action $\Gamma$ in the preceding part, we are now prepared to study its renormalization. The perturbative expansion of $\Gamma$ in the bare coupling constant $g$ contains divergencies when the cutoff $a$ tends to zero. In order to remove such divergencies from the theory it turns out that two renormalization constants suffice. They relate the initial coupling constants $\sigma$ and $g$ and the ``renormalized'' ones, $\sigma_R$ and $g_R$ respectively. Written in terms of renormalized quantities, the effective action will be free of divergencies (up to third order, which is the order we are working with). In the following we will calculate the divergencies of the effective action (\ref{Gamma-def}) in a double expansion in two small parameters, $g$ and $\tau$, meaning close to the critical temperature and for weak anharmonic terms of the model (\ref{H}).

We now write $\Gamma$ in the same form as the starting Hamiltonian with new coefficients, plus irrelevant terms.
That allows us to define $g_R$ and $\sigma_R$ below.

The first-order term (\ref{Gamma1}) is already in a proper form. The second-order term $\Gamma_2$ [Eq.~(\ref{Gamma2})] contains terms with two and three replica sums. The former gives contribution to the off-diagonal part of $H_0^{rep}$, while the latter changes $H_1^{rep}$, as we will see below. It is important to stress here, that even if we had started without the term $\sim h$ in (\ref{H}) [i.e.,~without the term $\sim\sigma$ in (\ref{H0})] this term would have been generated under the RG coarse graining procedure, as first noted by \citet{Cardy+82}. This term is very important for the behavior of correlation function at $T<T_c$ and its super-rough $\ln^2$ form.\cite{Toner+90} We will come to that point later when we investigate the correlation function of the model. After expanding the two-replica part of $\Gamma_2$ [Eq.~(\ref{Gamma2})] one obtains
\begin{align}\label{Gamma2-eval}
\Gamma_2=&\frac{1}{2}\left(-\frac{g}{2 \pi a^2}\right)^2\mathrm{e}^{-2G(0)} \sum_{\alpha\beta}\phantom{}^{'}\bigg[\frac{8 \pi a_1}{c^2m^2}\int\dif^2 x \cos(\theta_\alpha-\theta_\beta)\notag\\
&+\frac{2 \pi a_2}{4c^4m^4}\int\dif^2x \left(\nabla_x\theta_\alpha-\nabla_x\theta_\beta\right)^2
+\ldots\bigg],
\end{align}
where $\ldots$ stands for many irrelevant operators and we define the dimensionless integrals:
\begin{align}\label{a1}
&a_1=\frac{c^2m^2}{2 \pi} \int\dif^2y A(y,1),\\
\label{a2}
&a_2=\frac{c^4m^4}{2 \pi} \int\dif^2y y^2 A(y,2).
\end{align}

Finally the cubic term in $g$ after expansion produces the following terms
\begin{align}\label{Gamma3-eval}
\Gamma_3=&\frac{1}{2} \left(-\frac{g}{2 \pi a^2}\right)^3\mathrm{e}^{-3G(0)}(2 \pi)^2 \sum_{\alpha\beta} \phantom{}^{'} \bigg\{\notag\\
& \frac{b_1-8b_2+12a_1^2}{c^4m^4}\int\dif^2x \cos(\theta_\alpha-\theta_\beta)\notag\\ &+\frac{b_3}{2c^6m^6}\int\dif^2x \left(\nabla_x\theta_\alpha-\nabla_x\theta_\beta\right)^2
+\ldots
\bigg\}
\end{align}
where
\begin{align}
\label{b1}
b_1=&\frac{c^4m^4}{(2 \pi)^2} \int\dif^2 x\dif^2 y B(x+y,x,y,-2,2,2),\\
\label{b2}
b_2=&\frac{c^4m^4}{(2 \pi)^2} \int\dif^2 x\dif^2 y B(x,y,x+y,2,1,-1),\\
\label{b3}
b_3=&\frac{c^6m^6}{(2 \pi)^2} \int\dif^2 x\dif^2 y x^2B(x,x+y,y,1,1,1).
\end{align}
are dimensionless integrals. We emphasize here that in the above expressions (\ref{Gamma2-eval}) and (\ref{Gamma3-eval}) we already take into account the replica limit $n\to 0$. Had we kept $n$ we would have obtained $2(2-n)a_1$ instead of $4a_1$ in (\ref{Gamma2-eval}) and $b_1+4(n-2)b_2+2(n-2)(n-3)a_1^2$ instead of $b_1-8b_2+12a_1^2$ in (\ref{Gamma3-eval}). We should also mention that the term $a_1^2$ in (\ref{Gamma2-eval}) is basically the second-order term $\int\dif^2x\dif^2y B(x,y,0,1,1,0)$ that could be written as repetition of (\ref{a1}) term.

The final expression for the effective action reads
\begin{align}\label{Gamma-finaldiv}
\Gamma=&\sum_{\alpha\beta}\int\dif^2x\Big\{ \frac{\kappa}{2T}\delta_{\alpha\beta}\left[ (\nabla_x\theta_\alpha)^2+m^2(\theta_\alpha)^2\right]\notag\\
&-\frac{\sigma_R}{4 \pi}\nabla_x\theta_\alpha \cdot\nabla_x\theta_\beta-\frac{g_R}{2 \pi} c^2m^2\cos (\theta_\alpha-\theta_\beta)\Big\}
\end{align}
where
\begin{align}
\label{gR}
g_R=&\widetilde{g} -2a_1\widetilde{g}^2+\frac{1}{2}(b_1-8b_2+12a_1^2)\widetilde{g}^3,\\
\label{sigmaR}
\sigma_R=&\sigma+ \frac{a_2}{2} \widetilde{g}^2 -b_3\widetilde{g}^3,
\end{align}
with $\widetilde{g}=g\mathrm{e}^{-G(0)}/(cma)^2$. In the following it will be useful to have the inverse expression of (\ref{gR}) which reads
\begin{align}\label{gtilde}
\widetilde{g}=g_R+2a_1 g_R^2 - \frac{1}{2}(b_1- 8b_2 - 4 a_1^2)g_R^3.
\end{align}
In (\ref{Gamma-finaldiv}) we have returned to the unrestricted replica sum, since $\sum_{\alpha\beta}'(\nabla_x\theta_\alpha -\nabla_x\theta_\beta )^2=\sum_{\alpha\beta}(\nabla_x\theta_\alpha-\nabla_x\theta_\beta)^2$, while the cosine term (\ref{H1}) differs from the corresponding cosine in (\ref{Gamma-finaldiv}) by $\mathcal{O}(n)$ which goes to zero.

\subsection{Beta functions of the model}

In this subsection we obtain the general form of the beta function of the model (\ref{H}) in terms of the integrals $a_i,b_i$ defined in (\ref{a1}), (\ref{a2}), and (\ref{b1})-(\ref{b3}).
We are generally interested in the flow of the effective action when the cutoff is varied, and the
beta functions describe the flow of the terms (i.e.,~operators) which become relevant at the transition. Here they are thus defined by computing derivatives of $g_R$ and $\sigma_R$ with respect to the cutoff for a fixed microscopic model, i.e.,~keeping $g$ and $\sigma$ fixed, as a function of $g_R$ itself:
\begin{align}
&- m \partial_m g_R = \beta_g(g_R), \\
&- m \partial_m \sigma_R = \beta_\sigma(g_R).
\end{align}
We thus obtain from (\ref{Gamma-finaldiv}), (\ref{gR}), (\ref{sigmaR}):
\begin{align}\label{betag}
\beta_g(g_R) =&2\tau g_R-2(2a_1\tau-m\partial_m a_1)g_R^2\notag\\
&+\frac{1}{2}\big[4\tau(b_1-8b_2+4a_1^2)\notag\\
&- m\partial_m(b_1-8b_2+4a_1^2)\big]g_R^3 + \mathcal{O}(g_R^4) , \\
\label{betasigma}
\beta_\sigma(g_R) =&\left(2\tau a_2-\frac{1}{2}m\partial_m a_2\right)g_R^2\notag\\
&+\left[2\tau(4a_1 a_2-3b_3)-2a_1m\partial_m a_2+m\partial_m b_3\right]g_R^3\notag \\
& + \mathcal{O}(g_R^4)
\end{align}
When deriving the last two expressions we used that in the limit $m a \to 0$ one can replace \footnote{To be more precise, for the choice (\ref{G}) we have $- m \partial_m \ln \tilde g = 2 \tau_a$ with $\tau_a = \tau - \frac{1}{4} (1-\tau) m^2 a^2 [\ln(m^2 c^2 a^2)-1] + \mathcal{O}(m^4 a^4)$. One easily checks that this amount to replace $\tau$ with $\tau_a$ in the above equations, hence in the
limit $m a \to 0$ these terms have no effect on the beta functions.}
$\widetilde{g}=g(cma)^{-2\tau}$ which gives $m\partial_m\widetilde{g}=-2\tau\widetilde{g}$. We have also used the inverse relation (\ref{gtilde}).

Eqs.~(\ref{betag}) and (\ref{betasigma}) should be understood as an expansion of beta functions in powers of $g_R$ where the first three powers are taken into account. Anticipating that the fixed point is $g_R \sim \mathcal{O}(\tau)$ for $\tau>0$ we thus need to compute the coefficients of $g_R^2$ to $\mathcal{O}(\tau)$ and the ones of $g_R^3$ to order $\mathcal{O}(\tau^0)$ to study this equation in the vicinity of the fixed point consistently to the desired order in $\tau$.

\subsection{Universality and the beta function}\label{sec:universality}

From the above considerations we can thus surmise, and will check below by explicit calculation, that our
beta functions have the form
\begin{align}
\label{betagori}
&\beta_g(g_R) =2\tau g_R - Ag_R^2 - B\tau g_R^2 + Cg_R^3,\\
\label{betasigmaori}
&\beta_\sigma(g_R) =Dg_R^2+E\tau g_R^2 - Fg_R^3,
\end{align}
where the constants $A,B,C,D,E$, and $F$ are for now undetermined, and computed below.
We can already ask what is the amount of
universality in these coefficients. One way to address it is to allow for a class of changes in definitions of the renormalized parameters $\sigma_R$ and $g_R$ such that the new parameters ${\sigma}_R'$ and ${g}_R'$ are expressed in terms of the old ones as
\begin{align}
\label{gnew}
&g_R'=G g_R+ H \tau g_R+ I g_R^2+\ldots,\\
\label{sigmanew}
&\sigma_R'=\sigma_R + K g_R^2+\ldots,
\end{align}
where $\ldots$ stands for higher-order terms that do not interfere with beta functions to third order that we are considering. Note that a change of the scale of $\sigma_R$ is not permitted since it occurs in the quadratic part, hence is an observable. Although one can always consider a broader class of changes, this one is broad enough to account for changes in definitions of the small- and large-scale cutoff and cutoff functions while keeping the structure of (\ref{betagori}) unchanged. In particular, the coefficients in Eq. (\ref{betag}) and (\ref{betasigma}) contain some dependence on the details of the cutoff function, through the values of the integrals $a_i$ and $b_i$.

One easily finds that the beta functions for the new variables is the same as for the old ones (\ref{betagori}) with the change
$A \to A/G$, $C \to C/G^2$, $D \to D/G^2$, and
\begin{align}
&B \to \frac{B}{G} - \frac{A H + 2 I}{G^2},\quad E \to \frac{E}{G^2} - \frac{2 D H+ 4 K G}{G^3},\notag\\
&F \to \frac{F}{G^3} -  \frac{2 D I- 2 A G K}{G^4}.
\end{align}

Hence we find that there are the following three invariant combinations, which we define as:
\begin{align} \label{univcomb}
{\cal  D} = \frac{4 D}{A^2}, \quad \quad {\cal  C} = \frac{4 C}{A^2},\quad {\cal  I} = 8 \frac{F + B D - \frac{1}{2} A E}{A^3}.
\end{align}

To see what they mean we consider the value of the fixed point $\beta_g(g^*_R)=0$ of (\ref{betagori}):
\begin{align}
g^*_R = \frac{2}{A} \tau + \frac{2 (2 C-A B)}{A^3} \tau^2 + \mathcal{O}(\tau^3)
\end{align}
a value which is not universal. Then we find that one invariant combination is related to the correction-to-scaling exponent $\omega$, by definition \cite{ItzyksonDrouffe}:
\begin{align} \label{correctionscaling}
- \omega = \beta_g'(g^*_R) = - 2 \tau + {\cal C} \tau^2  + \mathcal{O}(\tau^3).
\end{align}

The other two invariant combinations enter into the expansion of
\begin{align}
\beta_\sigma(g^*_R) = {\cal D} \tau^2 + ( {\cal C} {\cal D}  - {\cal I}  )  \tau^3  + \mathcal{O}(\tau^4)
\end{align}
which will turn out to be related to the amplitude of the squared-logarithm, and are universal.

We now turn to the explicit calculation of the coefficients of the beta functions, and of their universal combinations.

\section{Evaluation of beta functions}

In this section we calculate the coefficients in the beta functions. The integrals defined in
(\ref{a1}), (\ref{a2}), (\ref{b1}), (\ref{b2}), and (\ref{b3}) are dimensionless numbers of the form:
\begin{align}
a_i = a_i(\tau, m a), \quad b_i = b_i(\tau, m a),
\end{align}
i.e.,~they depend only on $\tau$ and on the dimensionless ratio $m a$. This is easy to see
by the rescaling $y \to y/(m c)$ which means they can be computed setting $m=c=1$ and replacing $a^2 \to a^2 m^2 c^2$.
Note that for these
integrals one has $- m \partial_m = - a \partial_a$.

Hence we will now consider two alternative approaches. In the first one (finite ultraviolet-cutoff method) we keep finite $a$ in the propagator and evaluate the divergent parts of the integrals in an expansion in powers of $\tau$. In the second approach, close in spirit to the dimensional regularization method, we start at fixed $\tau>0$ in the glass phase. In that case we find that these integrals are ultraviolet convergent, hence one can set $a=0$ and compute these integrals directly in the continuum limit for $\tau>0$:
\begin{align}
a_i = a_i(\tau) = a_i(\tau, 0), \quad b_i = b_i(\tau) = b_i(\tau,0).
\end{align}
The divergent nature of this integrals then implies that they admit a Laurent series expansion, i.e.,~a pole expansion around $\tau=0$.

The renormalizability of the theory manifests itself by the fact that the coefficients of the above beta functions [e.g.~(\ref{betag}) and (\ref{betasigma})] will be found to be finite in the limit $m a \to 0$ and in the vicinity of the transition $\tau=0$, order by order in an expansion in powers of $\tau$.

\subsection{Finite-$a$ method}

For $\tau=0$ all the integrals (\ref{a1}), (\ref{a2}), (\ref{b1}), (\ref{b2}), and (\ref{b3}) contain divergencies when $a\to 0$ due to (\ref{G-smallx}). Our aim will be to calculate the coefficients close to the transition temperature $T=T_c$ (i.e.,~$\tau=0$) in the form of a $\tau$-expansion. Inspection of (\ref{betag}) and (\ref{betasigma}) shows that to obtain the beta functions up to cubic terms we need the term $a_1$ and $a_2$ evaluated to $\mathcal{O}(\tau)$ accuracy, while for the remaining terms it is sufficient to consider the limit $\tau=0$. The detailed procedure for evaluation of integrals is given in appendix \ref{appendix:finitea}. Here we only state the important results. We mention that a similar method is used for the SG model by \citet{Amit+80}.

The coefficient $a_1$ reads
\begin{align}\label{a1result}
a_1=-\frac{1}{4}\left\{2\lambda
+\tau\lambda^2+c_1\left[1+\mathcal{O}(\tau) \right]+\mathcal{O}(\tau^2)\right\},
\end{align}
where the constant $c_1$ is defined in (\ref{defc1}) and depends on the detailed form of the cutoff function. It is thus non-universal. It is important to keep that constant only for RG equations beyond $g_R^2$. We thus keep it here for our two loop calculation and will check that it does not enter the final result. We introduced the abbreviation
\begin{align}\label{lambda}
\lambda=\ln(c^2m^2a^2).
\end{align}
The other coefficient $a_2$ is
\begin{align}\label{a2result}
a_2=& -\frac{\lambda}{2}-\tau \lambda-\frac{\tau \lambda^2}{2}+c_2\left[1+\mathcal{O}(\tau)\right]+\mathcal{O}(\tau^2),
\end{align}
where the other non-universal constant is $c_2$ defined in (\ref{defc2}).

The evaluation of two-loop integrals $b_1$ and $b_2$ is somewhat complicated. The final results are given in Eqs.~(\ref{b1result}) and (\ref{b2result}). The important combination that appears in the beta functions now reads
\begin{align}\label{b1-8b2}
b_1-8b_2+4a_1^2=-\lambda-c_3+\mathcal{O}(\tau).
\end{align}
There are two important things to be mentioned. The first one is that the $\lambda^2$ divergence from all summands in (\ref{b1-8b2}) vanishes when combined. That is important for the renormalizability of the theory and leads to a finite beta function. The other point is that the non-universal term $c_1\lambda$ that appears in $b_2$ and $a_1^2$ is canceled in the combination, leaving the universal coefficient in front of $g_R^3$ in the first beta function (\ref{betag}).

The last two-loop integral reads
\begin{align}
b_3=\frac{1}{4} \left[\lambda^2-2(2c_2+1)\lambda+c_4\right],
\end{align}
with some constant $c_4$. Using $m\partial_m\lambda=2$ we finally obtain the beta functions
\begin{align}\label{betag-finitea}
&\beta_g(g)=2\tau g-2 g^2+c_1\tau g^2+g^3,\\
\label{betasigma-finitea}
&\beta_\sigma(g)=\frac{1}{2} g^2+(1+2c_2)\tau g^2-\frac{1}{2} (c_1+4c_2+2)g^3.
\end{align}
The beta functions as obtained here are independent of the details of the chosen function $G(x)$ to lowest one-loop order [the first two terms in (\ref{betag-finitea}) and the first one in (\ref{betasigma-finitea})] but not to the next two-loop order. The apparent one-loop universality is only due to our fixed choice of the definition of $g$ in terms of the effective action, which is an observable. It is spoiled by any change of scale in $g$ [coefficient $G$ in (\ref{gnew})], or any other change in definition of $g$.

One thus finds $A=2$, $C=1$, and $D=1/2$ and the universal combinations defined in (\ref{univcomb}) are
\begin{align} \label{univer}
&& {\cal D} = 1/2, \quad {\cal C} = 1,\quad{\cal I} = F + \frac{1}{2} B  -  E = 0.
\end{align}

\subsection{Dimensional method}

In this subsection we calculate the integrals by the dimensional method. As discussed above, $\lambda$ divergencies obtained using the finite-$a$ method now become poles in $\tau$. In addition all integrals in the present method become $m$ independent, so only the terms that do not involve $m\partial_m$ in beta functions (\ref{betag}) and (\ref{betasigma}) survive.
Hence, from (\ref{betag}) and (\ref{betasigma}) we see that the constants in the coefficients of the beta functions (\ref{betagori}) and (\ref{betasigmaori}) are
determined from the integrals $a_i$ and $b_i$ as follows.

The pole and finite part of the one-loop integrals determine the four coefficients:
\begin{align}\label{a1dim}
&4 a_1 = \frac{A}{\tau} + B + \mathcal{O}(\tau), \\
\label{a2dim}
& 2 a_2 = \frac{D}{\tau} + E + \mathcal{O}(\tau).
\end{align}
The details of the calculation are presented in appendix \ref{appendix:dimensionalmethod}, with the result
\begin{align}
& A= 2, \quad B = - 4 c'_1, \\
& D = 1/2, \quad E = - 2 c'_2, \label{Dhalf}
\end{align}
where $c_1'$ and $c_2'$ are again two non-universal constants
defined in (\ref{a1-dimensional}) and (\ref{a2-dimensional}).

The two-loop integrals come in two combinations which determine $C$ and $F$ as:
\begin{align}
& b_1 - 8 b_2 + 4 a_1^2 = \frac{1}{2} \frac{C}{\tau} + \mathcal{O}(\tau^0) \label{firstcomb}, \\
& 6 b_3 - 8 a_1 a_2 = \frac{F}{\tau} + \mathcal{O}(\tau^0). \label{secondcomb}
\end{align}
From appendix \ref{appendix:dimensionalmethod} both $1/\tau^2$ poles and some non-universal
$1/\tau$ terms cancel in the first combination, which is found to have the form (\ref{firstcomb})
with
\begin{align}
C = 1.
\end{align}
The third two-loop integral is
\begin{align}
 b_3= \frac{1}{6\tau^2}-\frac{c_2'}{\tau} +\mathcal{O}(\tau^0).
\end{align}
Forming the second combination (\ref{secondcomb}) one finds a cancellation of the $1/\tau^2$ pole
and
\begin{align}
F = 2  (c'_1-c'_2).
\end{align}
In summary, one finds the following beta functions
\begin{align}\label{betag-dimensional}
&\beta_g(g) =2\tau g-2 g^2+4 c_1' g^2\tau+ g^3,\\
\label{betasigma-dimensional}
&\beta_\sigma(g) =\frac{1}{2} g^2-2 c_2' g^2\tau+2(c_2'-c_1')g^3.
\end{align}
Note that the coefficients $A=2$ and $D=1/2$ needed for the lowest one-loop-order calculation
are the same as obtained above in (\ref{betag-finitea}) by a different scheme. Again, this is because the
definition of the coupling constant $g_R$ is the same in both cases and fixed by the effective action.
Although the other coefficients, needed for the next-order calculation (two-loop) are not the same,
one can check that the universal invariants ${\cal D}$, ${\cal C}$, and ${\cal I}$ yield the same values as in (\ref{univer}). The two calculations are thus consistent.

For sake of completeness we have performed two additional calculations, both using dimensional regularization. In appendix \ref{moreontwoloop} we have independently confirmed the calculation of the present paragraph (and of appendix \ref{appendix:dimensionalmethod})
for a specific choice of the cutoff function $G(x)$ which allows explicit calculation of all coefficients. In section \ref{sec:OPE} and appendices \ref{s:2-loop-OPE} and \ref{confmap} we have done a different calculation using the operator product expansion and a different cutoff scheme. The beta functions obtained there lead again to $A=2$, $D=1/2$, and $C=1$ and now $B=E=F=0$ which leads again to the same values as in (\ref{univer}) for the three invariants. All four calculations are thus consistent.

\section{Renormalization group equations and correlation function: results}

Let us summarize what has been achieved. First we have established the following RG equations, in terms of the scale
$\ell=-\ln m$:
\begin{align}
\label{RGtau}
&\frac{\dif\tau}{\dif\ell}=0, \\
\label{RGgfinal}
&\frac{\dif g_R}{\dif\ell}=2\tau g_R - 2 g_R^2-B\tau g_R ^2+ g_R^3,\\
\label{RGsigmafinal}
&\frac{\dif\sigma_R}{\dif\ell}=\frac{1}{2} g_R^2+E \tau g_R^2-F g_R^3,
\end{align}
where we recall that $\tau=1-T/T_c$. Although the more general expression are (\ref{betagori}) and (\ref{betasigmaori}), from now on, we use $A=2$, $D=1/2$ which, as discussed above, has been obtained
in several schemes. Here $B,E$, and $F$ are non-universal constant which we found satisfy
\begin{align}\label{inv}
{\cal I} = F + \frac{1}{2} B  -  E = 0.
\end{align}
These equations generalize to next order (two loop) the one-loop equations obtained in Refs.~\cite{Cardy+82,Toner+90,Hwa+94,Carpentier+97}. We have also clarified their universality to next order in section \ref{sec:universality}. The first equation (\ref{RGtau}) encodes the exact result (\ref{sts}) from STS. The first two equations
show that the model has a transition at $T=T_c$. For $T > T_c$ the renormalized coupling $g_R(\ell)$ flows to zero, while for $T<T_c$ it flows to a finite value $g_R^*$ which continuously depends on $\tau$:
\begin{align}
\label{g*}
g_R^*=\tau +\frac{1}{2} (1-B) \tau^2+\mathcal{O}(\tau^3),
\end{align}
i.e.,~the line of nonzero fixed points, shown in Fig.~\ref{Fig:flow-co}, is here computed to next order. Its precise value however is non-universal. What is universal however is its attractive character for $T<T_c$, together with the value of the leading attractive eigenvalue $- \omega$ (in the effective-action functional space) which defines the correction-to-scaling exponent (\ref{correctionscaling}):
\begin{align}\label{scalingexponent}
\omega =  - 2 \tau +  \tau^2  + \mathcal{O}(\tau^3)
\end{align}
for $\tau>0$, while it is $\omega = 2 \tau$ in the high temperature phase for $\tau<0$ in the vicinity of $T_c$.
As an example of application, we can expect that the dimensionless susceptibility fluctuation ratio
computed to first order in $\tau$ in Ref.~\cite{Hwa+94} will exhibit a $L^{-\omega}$ finite size correction
as a function of system size $L$.

We now consider the RG equation for $\sigma_R$. From it we can obtain the value of the universal amplitude
${\cal A}$ of the squared logarithm, by a simple but non-rigorous argument, as was done in \cite{Toner+90}.
Indeed the asymptotic solution of (\ref{RGsigmafinal}) is
\begin{align}
\sigma_R(\ell) \simeq \sigma_0 + \beta_\sigma(g^*_R) \ell
\end{align}
where $\sigma_0$ depends on all details of the initial condition and is unimportant as it leads only to a subdominant
single logarithmic growth. To estimate the off-diagonal correlation at a given wave-vector $q$, one may consider the limit of
small mass $m \ll q$ and argue that $q$ itself sets the scale $\ell^*= \ln[1/(aq)]$ at which one should stop the RG. At the same time one replaces $\sigma$ by its effective value at that scale, i.e., $\sigma \to \sigma_R(\ell)$. Hence from (\ref{G-fourier}) one writes:
\begin{align} \label{appro}
 {\cal G}_0(q)&\simeq 8 \pi (1-\tau)^2 \frac{\sigma(\ell^*)}{q^2} |_{\ell^* = \ln[1/(q a)]} \notag\\
 &\simeq_{q \to 0} 8 \pi (1-\tau)^2 \beta_\sigma(g^*_R) \frac{\ln[1/(q a)]}{q^2}.
\end{align}
We can now compute the variance of the phase fluctuations as
\begin{align}  \label{thetaG}
\overline{\langle\theta(x) - \theta(0)\rangle^2}
= 2 \int \frac{d^2 q}{(2 \pi)^2} (1-\cos q x) {\cal G}_0(q).
\end{align}
Using (\ref{appro}) and the following estimate of the momentum integral:
\begin{align}
& \int \frac{\dif^2 q}{(2 \pi)^2} (1-\cos q x) \frac{\ln[1/(q a)]}{q^2} \notag\\
& \simeq \int_0^{1/a} \frac{\dif q}{2 \pi} [1-J_0(q x)] \frac{\ln[1/(q a)]}{q} \notag\\
& = \frac{1}{4 \pi} \ln^2(x/a) + \mathcal{O}[\ln(x/a)]
\end{align}
we obtain the leading squared-logarithmic behavior
\begin{align}\label{correlationfunctionfinal}
\overline{\langle\theta(x) - \theta(0)\rangle^2}
= {\cal A} \ln^2 (x/a) + \mathcal{O}[\ln(x/a)].
\end{align}
The amplitude in the above equation is
\begin{align}
{\cal A} = & 4 (1- \tau)^2 \beta_\sigma(g^*_R)\notag\\
 =& 4 (1-\tau)^2 \tau^2 [ {\cal D} +  ( {\cal C} {\cal D}  - {\cal I}  )  \tau  + \mathcal{O}(\tau^2) ]
\end{align}
and using the values of the invariants computed above (\ref{univer}) we obtain our main result
\begin{align} \label{final result}
{\cal A} = 2 \tau^2 -2\tau^3 + \mathcal{O}(\tau^4).
\end{align}
In the following section we present a calculation of the correlation function from first principles which
confirms and complement the above more qualitative argument.

Before we do so, two comments are in order. The correlation function $\overline{\langle\left[\theta(x)-\theta(0)\right]^2\rangle}$ in the leading order has the same behavior as the one in (\ref{correlationfunctionfinal}). Their difference is the thermal correlation function [c.f.~(\ref{calG})] that is a logarithm for the present model due to STS. Another comment is
about the region above the critical temperature $T>T_c$. There
$\sigma_R(\ell)$ saturates to a finite value at large scales. This leads to simple logarithmic growth of the off-diagonal disorder averaged phase fluctuations (\ref{thetaG}) with a non-universal prefactor.

\section{Explicit calculation of the two-point function}

As is well known, the correlation function can be obtained from the inverse of the quadratic part of the effective action, i.e.,~in Fourier space
\begin{align}
{\cal G}_{\alpha \beta}(q) = \Gamma^{-1}_{\alpha \beta}(q).
\end{align}
The inversion is in replica space and $\Gamma_{\alpha \beta}(q)$ is defined by expanding the effective action (\ref{Gamma-def}) into powers of the fields to quadratic order, i.e.,
\begin{align}
\Gamma=\frac{1}{2}\sum_{\alpha\beta}\int\frac{\dif^2q}{(2\pi)^2} \Gamma_{\alpha\beta}(q)\theta_\alpha(q)\theta_\beta(-q)+
\mathcal{O}(\theta^4)
\end{align}
up to a constant (which encodes for the fluctuations of the free energy). It is convenient to decompose the replica matrix as
\begin{align}\label{gamma-decom}
\Gamma_{\alpha \beta}(q) = \Gamma_c(q) \delta_{\alpha \beta} + \Gamma(q).
\end{align}
Because of the STS we expect that
\begin{align}
\Gamma_c(q) = \frac{1}{G(q)} = \frac{1}{4\pi(1-\tau)}(q^2+m^2),
\end{align}
and one can indeed check explicitly on the expressions (\ref{Gamma1}), (\ref{Gamma2}) and (\ref{Gamma3'})
that there are no corrections to any order in $g$ to $\Gamma_c(q)$. This leads to the exact formula (\ref{sts})
for the disorder average of the thermal correlation.

We now turn to the off-diagonal part $\Gamma(q)$ which gives the second correlation defined in (\ref{calG}), in Fourier space
\begin{align}\label{GGamma0}
{\cal G}_0(q) = - G(q)^2 \Gamma(q) = - (1-\tau)^2 \frac{(4 \pi)^2}{(q^2 + m^2)^2} \Gamma(q)
\end{align}
from replica matrix inversion of (\ref{gamma-decom}) in the replica limit $n\to0$.\footnote{For the matrix $M_{\alpha\beta}=M_1\delta_{\alpha\beta}+M_2$ we have used the following inversion formula, valid for $n\to 0$, $(M_{\alpha\beta})^{-1}=\delta_{\alpha\beta}/M_1-M_2/M_1^2$.}
We now examine its perturbative expansion with respect to the anharmonic perturbation (\ref{H1}), i.e., as an expansion in $g$, which can be written as a sum
\begin{align}
\Gamma(q)=\sum_{i=0}^\infty\Gamma_{(i)}(q),
\end{align}
where $\Gamma_{(i)}(q)$ is coming from the corresponding term $\Gamma_i$ in (\ref{Gamma-def}).

The lowest-order term in $\Gamma(q)$ is the inverse propagator (\ref{G-fourier})
\begin{align}\label{gamma0res}
\Gamma_{(0)}(q)= -\frac{\sigma}{2 \pi} q^2,
\end{align}
while the next-order term is trivially obtained from (\ref{Gamma1}) and reads
\begin{align} \label{ag1}
\Gamma_{(1)}(q) = -2\frac{g}{2 \pi a^2}\mathrm{e}^{-G(0)}
\end{align}
and it is momentum independent. The first nontrivial term is obtained from $(\ref{Gamma2})$ and reads
\begin{align}\label{Gamma2q}
& \Gamma_{(2)}(q) = -4 \frac{g^2}{(2 \pi)^2 a^4}\mathrm{e}^{-2G(0)}
\int\dif^2 x \Big\{\mathrm{e}^{iq\cdot x}\big[2\sinh G(x)\notag\\
&-\sinh 2G(x)\big]+3-4\cosh G(x)+\cosh 2G(x)\Big\}.
\end{align}
Note that expanding this expression at small $q$ one finds that the coefficient
of $q^2$ does not yield exactly the renormalized $- \sigma_R/(2 \pi)$ as
defined for convenience in (\ref{Gamma-finaldiv}). The difference however is related
to corrections to irrelevant terms in the effective action, and vanishes in the limit of zero mass. Indeed the expansion of the second-order term of the effective action (\ref{Gamma2}) which produces (\ref{Gamma2-eval}) is somewhat different from the expansion (\ref{Gamma2q}). The reason for that is that the second cosine term multiplied by $A(x-y,1)$ from (\ref{Gamma2}) produced the cosine in (\ref{Gamma2-eval}), while the first cosine multiplied by $A(x-y,2)$ from (\ref{Gamma2}) produced the gradient in (\ref{Gamma2-eval}). On the other hand, all terms of (\ref{Gamma2}) contribute to (\ref{Gamma2q}). However, in the limit $m\to 0$ only the term multiplied by $A(x-y,2)$ gives a finite contribution that is evaluated below in (\ref{Gamma2q-eval}), and in that sense the terms (\ref{Gamma2q-eval}) and the gradient term of (\ref{Gamma2-eval}) are similar. The finite mass effects are
studied in appendix \ref{appendix:twopoint-finitem}.

The previous expressions can be evaluated in presence of a small-scale cutoff $a$ in the limit $m\to 0$.
Using that $\mathrm{e}^{-G(0)}=(m c a)^{2(1-\tau)}$ one finds from (\ref{ag1}) that
\begin{align} \label{gammaq1res}
\Gamma_{(1)}(q)=0.
\end{align}
To evaluate (\ref{Gamma2q}) in the limit $m \to 0$ (and $\tau$ small) we note that there is a factor
$\mathrm{e}^{-2G(0)} \sim m^{4(1-\tau)}$ in front and
the integrand can be split in a sum of terms of the form $\mathrm{e}^{p G}-1$. For $p=-2,-1$ one can rescale $x \to x/m$ and the corresponding integrals are convergent in the limit $m \to 0$, hence the original integrals are bounded by $\mathcal{O}(1/m^2)$. For $p=1$ (and $\tau=0$) the same holds up to a factor $\ln m a$. Hence we find that all terms except $\mathrm{e}^{2 G(x)}-1$ vanish as $m \to 0$. The limit can be computed by using (\ref{G-smallx}), equivalently written as
\begin{align}
\mathrm{e}^{2 G(x)} \simeq \mathrm{e}^{2 G(0)} (1 + x^2/a^2 )^{-2 (1-\tau)}.
\end{align}
The powers of $m$ (produced by of $\mathrm{e}^{G(0)}$) exactly match hence we are left with:
\begin{align}
\label{Gamma2q-eval}
 \Gamma_{(2)}(q) =& 2 \frac{g^2}{(2 \pi)^2 a^4}
\int\dif^2 x \frac{\mathrm{e}^{iq\cdot x}-1}{ \left(1 + x^2/a^2 \right)^{2(1-\tau)}} \notag \\
 =& - \frac{g^2}{2 \pi a^2}
\left[ \frac{1}{1-2 \tau} - \frac{4^\tau(q a)^{1-2 \tau} K_{-1+2 \tau}(q a) }{\Gamma(2-2\tau)}  \right] \nn \\
=& - \frac{g^2}{2 \pi a^2}\left[1-qaK_1(qa)\right] -2\frac{g^2\tau}{2 \pi a^2}\big\{1+K_0(aq)\notag\\
&+aq[\ln(caq)-1]K_1(aq)\big\}+\mathcal{O}(g^2\tau^2)
\end{align}
Expanding at small $aq$ we find
\begin{align} \label{gammaq2res}
\Gamma_{(2)}(q) =& - \frac{g^2}{2 \pi}
q^2 \ln (aq) \left\{ - \frac{1}{2} + \tau [2(\ln c-1)+\ln(a q)] \right\} \notag \\
&+ \mathcal{O}(g^2\tau^2,g^2 q^2).
\end{align}

The cubic term of the two-point function is obtained from (\ref{Gamma3'}). In the limit $m\to 0$ the only term that survives contains $B(x-y,y-z,z-x,1,1,1)$. This is similar to the simplification which occurs for $\Gamma_{(2)}(q)$ where only relevant terms in the effective action need to be considered in the limit $m \to 0$. After transforming the sum over three different replica indices into sums over unrestricted indices using $\sum_{\alpha\beta\gamma}'f(\alpha,\beta,\gamma)= \sum_{\alpha\beta\gamma}f(\alpha,\beta,\gamma)-2\sum_\alpha f(\alpha,\alpha,\alpha)-\sum_{\alpha\beta}[f(\alpha,\alpha,\beta) +f(\alpha,\beta,\alpha)+f(\alpha,\beta,\beta)]$, the only term that survives in the replica limit is
\begin{align}\label{Gamma3q}
&\Gamma_{(3)}(q)=2\frac{g^3}{(2 \pi)^3 a^6}\mathrm{e}^{-3G(0)}\sum_{\alpha\beta} \int\dif^2 x\dif^2 y\dif^2 z\notag\\ &B(x-y,y-z,z-x,1,1,1)\theta_\alpha(x) \left[\theta_\beta(x)-\theta_\beta(y)\right]+\ldots
\end{align}
where $\ldots$ stands for all other terms that vanish in the limits $m\to 0,n\to 0$ and for terms that are more than quadratic in the fields. After using (\ref{G-smallx}) in the limit $\tau\to 0$ the previous expression becomes
\begin{align}
&\Gamma_{(3)}(q)=2\frac{g^3}{(2\pi)^3}\sum_{\alpha\beta}\int\dif^2 x\dif^2 yf(x-y)\notag\\
&\quad\quad\quad\quad\times\theta_\alpha(x)\left[\theta_\beta(x)- \theta_\beta(y)\right],\\
&f(x-y)=\int\dif^2 z \frac{[(x-y)^2+a^2]^{-1}}{[(y-z)^2+a^2][(z-x)^2+a^2]}.
\end{align}
Doing a Fourier transform one finally obtains
\begin{align} \label{gamma3qexpr}
&\Gamma_{(3)}(q)=\frac{4 g^3}{(2 \pi)^3} \int\dif^2 x\frac{1-\mathrm{e}^{iq\cdot x}}{x^2+a^2}\frac{1}{a^2}g\left(\frac{x}{a}\right),\\
&g(x)=\frac{\pi \ln\left[1+x^2+\frac{x}{2}(2+x^2) \left(x+\sqrt{x^2+4}\right) \right]}{x\sqrt{x^2+4}}.
\end{align}
After evaluation of the previous integral we find
\begin{align}
\Gamma_{(3)}(q)=\frac{g^3}{2 \pi} q^2 \ln (aq)[2 (\ln c -1)+ \ln(a q)]+ \mathcal{O}(q^2). \label{gammaq3res}
\end{align}
To obtain this result one method is to perform the angular integral in $(\ref{gamma3qexpr})$, to
differentiate twice with respect to $q$ and then to use the following property:
\begin{align}
&\int_{x_0}^\infty \dif x f(q x) \frac{1}{x+..} \ln(x+..)\notag\\
 &=\frac{1}{2} f(0) \ln^2 q+ \ln q \int_0^\infty \dif z f'(z) \ln z + \mathcal{O}(1),
\end{align}
where the $..$ means subdominant terms in the large $x$ limit, $f(z)$ vanishes at infinity and $x_0$ is arbitrary.
This formula is obtained by rescaling $x \to x/q$ followed by a partial integration and an expansion at small $q$. It is then
applied to $f(z) = \frac{\dif^2}{\dif z^2} [1-J_0(z)]$ and gives $(\ref{gammaq3res})$.

We can now add formulas (\ref{gamma0res}), (\ref{gammaq2res}) and (\ref{gammaq3res}) and obtain
\begin{align}\label{gammaqfinal}
\Gamma(q)=-\frac{\sigma(q)q^2}{2\pi}
\end{align}
with
\begin{align}
\sigma(q) =&\sigma[1+\mathcal{O}(g^2)]-\frac{g^2}{2}\ln(aq)-(g^3-g^2\tau)[\ln^2(aq)\notag\\
&+2(\ln c-1)\ln(aq)]+\mathcal{O}(g^3\tau,g^2\tau^2).
\end{align}
Having obtained (\ref{gammaqfinal}) we should reexpress the bare coupling $g$ in terms of the renormalized one $g_R$. One has from (\ref{gtilde})
\begin{align}
g = \tilde g (c m a)^{2 \tau} = g_R (1 + \tau \lambda + 2 a_1 g_R) + \mathcal{O}(g_R^3,\tau^2 g_R, \tau g_R^2).
\end{align}
Since we want $\Gamma(q)$ to order $\tau^3$, considering that $g_R=\mathcal{O}(\tau)$, we obtain
\begin{align}
\sigma(q) =&\sigma[1+\mathcal{O}(g_R^2)]-\frac{1}{2}\ln(aq)g_R^2 (1+2\tau\lambda+4a_1 g_R)\notag\\
&-(g_R^3-g_R^2\tau)[\ln^2(aq)
+2(\ln c-1)\ln(aq)]\notag\\
&+\mathcal{O}(g_R^3\tau,g_R^2\tau^2),
\end{align}
where to this order we can replace $4 a_1 \to - 2 \lambda - c_1$ from (\ref{a1result}). We recall that $\lambda=\ln(c^2 m^2 a^2)$.
Now, the renormalized coupling $g_R$ depends a priori on the product $a m$ and on the bare value $g$. However we know that it
satisfies the flow equation (\ref{betag-finitea}) as a function of $m$. Since here we work in the limit $m \to 0$ it has thus
reached its fixed point $g_R^*$, hence we must set in the above calculation
\begin{align}
g_R = g_R^* = \tau + \frac{1+c_1}{2} \tau^2 + \mathcal{O}(\tau^3)
\end{align}
obtained from (\ref{betag-finitea}).
Remarkably, all non-universal constants cancel and we are left with
\begin{align}\label{last}
\sigma(q) =\sigma[1+\mathcal{O}(\tau^2)]+\frac{1}{2}[\tau^2+\tau^3+\mathcal{O}(\tau^4)] \ln[1/(aq)].
\end{align}
Another remarkable fact is the cancellation of the contribution from $\Gamma_{(3)}(q)$ with the
$\mathcal{O}(\tau)$ part of the $\Gamma_{(2)}(q)$ contribution. Taken together these cancellations
are likely to be equivalent to the vanishing of the invariant ${\cal I}=0$ found in the method using $\sigma_R$. One can notice that (\ref{gamma0res}) is the subdominant contribution in the small-$q$ limit. Stated differently, the presence of the (bare) disorder $h(x)$ in the starting model (\ref{H}), that is characterized by the bare disorder strength $\sigma$ present in (\ref{last}), is not important for the leading low-energy behavior of the effective action. However, the generated disorder $h(x)$ under the RG procedure [contained in terms $\propto \ln(aq)$ in (\ref{last})] determines the behavior of the correlation function.

We can now compute the amplitude ${\cal A}$. From inserting
\begin{align}
\mathcal{G}_0=8\pi(1-\tau)^2\frac{\sigma(q)q^2}{(q^2+m^2)^2}
\end{align}
into (\ref{thetaG}), in the limit $m\to 0$, we recover exactly the same result as (\ref{final result}) for the amplitude.

\section{RG via operator product expansion}
\label{sec:OPE}

\subsection{Operator product expansion as an efficient tool to extract the renormalization constants}\label{s:OPE=efftool}

The operator product expansion (OPE)  is a very efficient tool to extract the RG-functions for renormalizable field theories.
The first to construct a general theory of renormalization
were Bogoliubov and Parasiuk \cite{BogoliubovParasiuk1957}, followed by Hepp \cite{Hepp1966}.
They introduced what since then is called a {\bf R}-operation, which subtracts
the divergences from a given Feynman-diagram, and renders {\em all observables}, as e.g.\ correlations functions finite.
This was done by considering each  ordering of the
distances in the Feynman-integral, the since then so-called
Hepp-sectors, separately.
The {\bf R}-operation can be thought of as an OPE, or Taylor-expansion of
a diagram for all possible ways to contract the points, from which are retained
as counter-terms only the divergent contributions, {\em restricted} to the
sector in which they diverge.  Further it could be shown that the {\bf R}-operation can indeed be interpreted as a multiplicative
renormalization, i.e., to introducing $Z$-factors.
This was most clearly demonstrated by Zimmermann
\cite{Zimmermann1969}, who reformulated the {\bf R}-operation in
terms of forests, i.e., mutually disjoint or included sets.
An equivalent formulation, which in some respects is
technically more convenient uses  nests.
It is this formulation of the proof of perturbative renormalizability,
introduced by Berg\`ere and Lam \cite{BergereLam1975},
which finally has been generalized by David, Duplantier and Guitter
\cite{DDG1,DDG2,DDG3,DDG4}, and Wiese \cite{WieseHabil} to polymerized tethered membranes, and which we will use here.
It is a very generally applicable technique, which correctly treats distributions, and  allows for local as well as multi-local divergences. We state the
{\it general criterion for renormalizability}\cite{WieseHabil}\label{r:theorem:General
criterion for renormalizability}:

A statistical field theory is
perturbatively  renormalizable, if
{\begin{enumerate}
\setlength{\parskip}{-4mm}
\setlength{\partopsep}{-2cm}
\itemindent0pt
\itemsep4mm \parsep0pt
\item[\bf(i)] the theory is renormalizable by power-counting,
\item[\bf(ii)] divergences are short-ranged, i.e., no divergences appear
at finite distances,
\item[\bf(iii)]
the dilation operators commute,
\item[\bf(iv)]
there exist an operator product expansion,
which describes these divergences,
\item[\bf(v)]
the divergences of the  operator product expansion
do not have an accumulation point at dimension 0.
Especially, after subtracting them,
the integrand has to be convergent when the distances are
contracted.
\end{enumerate}
This leads us to the {\it theorem of renormalizability}
\label{r:theorem:Renormalizability}
\smallskip
{\begin{enumerate}
\setlength{\parskip}{-4mm}
\itemindent0pt
\itemsep4mm \parsep0pt
\item[\bf(i)] The renormalized integral
$$
\int_{\vec x_1,\ldots,\vec x_N} {\bf R} \, I(\vec x_1,\ldots,\vec x_N)
$$
is UV-finite at $\tau=0$.

\item[\bf(ii)] The renormalized  integral, which contributes to
the connected expectation value of the observable $O$ at $n$-th order,
$$
{O}_{{\bf R}}^{(n)}(\vec z_1,\ldots,\vec z_m):= \int_{\vec x_1,\ldots,\vec  x_N} {\bf R} \, I^\mathrm{conn}_{O}(\vec x_1,\ldots,\vec x_{N})  \ ,
$$
is UV-finite and IR-finite at $\tau=0$.

\item[\bf(iii)] In perturbation theory, the renormalized expectation value
of an observable is given by
$$
{O}_{{\bf R}}(\vec z_1,\ldots,\vec z_m):= \sum_{n=0}^{\infty} \frac{(-g)^n}{n!}
\,{O}_{{\bf R}}^{(n)}(\vec z_1,\ldots,\vec z_m) \ .
$$

\item[\bf(iv)] The subtraction operation ${\bf R}$ is equivalent to multiplicative
renormalization, i.e., to introducing $Z$-factors in the standard way.
\end{enumerate}
In practice, what these two theorems mean for a theory like the random-phase sine-Gordon model is summarized by the following remarks:\smallskip
{\begin{enumerate}
\setlength{\parskip}{-4mm}
\itemindent0pt
\itemsep4mm \parsep0pt
\item[\bf(i)] The microscopic model can be defined without a microscopic cutoff, as long as one is below the upper critical dimension ($\tau>0$). It may or may not contain a macroscopic cutoff, e.g.\ the system size.
\item[\bf(ii)] The macroscopic (renormalized theory) is defined via perturbation theory. The latter depends on a large-scale cutoff $L$. The choice which is implemented in the above theorem, is to bound all distances which appear in the space-integrals of the perturbative expansion by $L$. If there are strong UV divergences, they are to be treated via finite-part prescription.
\item[\bf(iii)]
Knowledge of the working of the proof of  renormalizability, i.e., usage of the operator product expansion, is a {\em useful} tool to identify and subtract subdivergences. The latter can in general easily be calculated analytically. Since the resulting (subtracted) integral is absolutely convergent, it can be evaluated in an expansion in $\tau$, i.e., for the order we  need at $\tau=0$. While the OPE is {\em useful} to analyze subdivergences, it is   {\em not necessary} to know its working to calculate the needed integrals.
\item[\bf(iv)] The theorem then ensures that with this renormalization all expectation values are finite, for {\em any} cutoff.
\item[\bf(v)] All critical exponents are universal.
\end{enumerate}
For the practical calculation, we rely on the techniques developed in
\cite{WieseDavid1997,DavidWiese1996,WieseDavid1995,WieseHabil}. The
simplest and most pedagogical example of a two-loop calculation in this
scheme is given in
\cite{PinnowWiese2001}. An example at four-loop order can be found in \cite{LudwigWiese2002}.

\subsection{Bare Model and renormalization conventions}

We consider the same bare model as in section \ref{s:model}. However for simplicity here and below
we make the choice $\kappa=1/(2\pi),$ hence $T_c=2$ and
\begin{align}
T=2(1-\tau).
\end{align}
The bare action  thus  is ${S}_{0} = H_0^{\rm rep}/T$ with
\begin{eqnarray}\label{K:a1}
{S}_{0} &=& \frac{1}{2 \pi}\sum_{\alpha} \int_{x}  \frac{\left[\nabla
\theta_{\alpha } (x) \right]^{2} }{2T } - \frac{ g_{0}}{2\pi} \int_{x} \sum_{\alpha  \beta}\phantom{ }^{'} :\! \rme^{i \left[\theta_{\alpha} (x)-\theta_{\beta} (x) \right]} :\! \nn
\end{eqnarray}
where we recall that $\sum\phantom{ }^{'}$ denotes sums where all replica indices take different values.
The notation $ :\rme^{i \theta_{\alpha} (x)}\!  : $ denotes normal ordering;
implicitly a vertex operator depending on a single point as $\rme^{i \theta_{\alpha} (x)}$ is considered normal-ordered, i.e., $\left<\rme^{i \theta_{\alpha} (x)}\right> = 1$.
The correspondence with (\ref{H1}) is that $g_0 \equiv \frac{g}{a^2} \mathrm{e}^{G(0)}$. The free correlation function for $g_0=0$ is
\begin{equation}\label{K:a2}
\left< \theta_{\alpha} (x)\theta_{\beta} (y) \right> = -T\delta_{\alpha
\beta}   \ln |x|\ .
\end{equation}
Noting the partition function as
${ Z} :=\int {\cal D} [\theta]\,
\rme^{{-S}_0}$, the effective energy is ${S} =- \ln {Z}$,   defined by expanding the partition function to a given order in presence of a background field $\theta$, taking the average over the fluctuations around this background field, and re-exponentiating, keeping only the relevant terms near $T_c$. It
is parameterized as
\begin{eqnarray}\label{a4}
{S} &=& \frac{1}{2 \pi}\sum_{\alpha \beta} \int_{x}  \frac{K_{\alpha \beta }\nabla
\theta_{\alpha } (x)  \nabla \theta_{\beta } (x) }{2T }  \nn\\
&& - \frac{g_R L^{T-2}}{2\pi}
\int_{x} \sum_{\alpha \beta}\phantom{ }^{'} :\!  \rme^{i \left[\theta_{\alpha} (x)-\theta_{\beta} (x) \right]} :\!
\\
\frac{K_{\alpha \beta}}{T} &=& \frac{\delta_{\alpha \beta}}T \left[1+ \mathcal{O}(g_{0}) \right] -\sigma_R
 \\ \sigma_R &=&\mathcal{O}(g_{0}^2) \quad , \quad g_R  L^{-2 \tau} =  g_{0} + \mathcal{O}(g_{0}^{2})
\end{eqnarray}
We now compute the renormalized couplings $g_R$ and $\sigma_R$ in perturbation
theory of the bare coupling $g_0$. We are working for $\tau>0$ and in the continuum limit, taking $a\to 0$.

\subsection{One-loop diagrams}
\subsubsection{First diagram (one loop)}
We use the graphical notation
\begin{equation}\label{a6}
{_{\alpha }}\!\!\diagram{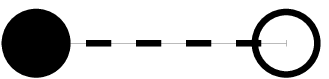}\!\!{_{\beta}} = \rme^{i[ \theta_{\alpha} (x) -\theta_{\beta} (x)]}\ .
\end{equation}
Further, an  ellipse will enclose same-replica
terms. Here and below we denote $\delta {S}_{i} \equiv \int_x \delta { s}_{i}$
the contributions to the effective energy at 1 loop (\(i=1,2\)) and 2 loop (\(i=3,...,6\)).

The first contribution is
\begin{eqnarray}\label{a7}
-\delta { s}_{1}(x) &=& \diagram{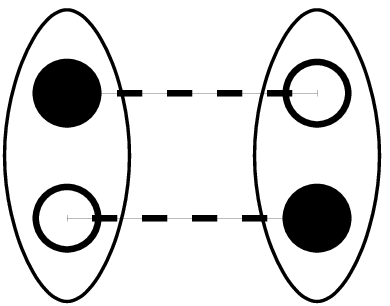}^{x}_{y} \nn\\
&=& \frac{C_1}{2!} \left(
\frac{ g_{0}}{2\pi} \right)^{2}\sum_{\alpha
\beta}\phantom{ }^{'} \int_{y}
:\!  \rme^{i [\theta_{\alpha} (x) -\theta_{\beta} (x) - \theta_{\alpha} (y)
+\theta_{\beta} (y)] }\!:\nn\\
&& \qquad \qquad  \qquad \quad\times  \rme^{-2 T\ln |x-y|}
\end{eqnarray}
where $C_1=1$ and here and below we use ellipse to show which same-replica terms are contracted. The
combinatorial factor is as follows: a factor of $1/2$ from the expansion of $\rme^{-{
H}}$; the factor $C_1=1$ follows from the fact that the second pair of replica sums has exactly one choice for this diagram.

This term contains a strongly divergent contribution to the free
energy (which we do not need) and the important sub-dominant term
\begin{eqnarray}\label{a8}
- \delta { s}_{1}& \approx&  -\frac{1}{4} \left(
\frac{ g_{0}}{2\pi} \right)^{2} \sum_{\alpha  \beta}\phantom{ }^{'}\int_{y} |x-y|^{-2 T  } \times \nn\\
&&
\times :\! \left[(x-y)\cdot \nabla \theta_{\alpha} (x) - (x-y)\cdot \nabla \theta_{\beta} (x) \right]^{2} \!:
  \nn\\
&=&  -\frac{1}{4} \left(
\frac{ g_{0}}{2\pi} \right)^{2} \sum_{\alpha  \beta}\phantom{ }^{'}\int_{y} |x-y|^{2-2 T  } \times \nn\\
&&\times:\! \left[ \nabla \theta_{\alpha} (x)- \nabla \theta_{\beta} (x)  \right]^{2}\!:
\end{eqnarray}
This corrects the quadratic term  (in the
the limit of $n\to 0$) as
\begin{align}\label{a9}
&\frac{\delta K_{\alpha \beta}}T
=-\frac{1}{2}g_{0}^{2}   \times {I}_1 , \\
&{I}_1 = \frac{1}{2\pi} \int \rmd y^{2}
|y|^{2-2T}\Theta(|y|<L) = \frac{L^{4\tau}}{4\tau}.
\qquad
\end{align}
Here and below we are using the prescription
that all distances are bounded by $L$, see remark (ii) in section \ref{s:OPE=efftool} above.
It thus produces the contribution to $\sigma_R$:
\begin{align}
{\delta^{(1)} \sigma_R}=\frac{1}{2}g_{0}^{2}   \times {I}_1.
\end{align}

\subsubsection{Second diagram (one loop)}
\begin{eqnarray}\label{a10}
-\delta {s}_{2}&=&\diagram{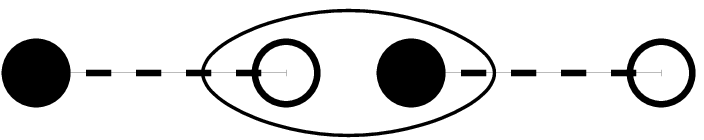} \nn\\
&=& \frac{{C}_2}{2!} \left(
\frac{ g_{0}}{2\pi} \right)^{2}\sum_{\alpha  \beta\gamma}\phantom{ }^{'} \int_{y}  \rme^{i [\theta_{\alpha} (x)- \theta_{\gamma} (y)]}
:\!  \rme^{ - i [\theta_{\beta} (x)
+\theta_{\beta} (y)] }\!: \nn\\
&& \qquad \qquad \qquad\qquad  \times\rme^{- T\ln |x-y|}.
\end{eqnarray}
The combinatorial factor is $C_2=2 $ from the two choices to contract the
ends.

Projecting onto the interaction yields
\begin{equation}\label{a11}
-\delta { s}_{2}\approx  \left(
\frac{ g_{0}}{2\pi} \right)^{2}\times (n-2) \sum_{\alpha  \gamma}\phantom{ }^{'} \rme^{i [\theta_{\alpha} (x)- \theta_{\gamma} (x)]}\,   {I}_2\ ,
\end{equation}
where the basic integral is  \begin{equation}
{I}_2 = \frac{1}{2\pi} \int \rmd^2 y\, |y|^{-T} \Theta(|y|<L)=\frac{L^{2\tau}}{2\tau}\ .
\end{equation}
Therefore
\begin{eqnarray}\label{a12}
 \delta^{(2)} g&=&  (n-2) g_{0}^{2}  \times {I}_2 \ .
\end{eqnarray}
where $n \to 0$ has to be taken at the end.

\subsection{Two-loop diagrams}
Expanding the partition sum to next order one finds the following four 2-loop
diagrams.

\subsubsection{Third diagram (two loops)}\label{s:D3}
The first diagram at 2-loop order  is
\begin{eqnarray}\label{a14}
-\delta { s}_{3} &=& \diagram{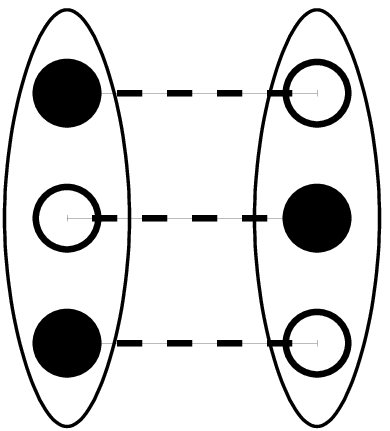}^{x}_{z} \!\!\! {}_{{y}} \\
& =& \frac{C_3}{3!} \times
\left(\frac{g_{0}}{2\pi} \right)^{3}  \sum_{\alpha  \beta}\phantom{ }^{'}  \int_{xyz}:\! \rme^{i \left[\theta_{\alpha} (x) -\theta_{\alpha} (y)
+\theta_{\alpha} (z) \right]} \!: \nn\\
&&\qquad\qquad\qquad \times \;:\, \rme^{-i \left[\theta_{\beta} (x)
-\theta_{\beta} (y) +\theta_{\beta} (z) \right]}\!: \nn\\
&& \qquad\qquad\qquad \times \rme^{-2T\left[\ln
|x-y|+\ln |y-z|-\ln |x-z| \right] }.\nn
\end{eqnarray}
The combinatorial factor is $C_3=3$ from the pattern with one dot within an ellipse differently colored from the rest.
This yields
\begin{eqnarray}\label{a15}
-\delta { s}_{3} &=&  \frac{C_3}{3!} \times
\left(\frac{g_{0}}{2\pi} \right)^{3}  \sum_{\alpha  \beta}\phantom{ }^{'} {I}_3,
\label{a16}\\
\delta^{(3)} g &=& \frac{C_3}{3!} {g_{0}^{3}} {I}_3.
\end{eqnarray}
The  non-trivial integral ${I}_3$ is, setting $y\to 0$:
\begin{equation}\label{I3d}
{I}_{3} =  \frac{1}{( 2\pi )^{2}} \int_{x}\int_{z}\left(\frac{|x-z|}{|x||z|} \right)^{2T} \Theta(|x-z|,|x|,|z|<L).
\end{equation}
We remind that finite-part prescription is used to define this integral.
In appendix \ref{s:I3} we show that
\begin{equation}\label{I3r}
{I}_3 =
\frac{(1-\tau )^2}{2} \frac{L^{4\tau}}{\tau^2}  -{2(1-\tau)}\frac{L^{4\tau}}{4\tau}+ \mathcal{O}(\tau^{0}) \ .\qquad
\end{equation}

\subsubsection{Fourth diagram (two loops)}\label{s:D4}
The fourth diagram is
\begin{eqnarray}\label{a25}
-\delta { s}_{4} &=&{}_{z} \diagram{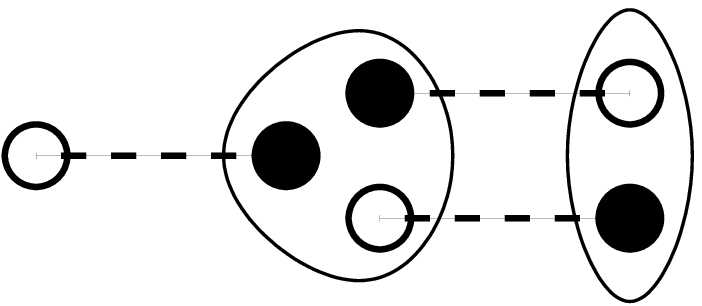}^{x}_{y} \nn\\
&=& \frac{C_4}{3!}
\frac{g_{0}^3}{2\pi} {I}_4 \sum_{\alpha  \beta}\phantom{ }^{'}\cos(\theta_\alpha-\theta_\beta)\\
\delta^{(4)} g &=& \frac{C_4}{3!} {g_{0}^{3}} {I}_4
\end{eqnarray}
The combinatorial factor is $C_4=12(n-2)=3\times 2 \times 2 \times (n-2)$ with a factor \(3\) for choosing the leftmost vertex; 2 for choosing which of its ends to put inside the left ellipse; 2 for choosing the second black dot within this same ellipse from the two possible interactions; then all combinatorial factors are fixed apart from a factor of \(n-2\) for the replica sum in the second ellipse.
Setting $y\to 0$, we find
\begin{equation}\label{I4d}
{I}_{4} =  \frac{1}{( 2\pi )^{2}}
\int_{x}\int_{z}\left(\frac{|x-z|}{|x|^{2} |z|} \right)^{T}\Theta(|x-z|,|x|,|z|<L).
\end{equation}
In appendix \ref{s:I4} we show that
\begin{equation}\label{I4r}
{I}_4 =\frac{(1-\tau )^2 L^{4 \tau }}{16 \tau ^2}
   +\frac{L^{4\tau}}{8\tau^2}-\frac{(1-\tau) L^{4\tau}}{8\tau} + \mathcal{O}(\tau^0)\ .\qquad
\end{equation}

\subsubsection{Fifth diagram (two loops)}\label{s:D5}
\begin{eqnarray}\label{lf1}-\delta { s}_5 &=&\!\!
\stackrel{x\quad \qquad~ y\qquad \quad~ z}{{} \diagram{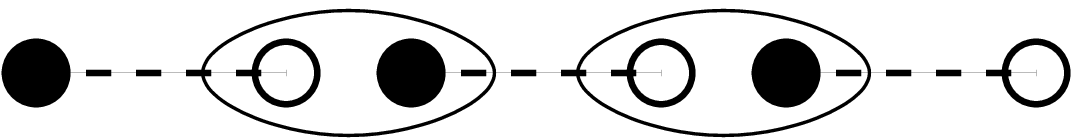}} \nn\\
&=& \frac{C_5}{3!}
\frac{g_{0}^3}{2\pi}   {I}_5  \sum_{\alpha  \beta}\phantom{ }^{'}\cos(\theta_\alpha-\theta_\beta)\\
\delta^{(5)} g &=& \frac{C_5}{3!} {g_{0}^{3}} {I}_5
\end{eqnarray}
The combinatorial factor is $C_5 = 6(n-2)(n-3) = 3\times 2 \times (n-2)\times (n-3)$ with a factor 3 for choosing the middle vertex; 2 for choosing the second white dot in the left ellipse -- then all vertices are placed; \((n-2)\times (n-3)\) for the replica sums within the ellipses.
The integral is
\begin{equation}\label{I5d}
{I}_5=\frac{1}{(2\pi)^2}\int_x\int_z \frac{\Theta(|x|,|z|,|x-z|<L)}{|x|^{T}|z|^T}
\end{equation}
In appendix \ref{s:I5} we show that
\begin{equation}\label{I5r}
{I}_5
=\frac{L^{4\tau}}{4\tau^2}  + \mathcal{O}(\tau^0)
\end{equation}

\subsubsection{Sixth diagram (two loops)}\label{lf2}

${C}_{6}=2$ is the combinatorial factor of this  diagram (corresponding to the 2 choices for ordering the 3 vertices, up to cyclic permutations).
\begin{align}\label{lf3}&-\delta { s}_{6} =
\parbox{2cm}{$\raisebox{13mm}{${}_{x}$}\hspace{-5mm} \stackrel{\diagram{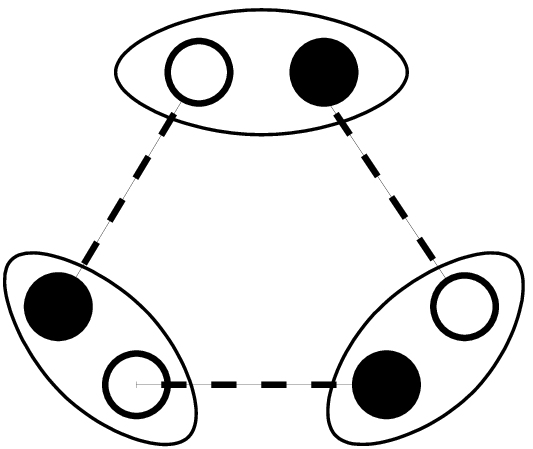}}{\rule{0mm}{1mm}^{\scriptstyle
y}}\hspace{-5.5mm}\raisebox{13mm}{${}_{z}$}$} \nn\\
  &= \frac{g_0^3}{3!}  \frac{{C}_{6}}{(2\pi)^{2}} \sum_{\alpha  \beta\gamma}\phantom{ }^{'}\int_{x} \int_{z}
\frac{1}{|x-y|^{T} |y-z|^{T} |z-x|^{T}}  \nn\\
& \qquad\qquad\quad \times :\! \rme^{i [\theta_{\alpha}
(x)-\theta_{\alpha} (z)]}  \rme^{i [\theta_\beta
(z)-\theta_{\beta } (y)]}  \rme^{i [\theta_\gamma
(y)-\theta_{\gamma } (x)]}  \!: \nn \\
 &=\frac{g_0^3}{3!}    \frac{{
 C}_{6}}{(2\pi)^{2}}  \frac{-1}{2} \sum_{\alpha  \beta\gamma}\phantom{ }^{'}\int_{x} \int_{z}
\frac{1}{|x-y|^{T} |y-z|^{T} |z-x|^{T}} \nn\\
& \qquad\qquad \times :\!
\left[(x-z)\nabla \theta_{\alpha}  + (z-y)\nabla\theta_{\beta} +
(y-x)\nabla \theta_{\gamma} \right]^{2} \!:\nn\\
& \quad  + \dots
\end{align}
In the analysis, let us distinguish between the resulting 1-replica
and 2-replica terms. We start with the 1-replica term:
\begin{align}\label{lf4}
&\left.\parbox{2cm}{$\raisebox{13mm}{${}_{x}$}\hspace{-5mm} \stackrel{\diagram{diag2L4}}{\rule{0mm}{1mm}^{\scriptstyle
y}}\hspace{-5.5mm}\raisebox{13mm}{${}_{z}$}$} \right|_{\mathrm{1~rep}} \nn\\
 &= \frac{g_0^3}{3!}   \frac{ {  C}_{6}}{(2\pi)^{2}}  \frac{-3}{2} \sum_{\alpha\beta\gamma}\phantom{ }^{'}\int_{x} \int_{z}
\frac{1}{|x-y|^{T} |y-z|^{T} |z-x|^{T}}  \nn\\
& \qquad\qquad\qquad\qquad\qquad\quad \times :\!
\left[(x-z)\nabla \theta_{\alpha}\right]^{2} \!: + \dots \nn  \\
&= \frac{g_0^3}{3!}  \frac{ {  C}_{6}}{(2\pi)^{2}}  \frac{-3}{2} \sum_{\alpha  \beta\gamma}\phantom{ }^{'}\int_{x} \int_{z}
\frac{(x-z)^{2}}{|x-y|^{T} |y-z|^{T} |z-x|^{T}}  \nn\\
& \qquad\qquad\qquad\qquad\qquad\quad \times :\!
\frac{1}{2}\left[\nabla \theta_{\alpha}\right]^{2} \!: + \dots \nn
\end{align}
In the last step we did the angular integral (average).
We need to calculate the  following integral:
\begin{equation}\label{I6ad}
{ I}_{6a}= \frac{1}{(2\pi)^{2}}
\int_{x} \int_{z}
\frac{(x-z)^{2}}{|x-y|^{T} |y-z|^{T} |z-x|^{T}}
\end{equation}
The calculation in appendix \ref{s:I6a} gives  \begin{equation}\label{I6ar}
{ I}_{6a}= \frac{L^{6\tau}}{6\tau^2}+\mathcal{O}(\tau^0)\ .
\end{equation}
This yields the correction to $K_{\alpha\beta}$
\begin{equation}-
\frac{\delta K^{(6a)}_{\alpha \beta}}{T} = \frac{ C_6(n-1)(n-2)}{3!} \frac{-3}{2}g_0^3 { I}_{6a}\delta_{\alpha\beta}\ .
\end{equation}
We now consider the term \textit{off-diagonal} in replica-space:
\begin{align}
&\left.
\parbox{2cm}{$\raisebox{13mm}{${}_{x}$}\hspace{-5mm} \stackrel{\diagram{diag2L4}}{\rule{0mm}{1mm}^{\scriptstyle
y}}\hspace{-5.5mm}\raisebox{13mm}{${}_{z}$}$} \right|_{\mathrm{2~rep}}\nn\\
&=\frac{g_0^3}{3!}3 {C}_6  (n-2) \sum_{\alpha  \beta}\phantom{}^{'} \  \frac12:\!
\nabla \theta_{\alpha}\nabla\theta_{\beta}  : \nn\\
& \qquad  \times\frac{1}{(2\pi)^{2}}
\int_{x} \int_{z}
\frac{(x-z)\cdot (y-z)}{|x-y|^{T} |y-z|^{T} |z-x|^{T}}\ .
\end{align}The integral to be calculated is
\begin{equation}\label{I6bd}
{ I}_{6b}= \frac{1}{(2\pi)^{2}}
\int_{x} \int_{z}
\frac{(x-z)\cdot (y-z)}{|x-y|^{T} |y-z|^{T} |z-x|^{T}}\ .
\end{equation}
In appendix \ref{s:I6b} we show that
\begin{equation}\label{I6br}
{ I}_{6b}=\frac{ L^{6 \tau }}{12 \tau ^2}+\mathcal{O}(\tau^0)\ .
\end{equation}
This yields the correction for the off-diagonal term\begin{equation}
-\frac{\delta K^{(6b)}_{\alpha \beta}}{T} =\frac{ C_6(n-2)}{3!} {3}g_0^3 { I}_{6b }(1-\delta_{\alpha\beta}).
\end{equation}
Taking together diagonal and off-diagonal terms yields in the limit of $n\to 0$
\begin{eqnarray}
\frac{\delta K^{(6a)}_{{\alpha\beta}}+\delta K^{(6b)}_{\alpha \beta}}{T} &=& \frac{C_6}{2} g_0^3 \left[\frac12 { I}_{6a}\delta_{\alpha\beta} +  { I}_{6b}(1-\delta_{\alpha\beta})\right]\nn\\
&=& \frac{L^{{6\tau}}}{{6\tau^{2}}} +\mathcal{O}(\tau^0)
\end{eqnarray}
This gives a correction of $\sigma_R$,
\begin{equation}
\delta^{(2)}\sigma_R = -
\frac{g_0^3}{6} \frac{L^{6 \tau }}{ \tau^2} +\mathcal{O}(\tau^0)\ .
\end{equation}
It is an important consistency check that only $\sigma$ gets renormalized, but not the diagonal term (temperature), as necessary due to the statistical tilt symmetry of the problem.

\subsection{Beta functions to two-loop order}\label{K:beta}

Summing all one- and two-loop contributions calculated above, we find
\begin{eqnarray}
g_R L^{-2\tau}&=& g_0 + \delta^{(1)}g+ \delta^{(3)}g +\delta^{(4)}g +\delta^{(5)}g \nn\\
&=& g_0 - g_0^2 \frac{L^{2\tau }}{\tau} + g_0^3 \frac{L^{4\tau }}{\tau^2}  + g_0^3 \frac{L^{4\tau }}{4\tau}  +\mathcal{O}(g_0^4) \qquad  \\
 \sigma_R  &=& {  \delta^{(1)} \sigma} + \delta^{(2)}\sigma=g_{0}^{2}    \frac{L^{4\tau }}{8\tau}
 -g_0^3 \frac{L^{6 \tau }}{6 \tau ^2}\ldots
\end{eqnarray}
Here the beta functions are defined as the variation with respect to the large-scale cutoff $L$, keeping
fixed the bare coupling $g_0$. The result,  reexpressed in terms of $g_R$, is
\begin{eqnarray}
\beta_g(g) &:=& L\frac{\partial }{\partial L} g_R\Big|_{g_0} = 2\tau g_R -2 g_R^2 +g_R^3+\mathcal{O}(g_R^4),\qquad\\
\beta_\sigma(g) &:=& L\frac{\partial }{\partial L} \sigma_R \Big|_{g_0} = \frac 12 g_R^2+\mathcal{O}(g_R^4
).
\end{eqnarray}
Comparing now with the general expression (\ref{betagori}) and (\ref{betasigmaori}) we have obtained the
coefficients:
\begin{align} \label{resope}
& A = 2, \quad B = 0, \quad C = 1, \notag\\
& D =1/2, \quad E = 0, \quad F = 0,
\end{align}
and we can now repeat the analysis of section \ref{sec:effectiveaction}. Remarkably, the universal
invariants in the beta functions (\ref{univcomb}) assume, using (\ref{resope}) the same values (\ref{univer}) as found above.
The various methods, while quite different, are thus mutually consistent and predict the same amplitude
${\cal A}$ as in (\ref{final result}) and finite-size-correction exponent $\omega$ as in (\ref{scalingexponent}).

\section{Conclusions}

In this paper we have reexamined the random-phase sine-Gordon model. We performed a perturbative RG calculation in the vicinity of the glass transition temperature, in a systematic expansion in $\tau=(T-T_c)/T$, to the next order (two loop) than was considered previously (one loop). We used several different RG schemes which yield consistent results. We have obtained the scaling equations of the model, i.e., the beta functions, given by Eqs.~(\ref{RGtau}), (\ref{RGgfinal}), and (\ref{RGsigmafinal}) to next order in the non-linearity. We elucidated the structure of these RG equations which contain several non-universal constants, and two temperature-dependent universal invariants. The first invariant yields the correction-to-scaling exponent (\ref{correctionscaling}) which control, e.g.~the finite-size dependence of the susceptibility fluctuations in the glass phase. We further calculated the correlation function in the low-temperature phase, and found that it has the super-rough squared-logarithm form given by Eq.~(\ref{correlationfunctionfinal}). Its amplitude, which is related to the second invariant in the beta functions, was obtained to be $\mathcal{A}= 2 \tau^2 -2\tau^3 + \mathcal{O}(\tau^4)$. To $\mathcal{O}(\tau^2)$ it agrees with the one-loop result first correctly obtained in \cite{Carpentier+97}. The next order $\mathcal{O}(\tau^3)$ obtained here is in discrepancy with the
prediction $\mathcal{A}= 2 \tau^2 (1- \tau)^2$ obtained in Ref.~\cite{LeDoussal+07} by a simple translation to the RPSG model of the exact results of \cite{Guruswamy+00} based on the fermionic version of the model (\ref{H}). The fact that this latter prediction could not be correct for all $0<\tau<1$, on physical grounds and inconsistency with the zero temperature numerics was pointed out in Ref.~\cite{LeDoussal+07}. Here however we find that the disagreement occurs already at two-loop level. Hence it would be important to perform the RG calculation for the model (\ref{H}) directly in a fermionic language and locate the origin of the discrepancies between our results and the ones from Ref.~\cite{Guruswamy+00} . This goes beyond the scope of this paper and is left for future work.

Another apparent discrepancy can be noted, since a calculation
taking into account corrections of orders $g$ and $g^2$ (which both lead to zero contribution) leads to the following result for the
Edwards-Anderson order parameter \cite{Cardy+82},
\begin{align}\label{EA}
\overline{\langle\mathrm{e}^{i\left[\theta(x)-\theta(0)\right]}\rangle \langle \mathrm{e}^{-i\left[\theta(x)-\theta(0)\right]}\rangle}\propto \left({x}/{a}\right)^{-4(1-\tau)},
\end{align}
which is different from the one quoted in \cite{Guruswamy+00} which has a temperature independent universal decay exponent equal to $-4$. More studies are then called for to clarify the full connection between the model considered in the present paper and the one from \cite{Guruswamy+00}. Note that the predictions  made in this paper have been compared to a numerical simulation, with excellent agreement. The numerical results and the comparison is presented in Ref.~\cite{Perret+12}.

\subsection*{Acknowledgements}

We acknowledge useful discussions with A. Petkovi\'{c}, G. Schehr, J. Troost, and a correspondence with J. Toner. This work is supported by the ANR grant 09-BLAN-0097-01/2.

\appendix

\section{Effective action}\label{appendix:effectiveaction}

In this appendix we derive an analytic formula of the effective action up to third order of perturbation theory. We have not found in the literature a systematic derivation of such an expression apart from the result in second order perturbation theory in Ref.~\cite{Neudecker82}. An advantage with respect to the traditional diagrammatic perturbation theory is that it gives all the terms in a certain order in the coupling constant without need to for determination of multiplicative prefactors for diagrams. Our aim is to calculate the effective action for a theory defined by the reduced action (or Hamiltonian in our case) $S(\varphi)=S_0(\varphi)+gV(\varphi)$, perturbatively in $g$ where $S_0$ is the quadratic part of the action and $V$ some perturbation. Denoting by ${W}(J)$ the generator of connected correlations\cite{Zinn-Justin}
\begin{align}\label{W}
\mathrm{e}^{{W}(J)}=\int\Dif\varphi \mathrm{e}^{-S(\varphi)+J\varphi}.
\end{align}
The effective action is defined as\cite{Zinn-Justin} $\Gamma(\varphi)=J\varphi-{W}(J)$. Using
\begin{align}
J(x)=\frac{\delta \Gamma}{\delta \varphi(x)},
\end{align}
after translating the field $\varphi\to\varphi+\chi$, Eq.~(\ref{W}) eventually becomes
\begin{align}\label{Gamma}
\mathrm{e}^{-\Gamma(\varphi)}=\int\Dif\chi \exp\left[-S(\varphi+\chi)+\int\dif x\chi(x)\frac{\delta \Gamma}{\delta \varphi(x)}\right].
\end{align}
Further we introduce $\widetilde{\Gamma}(\varphi)=\Gamma(\varphi)-S_0(\varphi)+\ln Z_0$, and Eq.~(\ref{Gamma}) is transformed into
\begin{align}\label{Gamma1appen}
\mathrm{e}^{-\widetilde{\Gamma}(\varphi)}=&\frac{1}{Z_0}\int\Dif\chi \exp\bigg[-S_0(\chi)-gV(\varphi+\chi)\notag\\
&+\int\dif x\chi(x)\frac{\delta \widetilde{\Gamma}}{\delta \varphi(x)}\bigg]\notag\\
=&\left\langle\exp\left[\int\dif x\chi(x)\frac{\delta \widetilde{\Gamma}}{\delta \varphi(x)}-gV(\varphi+\chi)\right]\right\rangle.
\end{align}
When deriving Eq.~(\ref{Gamma1appen}) we have used that $S_0$ is a quadratic action, so it satisfies
\begin{align}
S_0(\varphi+\chi)=&S_0(\varphi)+\int\dif x\frac{\delta S_0}{\delta\varphi(x)}\chi(x)\notag\\
&+\frac{1}{2}\int\dif x\dif y\frac{\delta^2 S_0}{\delta\varphi(x)\delta\varphi(y)}\chi(x)\chi(y)\notag\\
=&S_0(\varphi)+\int\dif x\frac{\delta S_0}{\delta\varphi(x)}\chi(x)+S_0(\chi).
\end{align}
Here one should notice that is important to define $\widetilde{\Gamma}$ as difference between $\Gamma$ and $S_0$ in order to avoid further complications when solving the implicit equation for $\widetilde{\Gamma}$.

The next step consists in solving Eq.~(\ref{Gamma1appen}) and extracting $\widetilde{\Gamma}$ out of it. This can be done by iterations. We write $\widetilde{\Gamma}$ in the form
\begin{align}
\widetilde{\Gamma}(\varphi)=\sum_{n=1}^\infty g^n V_n(\varphi)
\end{align}
and use the cumulant expansion
\begin{align}
&\ln\left\langle\mathrm{e}^{A}\right\rangle=\sum_{n=1}^{\infty}\frac{1}{n!} \left\langle A^n\right\rangle_c,
\end{align}
where the first few connected cumulants read\cite{Ma}
\begin{align}
&\left\langle A\right\rangle_c=\left\langle A\right\rangle,\\
&\left\langle A^2\right\rangle_c=\left\langle A^2\right\rangle-\left\langle A\right\rangle^2,\\
&\left\langle A^3\right\rangle_c=\left\langle A^3\right\rangle+2\left\langle A\right\rangle^3-3\left\langle A\right\rangle\left\langle A^2\right\rangle.
\end{align}
Then we easily obtain
\begin{align}
V_1(\varphi)=\langle V(\varphi+\chi)\rangle.
\end{align}
\begin{widetext}
Next, we get
\begin{align}
\langle A^2\rangle_c&=g^2\langle V^2(\varphi+\chi)\rangle_c+\int\dif x\dif y G(x-y)\frac{\delta\widetilde{\Gamma}}{\delta\varphi(x)} \left[\frac{\delta\widetilde{\Gamma}}{\delta\varphi(y)}-2g\frac{\delta\langle V(\varphi+\chi)\rangle}{\delta\varphi(y)}\right]\notag\\
&=g^2\langle V^2(\varphi+\chi)\rangle_c-g^2\int\dif x\dif y G(x-y)\frac{\delta\langle V(\varphi+\chi)\rangle}{\delta\varphi(x)}\frac{\delta\langle V(\varphi+\chi)\rangle}{\delta\varphi(y)}+\mathcal{O}(g^4),
\end{align}
where we used the notation $G(x-y)=G(y-x)=\langle\chi(x)\chi(y)\rangle$ and the identity
\begin{align}
&\langle V(\varphi+\chi)\chi(x)\rangle=\int\dif y G(x-y)\frac{\delta\langle V(\varphi+\chi)\rangle}{\delta\varphi(y)}.
\end{align}
Therefore,
\begin{align}
V_2(\varphi)=-\frac{1}{2}\langle V^2(\varphi+\chi)\rangle_c+\frac{1}{2}\int\dif x\dif y G(x-y)\frac{\delta\langle V(\varphi+\chi)\rangle}{\delta\varphi(x)}\frac{\delta\langle V(\varphi+\chi)\rangle}{\delta\varphi(y)}.
\end{align}
In a similar way we obtain
\begin{align}
V_3(\varphi)=\frac{1}{6}\langle V^3(\varphi+\chi)\rangle_c-\frac{1}{2}\int\dif x\dif y G(x-y)\frac{\delta \langle V(\varphi+\chi)\rangle}{\delta \varphi(x)}\frac{\delta\langle V^2(\varphi+\chi)\rangle_c}{\delta\varphi(y)}\notag\\ +\frac{1}{2}\int\dif x\dif y\dif z\dif t G(x-z)G(y-t)\frac{\delta^2\langle V(\varphi+\chi)\rangle}{\delta\varphi(z)\delta\varphi(t)} \frac{\delta\langle V(\varphi+\chi)\rangle}{\delta\varphi(x)} \frac{\delta\langle V(\varphi+\chi)\rangle}{\delta\varphi(y)},
\end{align}
where we used
\begin{align}
\langle V(\varphi+\chi)\chi(x)\chi(y)\rangle=\langle V(\varphi+\chi)\rangle G(x-y)+\int\dif z\dif t G(x-z)G(y-t)\frac{\delta^2 \langle V(\varphi+\chi)\rangle}{\delta\varphi(z)\delta\varphi(t)}.
\end{align}
Our final expression for the effective action up to third order in the potential $V$ reads
\begin{align}\label{Gammafinal}
\Gamma(\varphi)=&S_0(\varphi)-\ln Z_0+g \langle V(\varphi+\chi)\rangle -\frac{g^2}{2}\langle V^2(\varphi+\chi)\rangle_c+\frac{g^2}{2}\int\dif x\dif y G(x-y)\frac{\delta\langle V(\varphi+\chi)\rangle}{\delta\varphi(x)}\frac{\delta\langle V(\varphi+\chi)\rangle}{\delta\varphi(y)}\notag\\
&+\frac{g^3}{6}\langle V^3(\varphi+\chi)\rangle_c-\frac{g^3}{2}\int\dif x\dif y G(x-y)\frac{\delta \langle V(\varphi+\chi)\rangle}{\delta \varphi(x)}\frac{\delta\langle V^2(\varphi+\chi)\rangle_c}{\delta\varphi(y)}\notag\\ &+\frac{g^3}{2}\int\dif x\dif y\dif z\dif t G(x-z)G(y-t)\frac{\delta^2\langle V(\varphi+\chi)\rangle}{\delta\varphi(z)\delta\varphi(t)} \frac{\delta\langle V(\varphi+\chi)\rangle}{\delta\varphi(x)} \frac{\delta\langle V(\varphi+\chi)\rangle}{\delta\varphi(y)}+\mathcal{O}(g^4).
\end{align}

This formula straightforwardly extends to the case where the fields carry indices, such as replica indices.
In that case the propagator $G$ carries a double index.
These indices can be restored unambiguously using the spatial coordinate of the field by matching the field indices with the propagator ones.

The obtained formula (\ref{Gammafinal}) agrees with the one of Ref.~\cite{Neudecker82} to $\mathcal{O}(g^2)$ terms. (The formula of Ref.~\cite{Neudecker82} is written only to $\mathcal{O}(g^2)$). We have tested (\ref{Gammafinal}) on the sine-Gordon model to two-loop order (i.e.,~including $g^3$ terms) and found agreement with the effective action from \citet{Amit+80}. The final expression (5.1) of \citet{Amit+80} contains a typo: the argument of the last term in the third line of (5.1) should be $y$ instead of $x$.

\section{Transformation of sums}\label{appendix:sums}

When one calculates cumulants of (\ref{H1}) one needs to decompose the two sums over pairs of unequal indices, Eqs.~(\ref{sum4}) and (\ref{sum6}), into sums where all addends are sums over all unequal indices [denoted by superscript $\phantom{}^{'}$, i.e.,~$\sum_{\alpha\neq\beta} g(\alpha,\beta)=\sum_{\alpha\beta}^{'} g(\alpha,\beta)$]. These relations read:

\begin{align}\label{sum4}
\sum_{\alpha\neq\beta\atop\gamma\neq\delta} g(\alpha,\beta,\gamma,\delta)=&\sum_{\alpha\beta}\phantom{}^{'}\big[ g(\alpha,\beta,\alpha,\beta)+g(\alpha,\beta,\beta,\alpha)\big] +\sum_{\alpha\beta\gamma}\phantom{}^{'}\big[g(\alpha,\beta,\alpha,\gamma) +g(\alpha,\beta,\beta,\gamma)+ g(\alpha,\beta,\gamma,\alpha)+g(\alpha,\beta,\gamma,\beta)\big]\notag\\ &+\sum_{\alpha\beta\gamma\delta}\!\phantom{}^{'}g(\alpha,\beta,\gamma,\delta)
\end{align}
and
\begin{align}\label{sum6}
&\sum_{\alpha\neq\beta\atop{\gamma\neq\delta\atop{\mu\neq\nu}}} g(\alpha,\beta,\gamma,\delta,\mu,\nu)=\sum_{\alpha\beta}\phantom{}^{'}g_2 +\sum_{\alpha\beta\gamma}\phantom{}^{'}g_3 +\sum_{\alpha\beta\gamma\delta}\!\phantom{}^{'}g_4
+\sum_{\alpha\beta\gamma\delta\mu}\!\!\!\phantom{}^{'}g_5
+\sum_{\alpha\beta\gamma\delta\mu\nu}\!\!\!\!\phantom{}^{'}g_6
\end{align}
where $g_2=g(\alpha,\beta,\alpha,\beta,\alpha,\beta)+g(\alpha,\beta,\alpha,\beta,\beta,\alpha) +g(\alpha,\beta,\beta,\alpha,\alpha,\beta)+g(\alpha,\beta,\beta,\alpha,\beta,\alpha)$ and $g_6=g(\alpha,\beta,\gamma,\delta,\mu,\nu )$. The remaining terms from (\ref{sum6}) have somewhat lengthy form and we do not give them explicitly.

\section{The remaining term of $\Gamma_3$}

In this appendix we quote the remaining term of $\Gamma_3$, that is not important for purposes of renormalization:
\begin{align}\label{Gamma3''}
\Gamma_3''=&\frac{1}{6} \left(-\frac{g}{2 \pi a^2}\right)^3\mathrm{e}^{-3G(0)}\int\dif^2x\dif^2y\dif^2z \bigg\{\notag\\
&B(x-y,y-z,z-x,-2,-2,-2)\sum_{\alpha\beta}\phantom{ }^{'}\cos[\theta_\alpha(x)+\theta_\alpha(y)+\theta_\alpha(z) -\theta_\beta(x)-\theta_\beta(y)-\theta_\beta(z)]\notag\\
&+6B(x-y,y-z,z-x,-2,-1,-1)\sum_{\alpha\beta\gamma}\phantom{ }^{'}\cos[\theta_\alpha(x)+\theta_\alpha(y)+\theta_\alpha(z) -\theta_\beta(x)-\theta_\beta(y)-\theta_\gamma(z)]\notag\\
&+6B(x-y,y-z,z-x,-2,1,1)\sum_{\alpha\beta\gamma}\phantom{ }^{'}\cos[\theta_\alpha(x)+\theta_\alpha(y)-\theta_\alpha(z) -\theta_\beta(x)-\theta_\beta(y)+\theta_\gamma(z)]\notag\\
&+6B(x-y,y-z,z-x,-1,-1,1)\sum_{\alpha\beta\gamma}\phantom{ }^{'}\cos[\theta_\alpha(x)+\theta_\alpha(y)-\theta_\beta(x) +\theta_\beta(z)-\theta_\gamma(y)-\theta_\gamma(z)]\notag\\
&+6B(x-y,y-z,z-x,-1,-1,0)\sum_{\alpha\beta\gamma\delta}\phantom{ }^{'}\cos[\theta_\alpha(x)+\theta_\alpha(y)-\theta_\beta(y) -\theta_\beta(z)+\theta_\gamma(z)-\theta_\delta(x)]\notag\\
&+12B(x-y,y-z,z-x,-1,1,0)\sum_{\alpha\beta\gamma\delta}\phantom{ }^{'}\cos[\theta_\alpha(x)+\theta_\alpha(y)-\theta_\beta(y) +\theta_\beta(z)-\theta_\gamma(z)-\theta_\delta(x)]\notag\\
&+6B(x-y,y-z,z-x,1,1,-1)\sum_{\alpha\beta\gamma\delta}\phantom{ }^{'}\cos[\theta_\alpha(x)-\theta_\alpha(y)+\theta_\alpha(z) +\theta_\beta(y)-\theta_\gamma(z)-\theta_\delta(x)]\notag\\
&+2B(x-y,y-z,z-x,-1,-1,-1)\sum_{\alpha\beta\gamma\delta}\phantom{ }^{'}\cos[\theta_\alpha(x)+\theta_\alpha(y)+\theta_\alpha(z) -\theta_\beta(y)-\theta_\gamma(z)-\theta_\delta(x)]\bigg\}.
\end{align}
\end{widetext}

\section{Evaluation of integrals: finite-$a$ method}\label{appendix:finitea}

\subsection{One-loop integrals}

In this appendix we will evaluate the unknown integrals that appear in expressions for the beta functions (\ref{betag}) and (\ref{betasigma}). One should have in mind that we need the divergent contributions (when $a\to0$) from these integrals in a power law expansion with respect to the small parameter $\tau$. The divergent contribution to $a_1$ comes from the region of integration when the argument of $G(x)$ in $A(x,p)$ [Eq.~(\ref{A})] is around zero. After shifting the variable of integration and using the small-$x$ expansion of $G(x)$ given by Eq.~(\ref{G-smallx}) we obtain
\begin{align}\label{a1eval}
a_1=&\frac{c^2m^2}{2 \pi} \int\dif^2 y A(y,1)= \int_0^\Delta\dif y\frac{y}{y^2+a^2}\notag\\
&\times\left\{1+\tau\ln\left[c^2m^2(y^2+a^2)\right] +\mathcal{O}(\tau^2)\right\}+\text{f.t.}\notag\\
=&-\frac{1}{4}\left[2\lambda+\tau\lambda^2+ \mathcal{O}(\tau^2)+\text{f.t.}\right],
\end{align}
where $\lambda$ is defined in Eq.~(\ref{lambda}).
The introduced parameter $\Delta$ satisfies
\begin{align}\label{Deltacondition}
a\ll\Delta\ll (cm)^{-1},
\end{align}
and will further serve us to split the divergent part of $G(x)$ from the non-divergent one when $a\to0$ in integrals. In the region (\ref{Deltacondition}) one can always use the expansion of the propagator (\ref{G-smallx}). The abbreviation f.t.~stands for ``finite terms'' and denotes all terms that do not diverge in the limit $a\to 0$.

Using a similar reasoning as above, for the other contribution $a_2$ we get
\begin{align}\label{a2eval}
a_2=&\frac{c^4m^4}{2 \pi} \int\dif^2 y y^2A(y,2)=\int_0^\Delta\dif y\frac{y^3}{(y^2+a^2)^2}\notag\\
&\times\left\{1+2\tau\ln\left[c^2m^2(y^2+a^2)\right] +\mathcal{O}(\tau^2)\right\}+\text{f.t.}\notag\\
=&-\frac{1}{2}\lambda-\tau\lambda- \frac{1}{2}\tau\lambda^2+ \mathcal{O}(\tau^2)+\text{f.t.}
\end{align}
One may notice that divergent contributions come only from the term $\mathrm{e}^{G(y)}$ of $A(y,1)$ in (\ref{a1eval}) and only due to $\mathrm{e}^{2G(x)}$ from $A(y,1)$ in (\ref{a2eval}). The remaining terms in $A(y,2)$ and $A(y,1)$ determine the finite part of the integrals.

For our purpose of calculating the renormalization of the effective action up to third order in the coupling constant $g$ it turns out that finite parts in expressions (\ref{a1eval}) and (\ref{a2eval}) are important, since they contribute in the beta functions. It is important to notice that these finite parts, denoted by $c_1$ and $c_2$ in Eqs.~(\ref{a1}) and (\ref{a2}), multiply the renormalized parameters $g_R^3$ or $\tau g_R^2$, so for our order of accuracy of renormalization (that is third order in $g$) it is sufficient to evaluate them for $\tau=0$ in (\ref{a1eval}) and (\ref{a2eval}). Here we should have in mind that our general strategy of double expansion of the effective action (\ref{Gamma-finaldiv}) is in $g$ and $\tau$, which are assumed to be of the same order. That explains why we need constants $c_1$ and $c_2$ to order $\mathcal{\tau^0}$.

The constant $c_1$, which is a finite term in the expression (\ref{a1eval}) in the limit of $\tau=0$ and for $a\to 0$ is determined from the expression obtained from (\ref{a1eval}), \begin{align} \label{defc1}
\frac{c^2m^2}{2 \pi} \int\dif^2 y (\mathrm{e}^{G(y)}-G(y)-1)= -\frac{1}{4}\left(2\lambda+c_1\right),
\end{align}
taken in the limit $a\to 0$ and with the propagator $G(y)=2K_0(m\sqrt{y^2+a^2})$. This leads to
\begin{align}\label{c1}
c_1&=\lim_{\rho\to 0} \left\{-4\ln\left(c\rho\right)-4c^2\int_{\rho}^{\infty}\dif t t\left[\mathrm{e}^{2K_0(t)}-1\right]\right\}+8c^2\notag\\
&\approx 1.891.
\end{align}
We should mention that the contribution $8c^2$ in $c_1$ comes from the term $-G(x)$ that renders $A(x,1)$ one-particle irreducible.

Similarly, the finite part in (\ref{a2eval}) is determined from
\begin{align}  \label{defc2}
\frac{c^4m^4}{2 \pi}\int\dif^2 y y^2\left[\mathrm{e}^{2G(y)}-2G(y)-1\right]= -\frac{1}{2}\lambda+c_2 ,
\end{align}
taken in the limit $a\to 0$ for $\tau=0$. This leads to
\begin{align}\label{c2}
c_2=&-\frac{1}{2}+\lim_{\rho\to 0}\left\{\ln\left(c\rho\right)+c^4\int_{\rho}^{\infty}\dif t t^3 \left[\mathrm{e}^{4K_0(t)}-1\right]\right\}\notag\\
&-16c^4\approx 1.611.
\end{align}
The contribution $-16c^4$ in $c_2$ comes from the term $-2G(x)$ that makes $A(x,2)$ one-particle irreducible. With that, we arrive at the final expressions for $a_1$ and $a_2$ given in the main text, Eqs.~(\ref{a1result}) and (\ref{a2result}).

\subsection{Two-loop integrals}

We will now calculate the coefficients that stand in front of the operators in (\ref{Gamma3-eval}). To achieve that we use the procedure described in appendix \ref{An integral} for evaluating of the divergent parts of double integrals. We emphasize here that in all terms in $\Gamma_3$ we set $\tau=0$, since our purpose is to obtain the renormalized action to third order in the small parameters $g$ and $\tau$, and $\Gamma_3$ already contains a prefactor $g^3$. We emphasize here that for simplicity we calculate separately the part of $B$ without $B_1$ (terms with superscript $'$) and afterwards we evaluate $B_1$ (terms with superscript $''$), see Eq.~(\ref{Babc}). The first term of interest is
\begin{align}\label{b1def}
b_1=&\frac{c^4m^4}{(2 \pi)^2}\int\dif^2 x\dif^2 y B(x+y,x,y,-2,2,2)=b_{1}'-b_{1}'',
\end{align}
which has divergent contributions from three regions of integration, (a), (b), and (c), see Eq.~(\ref{regions}).
In the first region (a) after angular integration we have
\begin{align}
b_{1a}'=&\int_0^{\Delta}\dif x \int_0^{\Delta}\dif yxy\frac{4x^2y^2-a^4}{(x^2+a^2)^2(y^2+a^2)^2}\notag\\
=& \left[\frac{3}{4}-4\ln\frac{\Delta}{a} +4\ln^2\frac{\Delta}{a}+\mathcal{O}\left(\frac{a^2}{\Delta^2}\right)\right] +\text{f.t.}
\end{align}
Combining the contributions from regions (b) and (c) we easily obtain
\begin{align}\label{b1bc'}
b_{1bc}'=&-2 c^4m^4\int_0^\Delta\dif xx^3\mathrm{e}^{2G(x)}\int_\Delta^\infty\dif yyf(2,y)+\text{f.t.}\notag\\
=& -6\ln\frac{\Delta}{a}-8\ln\frac{\Delta}{a} \ln(cm\Delta) +\text{f.t.}
\end{align}
The contribution that comes from $B_1$ in (\ref{b1def}) reads
\begin{align}\label{b1''}
b_1''=\frac{c^4m^4}{(2 \pi)^2} \int\dif^2x G(x)\int\dif^2y\left[\mathrm{e}^{-2G(y)}+G(y)-1\right].
\end{align}
It does not contain divergencies and contributes only to finite terms. Therefore the final result reads
\begin{align}\label{b1result}
b_1=b_{1a}'+b_{1bc}'-b_{1}''= 5\lambda+\lambda^2 +\text{f.t.}
\end{align}
where $\lambda$ have been defined in Eq.~(\ref{lambda}).

The second term of interest is
\begin{align}\label{b2def}
b_2=&\frac{c^4m^4}{(2 \pi)^2} \int\dif^2 x\dif^2 y B(x,y,x+y,2,1,-1)=b_{2}'-b_{2}''.
\end{align}
As in the previous case we consider three regions of integration, (a), (b), and (c), see Eq.~(\ref{regions}).
In the first region (a) after angular integration we get
\begin{align}
b_{2a}'=& \int_0^{\Delta}\dif x \int_0^{\Delta}\dif y xy\frac{x^2}{(x^2+a^2)^2(y^2+a^2)}\notag\\
=& -\frac{1}{2}\ln\frac{\Delta}{a} +\ln^2\frac{\Delta}{a}+\mathcal{O}\left(\frac{a^2}{\Delta^2}\right) +\text{f.t.}
\end{align}
Combining the contributions from regions (b) and (c) we easily obtain
\begin{align}\label{b2bc'}
b_{2bc}'=&c^4m^4\int_0^\Delta\dif x\int_\Delta^\infty\dif y xy\bigg\{-x^2\mathrm{e}^{2G(x)}f(1,y)\notag\\
&+\mathrm{e}^{G(x)+G(y)}-\mathrm{e}^{G(x)}\bigg\} +\text{f.t.}\notag\\
=&  \left\{-1+c^2m^2\int_\Delta^\infty\dif y y\left[\mathrm{e}^{G(y)}-1\right]\right\} \ln\frac{\Delta}{a}\notag\\
&-\ln\frac{\Delta}{a}\ln(cm\Delta) +\text{f.t.}
\end{align}
The contribution that comes from $B_1$ in (\ref{b2def}) is
\begin{align}\label{b2''}
b_{2}''=&\frac{c^4m^4}{(2 \pi)^2} \int\dif^2x G(x)\int\dif^2y\left[\mathrm{e}^{G(y)}-1\right]\notag\\
=& -2c^2\ln(cma) +\text{f.t.}
\end{align}
Using the definition of $c_1$ in Eq.~(\ref{c1}) after simple algebra we finally obtain
\begin{align}\label{b2result}
b_{2}=b_{2a}'+b_{2bc}'-b_{2}''=\frac{1}{4}\lambda^2+\frac{6+c_1}{8}\lambda +\text{f.t.}
\end{align}
There we used the following result
\begin{align}
\int_\Delta^\infty \dif z z\left[\mathrm{e}^{G(z)}-1\right]
=&\frac{1}{c^2m^2}\bigg[2c^2-\frac{c_1}{4}
-\ln(cm\Delta)\bigg]\notag\\
&+\mathcal{O}\left(\frac{a^2}{\Delta^2},m^2\Delta^2\right),
\end{align}
where $c_1$ is defined in (\ref{c1}).

Finally we calculate the integral
\begin{align}\label{b3def}
b_3=\frac{c^6m^6}{(2 \pi)^2} \int\dif^2 x\dif^2 y x^2B(x,x+y,y,1,1,1)=b_3'-b_3''.
\end{align}
Contrary to the previous three cases where the region (d) [see Eq.~(\ref{regions})] has not yielded divergent terms, in the case of (\ref{b3def}) it produces divergent terms. First we will evaluate (\ref{b3def}) in region (a). We have
\vspace{-0.7cm}
\begin{widetext}
\begin{align}\label{b3-region a}
b_{3a}'&=\frac{c^6m^6}{(2 \pi)^2} \int_{|x|,|y|<\Delta}\dif^2 x\dif^2 y x^2 \left[\mathrm{e}^{G(x)+G(y)+G(x+y)}-\mathrm{e}^{G(x)} -\mathrm{e}^{G(y)}-\mathrm{e}^{G(x+y)}+2\right]\notag\\
&=\int_0^{\Delta/a}\dif x\int_0^{\Delta/a}\dif y\int_0^{2\pi}\frac{\dif\varphi}{2\pi}\frac{x^3y}{(x^2+1)(y^2+1)(y^2+x^2+2xy \cos\varphi+1)}+\text{f.t.}\notag\\
&=\int_0^{\Delta/a}\dif x\int_0^{\Delta/a}\dif y\frac{xy}{(y^2+1)\sqrt{(y^2-x^2)^2+2(x^2+y^2)+1}} \left(1-\frac{1}{x^2+1}\right)+\text{f.t.}\notag\\
&=\int_0^{\Delta/a}\dif y\frac{y} {2\left(y^2+1\right)}\left\{\ln \left[\frac{\Delta^2}{a^2}+1-y^2+\sqrt{\left(\frac{\Delta^2}{a^2}+1\right)^2+y^4-2 \left(\frac{\Delta^2}{a^2}-1\right)y^2}\right]-\ln2\right\}+\text{f.t.}\notag\\
&=\ln^2\frac{\Delta}{a}+\text{f.t.}
\end{align}
\end{widetext}
Only the first term on the right-hand-side of the first line in Eq.~(\ref{b3-region a}) contributes divergencies. After using the expansion (\ref{G-smallx}), rescaling the variables $x\to ax,y\to ay$  and doing the angular integration one ends up with two terms given in the third line of (\ref{b3-region a}). Only the first term gives a divergent contribution when $a\to 0$. By doing one more integration over $x$ we end up with only one integration over $y$, see the fourth line of (\ref{b3-region a}). Expanding the obtained integral with respect to the large parameter $\Delta/a$ followed by integration over $y$ leads to the final result.

The sum of contributions in regions (b), (c) and (d) is
\begin{align}\label{b3bcd''}
b_{3bcd}'=&2 c^6m^6\int_0^\Delta\dif x x\left[1+\mathcal{O}(x^2)\right]\mathrm{e}^{G(x)}\notag\\
&\times\int_\Delta^\infty\dif y y^3\left[\mathrm{e}^{2G(y)}-1\right]\notag\\
=&{2}{c^4m^4}\ln\frac{\Delta}{a}\int_\Delta^\infty\dif y y^3\left[\mathrm{e}^{2G(y)}-1\right]+\text{f.t.}
\end{align}
The contribution from $B_1$ in (\ref{b3def}) is
\begin{align}\label{b3''}
b_3''=&2 \frac{c^6m^6}{(2 \pi)^2} \int\dif^2x\dif^2y y^2\left[G(y)+G(x+y)\right]\left[\mathrm{e}^{G(x)}-1\right]\notag\\ &+\text{f.t.}
=-32c^4\ln(cma)+\text{f.t.}
\end{align}
At the end we get
\begin{align}
b_3=b_{3a}'+b_{3bcd}'-b_{3}''=\frac{1}{4}\lambda^2 -\frac{2c_2+1}{2}\lambda+\text{f.t.}
\end{align}
There we used the following result
\begin{align}
\int_\Delta^\infty \dif z z^3\left[\mathrm{e}^{2G(z)}-1\right]
=\frac{1}{c^4m^4}\bigg[c_2+\frac{1}{2}+16c^4\notag\\
-\ln(cm\Delta)\bigg]+\mathcal{O}\left(\frac{a^2}{\Delta^2},m^2\Delta^2\right),
\end{align}
where $c_2$ is defined in (\ref{c2}).

\section{Important integrals}\label{An integral}

The coefficient in front of operators in $\Gamma_3$ contain integrals of the common type that will be calculated in this appendix. They can be written in the following form
\begin{align}\label{I}
I=\int\dif^2x\dif^2y y^p \mathrm{e}^{\alpha G(x)-\beta G(x+y)+\gamma G(y)},
\end{align}
where the propagator $G(x)=2K_0(m\sqrt{x^2+a^2})$ is obtained by setting $\tau=0$ into Eq.~(\ref{G}). The parameters in Eq.~(\ref{I}) are assumed to belong to the set $\alpha,\beta,\gamma\in \{0,1,2\}$, $p\in \{0,2\}$ that occur in unevaluated expressions in Eqs.~(\ref{b1}), (\ref{b2}) and (\ref{b3}). The (logarithmic) divergence of integral (\ref{I}) arises because of the behavior of $G(x)$ at $x<a\ll m^{-1}$. In order to isolate divergent from non-divergent parts we split the range of integration in (\ref{I}) by a parameter $\Delta$ which is introduced in appendix \ref{appendix:finitea} and satisfies (\ref{Deltacondition}). We distinguish four regions of integration
\begin{align}\label{regions}
&\text{(a)}\quad |x|,|y|<\Delta,\quad &&\text{(b)}\quad |x|<\Delta,|y|>\Delta,\notag\\
&\text{(c)}\quad |x|>\Delta,|y|<\Delta, &&\text{(d)}\quad |x|,|y|>\Delta,
\end{align}
and analyze integral (\ref{I}) in these regions.

In region (a) one could use expansion (\ref{G-smallx}) for all correlation functions in (\ref{I}) and evaluate the integral. We will not do it explicitly here for the most general case. Particular cases are calculated in appendix \ref{appendix:finitea}.

In region (b) we can expand the correlation function around $y$:
\begin{align}\label{G(x+y)}
&G(|x+y|)=G(|y|)+\frac{1}{2}h\frac{1}{|y|}G'(|y|)\notag\\
&+\frac{1}{8}h^2 \left[\frac{1}{y^2}G''(|y|)-\frac{1}{|y|^3}G'(|y|)\right]+\mathcal{O}(h^3),
\end{align}
with $h={x^2+2x\cdot y}$. Then after expanding the term $\mathrm{e}^{-\beta G(x+y)}$ for small $h$ and doing the angular integration one gets
\begin{align}
I_b=&(2\pi)^2\int_0^\Delta\dif x\int_\Delta^\infty\dif y xy^{1+p}\mathrm{e}^{\alpha G(x)+(\gamma-\beta) G(y)}\notag\\
&\times\left[1-x^2f(\beta,y)+\mathcal{O}(x^4)\right],
\end{align}
where for convenience we have introduced a function
\begin{align}
f(\beta,y)=\frac{\beta}{4}\left[G''(y)+\frac{1}{y}G'(y)-\beta G'(y)^2\right].
\end{align}
Using similar manipulations as above in region (c), with the difference that we expand $G(x+y)$ around $x$, after exchanging the integration variables we get
\begin{align}
I_c=&(2\pi)^2\int_0^\Delta\dif x\int_\Delta^\infty\dif y x^{1+p}y\mathrm{e}^{\gamma G(x)+(\alpha-\beta) G(y)}\notag\\
&\times\left[1-x^2f(\beta,y)+\mathcal{O}(x^4)\right].
\end{align}

While the divergent terms of integral (\ref{I}) in regions (b) and (c) arise only when one of the variables is around zero and the corresponding propagator diverges, in region (d) both $|x|$ and $|y|$ are large. However, their sum $|x+y|$ could be a small number which may in certain cases produce divergencies. Therefore, region (d) may contain divergencies when $|x+y|<\Delta$ and $\beta<0$. Changing the variables of integration $x+y\to x$ and after using expansion (\ref{G(x+y)}) one ends up with
\begin{align}\label{Id}
I_d=&\int_{|x|,|y|>\Delta}\dif^2x\dif^2y y^p \mathrm{e}^{\alpha G(x)-\beta G(x+y)+\gamma G(y)}\notag\\
=&(2\pi)^2\int_0^\Delta\dif x\int_\Delta^\infty\dif y x y^{1+p}\mathrm{e}^{-\beta G(x)+(\alpha+\gamma)G(y)}\notag\\
&\times\left[1+\mathcal{O}(x^2)\right]+\text{f.t.}
\end{align}
The last expression has divergences for $\beta=-1$. That type of integral appears during evaluation of $b_3$, see Eq.~(\ref{b3}). The case $\beta<-1$ could be also analyzed, but it is of no interest for us.

We close this appendix by explicitly evaluating three integrals. The first one is
\begin{align}\label{fintegral}
\int_\Delta^\infty \dif y y f(\beta,y)=&\frac{\beta^2}{2}\left[1+\frac{1}{\beta}+\ln(c^2 m^2 \Delta^2)\right]\notag\\
&+\mathcal{O}\left(m^2\Delta^2,{a^2}/{\Delta^2}\right).
\end{align}
It could be done by expanding the propagator $G(x)=2K_0(m\sqrt{x^2+a^2})$ around $a=0$ to zeroth order, since $\Delta\gg a$. The remaining terms of that expansion produce after integration a result which is at least $\sim a^2/\Delta^2$. We remind the reader that we have already set $\tau=0$ in all terms that come with the overall prefactor $g^3$, i.e.,~in all terms that arise from $\Gamma_3$.

The second and the third one can be easily done by using the expansion of the propagator (\ref{G-smallx}) followed by simple integrations. They read
\begin{align}\label{intzG}
\int_0^\Delta\dif z z\mathrm{e}^{G(z)}=\frac{1}{c^2m^2}\ln\left(\Delta/a\right)+\mathcal{O}(a^2/\Delta^2)
\end{align}
and
\begin{align}\label{intz32G}
\int_0^\Delta\dif z z^3\mathrm{e}^{2G(z)} =\frac{1}{c^4m^4}\left[\ln\left(\Delta/a\right)-1/2\right]+\mathcal{O}(a^2/\Delta^2).
\end{align}

\section{Dimensional method}\label{appendix:dimensionalmethod}

In this appendix we calculate integrals (\ref{a1}), (\ref{a2}), (\ref{b1}), (\ref{b2}), and (\ref{b3}) by a dimensional method. The main idea is to consider $\tau>0$ (i.e.,~$T<T_c$) where one can set the short-distance cutoff $a$ to zero in all correlation functions that appear in the above-mentioned integrals. The logarithmic divergencies contained in the parameter $\lambda$ [see Eq.~(\ref{lambda})] will become poles with respect to $\tau$ in final expressions. The calculation is straightforward once one is acquainted with the techniques and ideas presented in appendix \ref{appendix:finitea}.

While some results at intermediate steps in appendix \ref{appendix:finitea} are calculated using the Bessel function (\ref{G}) for the propagator, here we show that this is not necessary because the universal parts of the integrals come from the short-distance behavior of the propagator, that is universal. For distances
 $|x|\ll (cm)^{-1}$, it reads
\begin{align}\label{Gsmallx-dimmensional}
G(x)=-(1-\tau)\ln(c^2m^2x^2).
\end{align}
In addition the limiting behavior $G(\infty)=0$ is necessary for the calculation, which is a quite weak assumption. Later we will see that we need one more condition and it is $\lim_{y\to\infty}yG'(y)=0$.

The first term of interest is (\ref{a1}) and it can be evaluated by splitting the integration range by a parameter $\Delta$ that satisfies $\Delta\ll (cm)^{-1}$ [c.f.~(\ref{Deltacondition})]. The divergence arises from the region of integration $0\le x\le\Delta$, while the remaining region $x>\Delta$ delivers a constant $c_1'$. The final result can be written in the form
\begin{align}\label{a1-dimensional}
a_1=\frac{1}{2\tau}-c_1'+\mathcal{O}(\tau).
\end{align}
Similarly we get
\begin{align}\label{a2-dimensional}
a_2=\frac{1}{4\tau}-c_2'+\mathcal{O}(\tau),
\end{align}
where again we have a constant $c_2'$ that only depends on the precise form of the correlation function $G(x)$. We should mention that for the special choice of propagator (\ref{G}) the constants $c_1'$ and $c_2'$ are well defined, however their precise value is immaterial for our purposes.

Further we compute two-loop integrals by the new method (they have been already calculated in appendix \ref{appendix:finitea} by another method), and we closely follow the notation from that appendix. In the following we are only interested in the divergent parts of the expressions in the limit $\tau\to 0$. It is convenient to introduce the abbreviation
\begin{align}
&g_n(\Delta)=\int_0^\Delta \dif x x^{2n-1}\mathrm{e}^{nG(x)},
\end{align}
which after using (\ref{Gsmallx-dimmensional}) yields
\begin{align}
&g_1(\Delta)=\frac{1}{c^2m^2}\left[\frac{1}{2\tau}+\ln (cm\Delta)\right],\\
&g_2(\Delta)=\frac{1}{c^4m^4}\left[\frac{1}{4\tau}+\ln (cm\Delta)\right],
\end{align}
valid in the limit $cm\Delta\to 0$.

The first term is defined in (\ref{b1}). It has divergent contributions from three regions of integration, (a), (b), and (c), see Eq.~(\ref{regions}) for the definition. In the first region all correlation functions have the logarithmic form (\ref{Gsmallx-dimmensional}) and we obtain
\begin{align}
b_{1a}'=\frac{1}{2\tau^2}- \frac{1-2\ln(cm\Delta)}{\tau}+\mathcal{O}(\tau^0).
\end{align}
Regions (b) and (c) combine into the form given in the first line of Eq.~(\ref{b1bc'}). After evaluation one gets
\begin{align}
b_{1bc}'=- \frac{1}{2\tau} \left[1-h(\Delta)\right],
\end{align}
where $h(\Delta)=\int_\Delta^\infty\dif y y G'(y)^2$. We also used
\begin{align}
\int_\Delta^\infty \dif y y f(\beta,y)=&-\frac{1}{4}y G'(y)|_\Delta^\infty -\frac{\beta^2}{4}h(\Delta)\notag\\
=&\frac{\beta}{2}-\frac{\beta^2}{4}h(\Delta),
\end{align}
where the assumption $\lim_{y\to\infty}yG'(y)=0$ has been used. The remaining term (\ref{b1''}) does not contain divergencies, and we finally obtain
\begin{align}\label{b1dimensional}
b_1=\frac{1}{2\tau^2}- \frac{3}{2\tau}+\frac{4\ln (cm\Delta)+h(\Delta)}{2\tau}+\mathcal{O}(\tau^0).
\end{align}

Further we calculate the term given in (\ref{b2}). In region (a) we obtain
\begin{align}
b_{2a}'=\frac{3}{16\tau^2}- \frac{1-6\ln(cm\Delta)}{8\tau}+\mathcal{O}(\tau^0).
\end{align}

The first two lines of (\ref{b2bc'}) after simple manipulations become
\begin{align}
\frac{b_{2bc}'}{c^4m^4}=&- g_2(\Delta) \int_\Delta^\infty\dif y y f(1,y)
+ g_1(\Delta)\left[a_1-g_1(\Delta)\right]\notag\\
&+g_1(\Delta)\frac{1}{2 \pi} \int\dif^2 x G(x),
\end{align}
while the divergent part of the contribution (\ref{b2''}) reads
\begin{align}
b_2''=c^4m^4g_1(\Delta)\frac{1}{2 \pi} \int\dif^2 x G(x).
\end{align}
Combining the previous expressions one obtains
\begin{align}\label{b2dimensional}
b_2=\frac{3}{16\tau^2} -\frac{4+8c_1'-4\ln(cm\Delta)-h(\Delta)}{16\tau} +\mathcal{O}(\tau^0).
\end{align}
The important combination that appears in the beta function (\ref{betag}) now reads
\begin{align}\label{b1-8b2dimensional}
b_1-8b_2+4a_1^2=\frac{1}{2 \tau}+\mathcal{O}(\tau^0).
\end{align}
There are several important things to mention about the last result. First, all non-universal terms connected with $\Delta$ and $c_1'$ from (\ref{b1dimensional}) and (\ref{b2dimensional}) have canceled in (\ref{b1-8b2dimensional}). Also there is no $1/\tau^2$ divergence in the combination. This is quite reminiscent to the situation we had in the same term evaluated by another method, see Eq.~(\ref{b1-8b2}).

The last integral we evaluate is defined in (\ref{b3}). However, it is more convenient to rewrite it in an equivalent form
\begin{align}\label{b3-dimensional}
b_3=\frac{c^6m^6}{(2 \pi)^2} \int\dif^2 x\dif^2 y (x+y)^2B(x,x+y,y,1,1,1).
\end{align}
In region (a) we obtain
\begin{align}
b_{3a}'=\frac{1}{6\tau^2} +\frac{\ln(cm\Delta)}{\tau}+\mathcal{O}(\tau^0).
\end{align}
In regions (b), (c), and (d) now there are two contributions. The terms multiplied by $x^2+y^2$ from (\ref{b3-dimensional}) give a contribution that is two times larger than the result stated in the first two lines of (\ref{b3bcd''}). The remaining term $2x\cdot y$ has divergent contributions only in the region (d) that equals (\ref{b3bcd''}) multiplied by minus one. Overall we obtain the same contribution as given in the first two lines of (\ref{b3bcd''}). After simple regrouping we obtain
\begin{align}
\frac{b_{3bcd}'}{c^6m^6} =2g_1(\Delta)\left[a_2-g_2(\Delta)\right]+4g_1(\Delta)\frac{1}{2 \pi} \int\dif^2y y^2 G(y).
\end{align}
The divergent part of (\ref{b3''}) reads
\begin{align}
b_3''=4\frac{c^6m^6}{2 \pi}g_1(\Delta)\int\dif^2y y^2G(y),
\end{align}
and finally we obtain
\begin{align}
b_3=\frac{1}{6\tau^2}-\frac{c_2'}{\tau} +\mathcal{O}(\tau^0).
\end{align}

\section{Two point function at finite mass and one loop in the continuum limit $a \to 0$}\label{appendix:twopoint-finitem}

In this appendix we study the one loop result (\ref{Gamma2q}) for the two-point function derived in the main text.
We focus on the continuum limit $a \to 0$, keeping $m$ finite, which exists for $\tau>0$. To recover this limit we must express the result in terms of $\tilde g$, or equivalently of $g_R$ using (\ref{gtilde}). We
use $G(x) = 2 (1-\tau) K_0(m x)$, perform the rescaling $x \to x/(m c)$ and get
by adding the first and second orders in $g$ contributions (\ref{ag1}) and (\ref{Gamma2q}):
\begin{align}
&\Gamma(q) = \Delta\left(\frac{q}{m c}\right) + \Delta_0 \\
&\Delta_0= - \frac{1}{\pi} c^2 m^2  g_R \left[1 + 2 (a_1 + \tilde a_1) g_R\right]
\end{align}
with
\begin{align}
\tilde a_1 = \int_0^\infty x dx \left[ \mathrm{e}^{-2 \tilde G(x)}-3 \mathrm{e}^{-\tilde G(x)}-\mathrm{e}^{\tilde G(x)}+3 \right].
\end{align}
Here $\tilde G(x)=2 (1-\tau) K_0(x/c)$, and we note that in the combination
\begin{align}
a_1+ \tilde a_1 &= \int_0^\infty x dx \left[ \mathrm{e}^{-2 \tilde G(x)}-3 \mathrm{e}^{-\tilde G(x)}-\tilde G(x) +2 \right] \nn \\
 &= -0.0473276, \quad {\rm for} ~ \tau=0
\end{align}
the logarithmic divergencies cancel yielding a finite result for $\Delta_0$ at $\tau=0$. The momentum
dependent part is:
\begin{align}
\Delta (p) =& - \frac{1}{ \pi^2} c^2 m^2 g_R^2
\int\dif^2 x (\mathrm{e}^{i p \cdot x}-1)\big[2\sinh \tilde G(x) \nn \\
& -\sinh 2 \tilde G(x)\big]
\end{align}

Let us now consider its limit for large $p=q/m$ limit obtained by rescaling $x=y/p$. Using
\begin{align}
&\lim_{p \to \infty} p^{-4(1-\tau)} \sinh[2 G(y/p)] = 1/\left[2 y^{4(1-\tau)}\right], \\
& \lim_{p \to \infty} p^{-4(1-\tau)} \sinh[G(y/p)] = 0,
\end{align}
one finds
\begin{align}
\Delta(p) &=  \frac{1}{ \pi} c^2 m^2 g_R^2 p^{2-4 \tau}
\int_0^\infty y dy (J_0(y)-1) y^{- 4 (1-\tau)}   \nn \\
& = \frac{1}{ \pi} c^2 m^2 g_R^2 p^{2-4 \tau}  \frac{2^{4 \tau-3} \Gamma (2 \tau-1)}{\Gamma (2-2 \tau)} \\
& = \frac{1}{ \pi} c^2 m^2 g_R^2 \left[ -\frac{p^2}{16 \tau} + \frac{1}{4} p^2 \ln(p c/e) + \mathcal{O}(\tau) \right].
\end{align}
Let us now comment on the various regimes for $\Gamma(q)$ as a function of $q$. By comparing with
(\ref{gammaq2res}) we see that the $\ln q$ behavior (that leads to super-rough correlation) extends from
the region $q \sim 1/a$ up to the region $q \sim m$. However for very small $q$, $q \sim m \mathrm{e}^{-1/(4 \tau)}$
the coefficient $- 2 \pi \Gamma(q)/q^2$ saturates to $1/8 \tau$ (plus a finite part). The pole in $\tau$ has
precisely the value obtained
in (\ref{sigmaR}) together with (\ref{a2dim}) and (\ref{Dhalf}) by considering the limit $q \to 0$ first.
Finally the above result allow to determine the correlation function ${\cal G}_0(q)=f(q/m)/m^4$ as a scaling function of $q/m$ using (\ref{GGamma0}).

\section{More on two loop integrals: simplifications and explicit evaluation for a simple cutoff function}\label{moreontwoloop}

In this appendix we calculate the integrals (\ref{a1}), (\ref{a2}), and (\ref{b1})--(\ref{b3}) using a simple cutoff function.

\subsection{Simplifications}

First we show that the terms linear in $G(x)$ in all integrals, which make the effective action a sum of one particle irreducible graphs,
cancel in universal invariants in the beta functions. We also provide simpler expressions for evaluation of two-loop integrals.

Let us start with one-loop integrals. We define:
\begin{align}
& a_1^{(p)} = \frac{m^2 c^2}{2 \pi} \int_y \mathrm{e}^{p G(y)} - p G(y) -1, \\
& a_2^{(p)} =  \frac{m^2 c^2}{2 \pi} \int_y y^2 \left[\mathrm{e}^{p G(y)} - p G(y) -1\right],
\end{align}
so that $a_1=a_1^{(1)}$ and $a_2=a_2^{(2)}$. Here we denote $\int_y = \int \dif^2 y$. Analogously we define expressions without the term linear in $G(x)$:
\begin{align}
& \hat a_1^{(p)} = \frac{m^2 c^2}{2 \pi} \int_y \mathrm{e}^{p G(y)} -1, \\
& \hat a_2^{(p)} =  \frac{m^2 c^2}{2 \pi} \int_y y^2 \left[\mathrm{e}^{p G(y)} -1\right].
\end{align}
This changes only the finite parts in the one-loop integrals since the integrals linear in $G(y)$ are
finite. More precisely (\ref{a1dim}) and (\ref{a2dim}) are changed into
\begin{align}
&4 \hat a_1=4 \hat a_1^{(1)}=\frac{A}{\tau} + \hat B + \mathcal{O}(\tau), \\
&2 \hat a_2=2 \hat a_2^{(1)}=\frac{D}{\tau} + \hat E + \mathcal{O}(\tau),
\end{align}
with $B = \hat B - 4 \frac{m^2 c^2}{2 \pi} \int_x G(x)$ and $E= \hat E - 4 \frac{m^2 c^2}{2 \pi} \int_x x^2 G(x)$
(in other words $\hat A=A$ and $\hat D=D$).

For the two-loop integrals we will use an expression equivalent to (\ref{Babc}):
\begin{widetext}
\begin{align} \label{other}
 B(x,y,z,a,b,c) = &\left[\mathrm{e}^{a G(x)} -1\right] \left[\mathrm{e}^{b G(y)} -1\right] \left[\mathrm{e}^{c G(z)} -1\right] + \left[\mathrm{e}^{a G(x)} - a G(x) -1\right] \left[\mathrm{e}^{b G(y)} - b G(y) -1\right]
\notag\\
& + \left[\mathrm{e}^{b G(y)} - b G(y) -1\right] \left[\mathrm{e}^{c G(z)} - c G(z) -1\right] +  \left[\mathrm{e}^{a G(x)} - a G(x) -1\right] \left[\mathrm{e}^{c G(z)} - c G(z) -1\right].
\end{align}
Then one can rewrite the needed combinations of two-loop integrals:
\begin{align}
&b_1 - 8 b_2 + 4 a_1^2 = \frac{m^4 c^4}{(2 \pi)^2} \int_{x+y+z=0} \left[\mathrm{e}^{-2 G(z)} -1\right]\left[\mathrm{e}^{2 G(x)} -1\right]\left[\mathrm{e}^{2 G(y)} -1\right] + (\hat a_1^{(2)})^2 + 2 \hat a_1^{(2)} \hat a_1^{(-2)}+ 4 (\hat a_1^{(1)})^2\notag
\\
& - 8 \left\{ \frac{m^4 c^4}{(2 \pi)^2} \int_{x+y+z=0} \left[\mathrm{e}^{-G(z)} -1\right]\left[\mathrm{e}^{2 G(x)} -1\right]\left[\mathrm{e}^{G(y)} -1\right] + \hat a_1^{(2)} \hat a_1^{(1)}
+ \hat a_1^{(2)} \hat a_1^{(-1)} + \hat a_1^{(1)} \hat a_1^{(-1)} \right\}  + {\rm finite}
\end{align}
the same expression without the hats and the additional finite part being exact [by definition from (\ref{other})]. Comparing with
(\ref{firstcomb}) it means $\hat C=C$, i.e., $C$ can be computed discarding linear terms in $G(y)$. The second combination (\ref{secondcomb}) can be rewritten:
\begin{align}\label{new1}
 6 b_3 - 8 a_1 a_2 =& 6  \frac{m^4 c^4}{(2 \pi)^2}  \int_{x+y+z=0} x^2 \left[\mathrm{e}^{G(x)} -1\right] \left[\mathrm{e}^{G(y)}-1\right]\left[\mathrm{e}^{G(z)} -1\right] \notag\\
& + 24
\hat a_1 \hat a_2 -8 \hat a_1 \hat a_2
 + 8 \frac{m^2 c^2}{2 \pi}  \int_x \left[\hat a_2  G(x) -  \hat a_1 x^2 G(x) \right] + {\rm finite}
\end{align}
\end{widetext}
again the same expression without the hats, the terms linear in $G$, and the additional finite part being exact. Comparing with (\ref{secondcomb}) one sees that now the coefficient $F$ is changed, i.e.,
$F = \hat F + 4 D \frac{m^2 c^2}{2 \pi}  \int_x G(x) - 2 A \frac{m^2 c^2}{2 \pi}  \int_x x^2 G(x)$. However the
combination
\begin{align}
F + B D - \frac{1}{2} A E = \hat F + \hat B D - \frac{1}{2} A \hat E
\end{align}
does not change and since $A$ and $D$ are also unchanged the three universal ratios (\ref{univcomb}) are unchanged.
Hence one can just suppress the terms $\int_x G(x)$ and $\int_x x^2 G(x)$ since they cancel in
universal quantities.

\subsection{Calculation with a simple cutoff function}

We consider the continuum (or dimensional) limit scheme where $a=0$ and $\tau>0$. Until now we have used the Bessel cutoff function $G(x)=2 (1-\tau) K_0(m x)$, and noted that all factors $m c$ in the definitions of the integrals can be set to unity by rescaling $x \to x/(m c)$. This is equivalent to using $\tilde G(x)=2(1-\tau) K_0(x/c)$. Here we will test further the universality with respect to the choice of the cutoff function by considering
\begin{align}
\mathrm{e}^{\tilde G(x)} = \Theta(x<1) x^{2 \tau - 2} + \Theta(x>1)
\end{align}
such that $\tilde G(x)$ is continuous and has the same logarithmic behavior at small $x$ as the Bessel function cutoff
and vanishes at infinity. For convenience we also suppress the tilde (and set $m=c=1$ in all definitions).

The one-loop integrals are:
\begin{align}\label{res1lpp}
&\hat a_1 =\int_0^1 \dif x x  (x^{2 \tau - 2} -1) = \frac{1}{2 \tau} - \frac{1}{2},\\
& \hat a_2 =  \int_0^1 \dif x x^3  (x^{4 \tau - 4} -1) = \frac{1}{4 \tau} - \frac{1}{4},
\end{align}
so that
\begin{align} \label{res1l}
A = 2, \quad \hat B = - 2,\quad D =1/2,\quad \hat E = - 1/2.
\end{align}

\subsubsection{Coefficient $\hat F$}

\begin{widetext}
From (\ref{new1}) and (\ref{secondcomb}) the coefficient $\hat F$ can be extracted as
\begin{align}
\frac{\hat F}{\tau} + \mathcal{O}(1) =
6 \int_{x+y+z=0} x^2 (\mathrm{e}^{G(x)} -1)(\mathrm{e}^{G(y)} -1)(\mathrm{e}^{G(z)} -1) + 24 \hat a_1 \hat a_2 -8 \hat a_1 \hat a_2  = N_1  -\frac{1}{\tau^2} + \frac{5}{\tau}
\end{align}
where we have used (\ref{res1lpp}) and computed the one loop integral $\hat a_1 \hat a_2 =
\frac{1}{2 t} \int_0^1 dx x^3 (x^{-2}-1) =1/(8 \tau)$, keeping only divergent parts. We have defined
\begin{align}
&& N_1 = \frac{1}{(2 \pi)^2} \int_{x<1,y<1,z<1} 2 (x^2+y^2+z^2) (x^{2\tau-2}-1)(y^{2\tau-2}-1)(z^{2\tau-2}-1).
\end{align}
To compute this integral (which is ultraviolet convergent) we use symmetry to restrict to the domain
$0<y<x<z<1$ and write $y=w x$ and $0<w<1$ and $z^2=x^2+y^2+ 2 x y c$ with $c = \cos(\phi)$.
The measure is then $12 x^3 w \dif x \dif w \dif c/\sqrt{1-c^2}$ with $c \in [-1,1]$. The condition $x< z<1$,
i.e., $x^2 < z^2 = x^2 (1 + 2 c w + w^2) < 1$ implies that $2 c + w>0$, hence we use that
\begin{align}
\int_{x<1,y<1,z<1}\ldots =  2 \pi
\int_{-1}^1 \frac{\dif c}{\sqrt{1-c^2}} \int_0^1 \dif w 12 w \int_0^{1/\sqrt{1+ 2 c w + w^2}} \dif x x^3
\Theta(w+2 c>0)\ldots
\end{align}
\end{widetext}

After integration over $x$ for
$0< x< 1/\sqrt{1+ 2 c w + w^2}$ one finds
\begin{align}
N_1 = \frac{1}{2 \pi} \int_0^1 \dif w \int_{-1}^1 \frac{\dif c}{\sqrt{1-c^2}} g_\tau(w,c) \Theta(w+ 2 c>0) \label{N1}
\end{align}
with
\begin{align}
g_\tau(w,c) = \frac{1}{\tau} \frac{8 [w (c+w)+1]}{ w \left(2 c w+w^2+1\right)} + \mathcal{O}(\tau^0)
\end{align}
That would naively give $N_1 \simeq A/\tau$ but there is a pole at $w=0$ \footnote{Note that the pole
at $w=1,c=-1$ is suppressed by the factor $\Theta(w+ 2 c>0)$.}. To treat this pole one checks that
replacing in (\ref{N1}) $g_\tau \to g_\tau^{reg}= g_\tau - g^{p}_\tau$ with
\begin{align}
 g^{p}_\tau(w,c) = \frac{16 (\tau-1)^2 w^{2 \tau-1}}{ \tau \left(2 \tau^2+5 \tau+2\right)}
 \Theta(0<w<1)
\end{align}
one finds a finite integral $\simeq A/\tau$ as $\tau \to 0$. Hence we can write $N_1=N_1^{p} + N_1^{reg}$ where the contribution of the pole is
\begin{align}
 N_1^{p} &= \frac{1}{2 \pi} \int_0^1 \dif w \int_{-w/2}^1 \frac{\dif c}{\sqrt{1-c^2}}  g^{p}_\tau(w,c) \nn \\
&  = \frac{1}{2 \pi} \frac{4 (\tau -1)^2 \left[4 \pi -3\times4^\tau B_{\frac{1}{4}}\left(\tau+\frac{1}{2},\frac{1}{2}\right)\right]}{3 \tau^2 \left(2 \tau^2+5 \tau+2\right)}\notag  \\
& = \frac{1}{\tau^2}+\frac{\alpha- \frac{9}{2} }{\tau} + \mathcal{O}(1).
\end{align}
Here $\alpha = \frac{3 \psi ^{(1)}\left(1/3\right)-\psi ^{(1)}\left(5/6\right)}{8 \pi \sqrt{3}}$ and the regularized part is
\begin{align}
N_1^{reg} =&-\frac{1}{2 \pi}
\int_0^1 \dif w \int_{-w/2}^1 \frac{\dif c}{\sqrt{1-c^2}} \frac{1}{\tau}\frac{8 c}{2 c w+w^2+1} \notag
\\
&+ \mathcal{O}(\tau^0) = - \frac{\alpha}{\tau} + \mathcal{O}(\tau^0)
\end{align}
where we have checked that the remainder $\mathcal{O}(\tau^0)$ is integrable.
In total we have $N_1=1/\tau^2 - 9/(2\tau)$ which gives $\hat F =1/2$ and the combination $\hat F + \hat B D - \frac{1}{2} A \hat E = 0$ using (\ref{res1l}). Hence
we find that the universal combination is ${\cal I} = 0$, in agreement with the other calculations.

\subsubsection{Coefficient $C$}

To treat the other two-loop coefficients one should group the terms so that the integral is ultraviolet finite
for $\tau>0$. This leads to:
\begin{align} \label{CC}
&&  \frac{C}{2 \tau} + \mathcal{O}(\tau^0) = b_1 - 8 b_2 + 4 a_1^2 = M + Q + \frac{1}{\tau^2} - \frac{2}{\tau}
\end{align}
with
\begin{align}
M =&  \frac{1}{(2 \pi)^2} \int_{x+y+z=0}
\mathrm{e}^{- 2 G(z)} \left[\mathrm{e}^{2 G(x)}-1\right]\left[\mathrm{e}^{2 G(y)}-1\right]  \nn \\
& + \left\{ \left[\mathrm{e}^{2 G(x)}-1\right] + \left[\mathrm{e}^{2 G(y)}-1\right] \right\} \left[\mathrm{e}^{-2 G(z)}-1\right],  \\
Q =& - 8 \frac{1}{(2 \pi)^2}  \int_{x+y+z=0}
\mathrm{e}^{- G(z)} \left[\mathrm{e}^{2 G(x)}-1\right]\left[\mathrm{e}^{G(y)}-1\right] \nn \\
& + \left\{ \left[\mathrm{e}^{2 G(x)}-1\right] + \left[\mathrm{e}^{G(y)}-1\right] \right\} \left[\mathrm{e}^{-G(z)}-1\right].
\end{align}
After symmetrization on the arguments $x,y,z$ one obtains (with $x+y+z=0$ implicit):
\begin{align}
& M+Q = N_1 + N_2, \quad N_1 =  \int_{x<1,y<1,z<1} f_1(x,y,z), \notag \\
& N_2 = 3 \int_{x<1,y<1,z>1} f_2(x,y,z),
\end{align}
since when two or more distances are larger than $1$ the integrand is identically zero. The functions $f_1$ and $f_2$ are easily obtained using Mathematica. For the sector $x<1,y<1,z<1$ we perform the same manipulations as above. We find
again (\ref{N1}) with now
\begin{align}
g_{\tau}(w,c) =& \frac{- 8 \left\{w \left[w^3 -2 c^4 w+6 c^2 w+3 c \left(w^2+1\right) \right]+1\right\}}{\tau w
   \left(2 c w+w^2+1\right)^2} \nn \\
   &+ \mathcal{O}(\tau^0).
\end{align}
The pole at $w=0$ can again be treated via a substraction with now:
\begin{align}
g^{p}_\tau(w,c)  = \frac{8 (\tau-1) w^{2 \tau-1}}{\tau (\tau+1)}  \Theta(0<w<1).
\end{align}
To see this we note that $g_\tau$ can be split in three terms: (i)
the coefficient of $w^{-1+2\tau}$ which gives the above $g^{p}$ (ii) the coefficient of $w^{-3+4 \tau}$ which
turns out to simplify into an expression yielding no pole in $w$ (due to a cancellation of poles between $M$ and $Q$) (iii) an expression regular at $w=0$. Hence we have again $N_1=N_1^{p} + N_1^{reg}$ where the contribution of the pole is:
\begin{align}
 N_1^{p} = & \frac{1}{2 \pi} \frac{2 (\tau-1) \left[4 \pi -3\times 4^\tau
   B_{\frac{1}{4}}\left(\tau+\frac{1}{2},\frac{1}{2}\right)\right]}{3 \tau^2 (\tau+1)} \notag \\
   = &- \frac{1}{\tau^2}+
   \frac{2  - \alpha}{\tau}  + \mathcal{O}(\tau^0).
\end{align}
and the regularized part is
\begin{align}
N_1^{reg} =&  \frac{1}{2 \pi}
\int_0^1 dw \int_{-w/2}^1 \frac{\dif c}{\sqrt{1-c^2}} \bigg\{ \notag\\
& \times \frac{1}{\tau} \left[ \frac{8 \left(2 c^4 w-2 c^2 w+c w^2+c+2 w\right)}{\left(2 c w+w^2+1\right)^2} \right]
 \notag\\
 & + \mathcal{O}(\tau^0) \bigg\} =  \frac{\alpha + \frac{1}{2}}{\tau} + \ldots
\end{align}
In total we have $N_1 = -1/\tau^2 + 5/(2\tau)$. Next, one easily checks that $N_2$ is finite, as the
expansion of the
integrand ($g_\tau$) up to $\mathcal{O}(\tau)$ at small $w$ is regular. Hence from (\ref{CC}) and below we find
cancellation of $1/\tau^2$ poles and the residue $C=1$. This is again in agreement with the other calculations.

\section{2-loop integrals for the  OPE method}
\label{s:2-loop-OPE}
\begin{figure}
\includegraphics[width=0.3\textwidth]{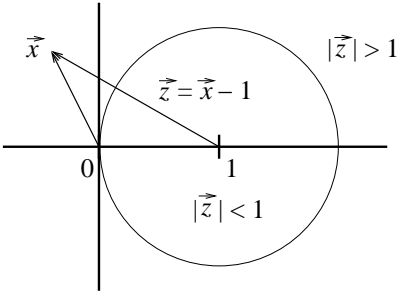}
\caption{The geometry used for \({J}_3\).}
\label{f:sec}
\end{figure}
\subsection{Integral ${I}_3$}\label{s:I3}
The first integral at 2-loop order  is
\begin{equation}\label{I3dbis}
{I}_{3} :=  \frac{1}{( 2\pi )^{2}} \int_{x}\int_{z}\left(\frac{|x-z|}{|x||z|} \right)^{2T} \Theta(|x-z|,|x|,|z|<L)
\end{equation}
We remind that this and other similar integrals are defined via finite-part prescription. Integral
(\ref{I3dbis}) is  a counter-term of the theory, which has itself sub-divergences. It can  be calculated by brute force, but the task is  simplified by  recognizing that  the structure of the renormalization group dictates these sub-divergences, or equivalently lower-order counter-terms. After subtraction of these counter-terms, which themselves are easy to calculate, the rest can be written as a  convergent integral and  be calculated.    This is the   road we will follow. To show how this works, consider the OPE-coefficient associated to $I_3$, which is the integral of three interactions projected onto a single one (not writing the replica content), which we note as (not yet setting $y=0$)\begin{equation}
\left.\left(\diagram{diag2L1diag2L1}\! \raisebox{-4mm}{\rule{0mm}{10mm}}^{x}_{z} \!\!\! {{\scriptstyle y}}\,\right| \diagram{vertex}\right) = \left(\frac{|x-z|}{|x-y||z-y|}
\right)^{2T}
\end{equation}
Subdivergences occur for $x\to y$ or \(z\to y\). Let us consider $x\to y$, keeping in mind that a similar term appears for $z\to y$. Then, we use the OPE to extract the small-$x$ behavior: \begin{eqnarray}\nn
\!\!\lefteqn{\left.\left(\diagram{diag2L1diag2L1}\! \raisebox{-4mm}{\rule{0mm}{10mm}}^{x}_{z}
\!\!\! {{\scriptstyle y}}\,\right| \diagram{vertex}\right)}  \\
&&= \left.\left( \diagram{diag1}\! \raisebox{-1.5mm}{\rule{0mm}{5mm}}^{x}_{y}
\,\right|{\bf 1}_y\right) \left.\left( {\bf 1} _y\diagram{vertex}{\!\!\scriptstyle z}
\,\right| \diagram{vertex}\right) \nn\\
&&+ \left.\left( \diagram{diag1}\! \raisebox{-1.5mm}{\rule{0mm}{5mm}}^{x}_{y}
\,\right| \nabla_j \theta(y)\right) \left.\left( \nabla_j \theta(y)\diagram{vertex} {\!\!\scriptstyle z}
\,\right| \diagram{vertex}\right) \nn\\
&&+ \left.\left( \diagram{diag1}\! \raisebox{-1.5mm}{\rule{0mm}{5mm}}^{x}_{y}
\,\right| \nabla_j \nabla_k\theta(y)\right)\times \nn\\
&&\ \quad\times   \left.\left( \nabla_j \nabla_k\theta(y)\diagram{vertex}
{\!\!\scriptstyle z}
\,\right| \diagram{vertex}\right) \nn\\
&&+ \left.\left( \diagram{diag1}\! \raisebox{-1.5mm}{\rule{0mm}{5mm}}^{x}_{y}
\,\right|  \nabla_j \theta(y) \nabla_k \theta(y)\right)\times \nn\\
&&\quad\  \times \left.\left( \nabla_j \theta(y) \nabla_k \theta(y)\diagram{vertex}
{\!\!\scriptstyle z}
\,\right| \diagram{vertex}\right) \nn\\
&&+\ldots  \label{cI3}
\end{eqnarray}
The above symbolic notations mean  that we first consider the OPE of the upper two interactions, and projecting in the order of their appearance: on the identity operator at position $y$,  ${\bf 1}_y$ (termed relevant counter-term below) times the contraction of ${\bf 1}_y$ with the interaction; then on $\nabla_j \theta(y)$ (termed relevant odd counter-term below) times the contraction of $\nabla_j \theta(y)$ with the interaction; then on
$\nabla_j\nabla_k \theta(y)$ times the contraction
of the latter with an interaction; and finally on the interaction itself times the projection of the interaction with the final interaction, termed marginal counter-term, or RG below, since this is the repeated counter-term from 1-loop calculations.
The coefficients in the order of their appearance are
\begin{align}
&\left.\left( \diagram{diag1}\! \raisebox{-1.5mm}{\rule{0mm}{5mm}}^{x}_{y}
\,\right|{\bf 1}_y\right) = |x-y|^{-2T} \label{ci4}\\
&\left.\left( {\bf 1} _y\diagram{vertex}{\!\!\scriptstyle
z}
\,\right| \diagram{vertex}\right)=1 \label{ci5}\\
&\left.\left( \diagram{diag1}\! \raisebox{-1.5mm}{\rule{0mm}{5mm}}^{x}_{y}
\,\right| \nabla_j \theta(y)\right) =\frac{2i(x-y)_j}{|x-y|^{2T}}\label{ci6} \\
& \left.\left( \nabla_j\theta(y)\diagram{vertex}
{\!\!\scriptstyle z}
\,\right| \diagram{vertex}\right) =\frac{i T (z-y)_j}{|z-y|^{2T}} \label{ci7}\\
&\left.\left( \diagram{diag1}\! \raisebox{-1.5mm}{\rule{0mm}{5mm}}^{x}_{y}
\,\right| \nabla_j \nabla_k\theta(y)\right) =\frac{i(x-y)_j(x-y)_k}{|x-y|^{2T}}
\label{cI8}\\
& \left.\left( \nabla_j\nabla_k\theta(y)\diagram{vertex}
{\!\!\scriptstyle z}
\,\right| \diagram{vertex}\right)\nn \\
&\qquad=\frac{i T [\delta_{jk}(z-y)^2-2(z-y)_j(z-y)_k]}{|z-y|^4} \label{cI9}\\
& \left.\left( \diagram{diag1}\! \raisebox{-1.5mm}{\rule{0mm}{5mm}}^{x}_{y}
\,\right|  \nabla_j \theta(y)   \nabla_k \theta(y)\right)=\frac{-2(x-y)_j(x-y)_k}{|x-y|^{2T}}\label{cI10} \\
& \left.\left( \nabla_j \theta(y) \nabla _k\theta(y)\diagram{vertex}
{\!\!\scriptstyle z}
\,\right| \diagram{vertex}\right)  \nn\\
&\qquad=\frac{-T^2(y-z)_j(y-z)_k}{|y-z|^4}\label{cI11}
\end{align}
Some remarks are in order: In a first-principle calculation,  the replica content would have to be written explicitly. Thus, e.g.\ the factor
of 2 in (\ref{ci6}) comes from the 2 possibilities to expand either the left our right replica.
In the effective action calculation presented earlier in this article, the first three terms of (\ref{cI3}) are absent, since they correspond to 1-particle reducible diagrams, which are automatically subtracted.

This allows us to give the list of
counter-terms. Note that if we only know that such an expansion exists, but do not know the coefficients, Eq.~(\ref{cI3}) can be obtained via simple Taylor-expansion.  (That is actually what we did, except to check the working of the OPE on an example.)

\paragraph*{Counter-terms for integral $I_3$:}
To simplify notations, we set \(y=0\).
First, the relevant counter-terms, minus the product of (\ref{ci4}) times (\ref{ci5}), since we want $I_3$ plus the counter-terms to be finite, (plus an analogous term with \(x\) and $z$ exchanged) are\begin{equation}\label{a18}
{I}_{3}^{\mathrm{c, rel}} =-  \frac{1}{( 2\pi )^{2}}
\int_{x}\int_{z}\left[\frac{1}{|x|^{2T}} +  \frac{1}{|z|^{2T}} \right]\Theta(|z-x|<L)\ .
\end{equation}This expression is   zero due
to analytical continuation.

Second, the odd relevant counter-terms, minus the product of (\ref{ci6}) times  (\ref{ci7}) (plus an analogous term with \(x\) and $z$ exchanged),\begin{eqnarray}\label{a19}
{I}_{3}^{\mathrm{c, odd}} &=&  \frac{2T}{( 2\pi )^{2}}
\int_{x}\int_{z}\left[ \frac{z \cdot x}{|z|^{2} |x|^{2T}}+ \frac{z \cdot x}{|x|^{2} |z|^{2T}}  \right]\nonumber \\
&&\qquad \qquad \qquad\times \Theta(|x|,|z|<L)\ .
\end{eqnarray}
These terms are
 zero due to analytical continuation, and zero due to parity.
The marginal counter-terms reads
\begin{eqnarray}\nn
{I}_{3}^{\mathrm{c, RG}} &=& -  \frac{T}{( 2\pi )^{2}}
\int_{x}\int_{z} \\
&& \bigg[ \frac{x^{2} z^{2}+2 (T-1) (z\cdot x)^{2}}{|x|^{2T} z^{4}} \Theta
(|x|<|z|<L)  \nn\\
&& +   \frac{x^{2} z^{2}+ 2 (T-1) (z\cdot x)^{2}}{|z|^{2T} x^{4}}  \Theta
(|z|<|x|<L) \bigg].  \nn \\\label{a22}
\end{eqnarray}
We have written two contributions:
the first \(\sim T^2\) is the repeated counter-term from RG, equal to minus $\rm(\ref{cI10})\times (\ref{cI11})$ (plus an analogous term with \(x\) and $z$ exchanged). Note that we have put $\Theta$-functions
to restrict the counter-term to the sector in which it is needed to subtract
the divergence.  The second contribution is $\rm (\ref{cI8})\times  (\ref{cI9})$   (plus an analogous term with \(x\) and $z$ exchanged). It is zero due to radial integration, but makes the integral absolutely convergent (it can e.g.\  be put on a computer).

Consider ${J}_3=\textstyle {L \frac{\partial}{\partial L} [{I}_3+ {
I}_{3}^{\mathrm{c, rel}} +
{I}_{3}^{\mathrm{c, odd}}+ {I}_{3}^{\mathrm{c, RG}}} ]_{L=1}$.
Using the mapping theorem of  appendix \ref{confmap}, to map  onto $|x-z|=L\stackrel{!}{=}1$,  using $z=x- \mathbf 1$ we can write with the respective counter-terms regrouped in one line
\begin{eqnarray}J_3&=&4\tau \left[{I}_3+ {I}_{3}^{\mathrm{c, rel}}
+
{I}_{3}^{\mathrm{c, odd}}+ {I}_{3}^{\mathrm{c, RG}} \right]_{L=1} \nn\\
&=&L \frac{\partial}{\partial L} \left[{I}_3+ {I}_{3}^{\mathrm{c, rel}} +
{I}_{3}^{\mathrm{c, odd}}+ {I}_{3}^{\mathrm{c, RG}} \right]_{L=1} \nn\\
&\equiv& \frac{1}{2\pi}\int_x \bigg[\frac{\max(1,|x|,|{\bf 1}-x|)^{-4\tau}}{|x|^{2T}|{\bf 1}-x|^{2T}}\nn\\
&&\qquad\qquad   -\frac1{|x|^{2T}}-\frac1{|{\bf 1}-x|^{2T}}
\nn\\
&& \qquad \qquad
-\frac{2T[x\cdot ({\bf 1}-x)] \max(|x|,|{\bf 1}-x|)^{-4\tau} }{|x|^{2T} |{\bf 1}-x|^{2} }
\nn\\
&& \qquad \qquad
-\frac{2T [x \cdot ({\bf 1}-x)]\max(|x|,|{\bf 1}-x|)^{-4\tau}}{|x|^{2}|{\bf 1}-x|^{2T}}
\nn\\
&&\qquad \qquad
-T \frac{x^{2} ({\bf 1}-x)^{2}+2 (T-1) [({\bf 1}-x)\cdot x]^{2}}{|x|^{2T} ({\bf 1}-x)^{4}}\nn\\
&& \qquad \qquad
\times  \Theta
(|x|<|{\bf 1}-x|) \,|{\bf 1}-x|^{-4\tau} \nn\\
&& \qquad \qquad -  T \frac{x^{2} ({\bf 1}-x)^{2}+ 2 (T-1) [({\bf 1}-x)\cdot x]^{2}}{|{\bf 1}-x|^{2T} x^{4}} \nn\\
&& \qquad \qquad
\times  \Theta
(|{\bf 1}-x|<|x|)\,|x|^{-4\tau} \bigg] \label{4.10}\\
&&= {J}_3^{\rm normal} + {J}_3^{\rm anomal}. \nn
\end{eqnarray}
There are {\em normal} and {\em anomalous} terms, where the latter are those for which  the difference between the above max-functions matters, even though in the limit of $\tau \to 0$ they all become 1.

The normal term is
\begin{align}
&{J}_3^{\rm normal}
= \frac{1}{2\pi}\int_x \bigg[\frac{1}{|x|^{2T}|{\bf 1}-x|^{2T}}-\frac1{|x|^{2T}}-\frac1{|{\bf 1}-x|^{2T}}\nn\\
&
-\frac{2T[x\cdot ({\bf 1}-x)]  }{|x|^{2T} |{\bf 1}-x|^{2} }
-\frac{2T [x \cdot ({\bf 1}-x)]}{|x|^{2}|{\bf 1}-x|^{2T}}
\nn\\
&
-T \frac{x^{2} ({\bf 1}-x)^{2}+2 (T-1) [({\bf 1}-x)\cdot x]^{2}}{|x|^{2T} ({\bf 1}-x)^{4}} \Theta
(|x|<|{\bf 1}-x|) \, \nn\\
&  -  T \frac{x^{2} ({\bf 1}-x)^{2}+ 2 (T-1) [({\bf 1}-x)\cdot x]^{2}}{|{\bf 1}-x|^{2T} x^{4}}  \Theta
(|{\bf 1}-x|<|x|)\, \bigg]\nn\\
&\times \max(1,|x|,|{\bf 1}-x|)^{-4\tau}. \label{4.11}
\end{align}
Since we have constructed the counter-terms such  that this  integral is  convergent in each sub-domain, one can take the limit $T\to 2$, i.e., $\tau\to0$.
This yields, up to terms of order $\tau$,
\begin{equation}\label{a20}
{J}_3^{\rm normal}+ \mathcal{O}(\tau)  = 0 \ ,
\end{equation}
since in that limit the integrand  vanishes identically.

We now turn to the anomalous terms. Those come from the regions where the divergences do not cancel exactly, either $x\to 0$, or $x\to 1$. Due to symmetry, we consider $x\to 0$ only (this gives a factor of 2). We write the different contributions as follows:
\begin{equation}{ J}_3^{\rm anomal} = (\rm\ref{4.10})-(\ref{4.11})= { J}_{3a}^{\rm anomal}+{ J}_{3b}^{\rm anomal} +\mathcal{O}(\tau)
\end{equation}
The first term is
\begin{eqnarray}
 { J}_{3a}^{\rm anomal}&=&-2 \frac{1}{2\pi}\int_{x}\left[  \frac{1-\max(1,|x|,|{\bf 1}-x|)^{-4\tau}}{|x|^{2T}}\right]\nn\\
 &&\qquad \qquad \times \Theta(|x|<1/2)\ ,
\label{4.14}\end{eqnarray}
but one could have taken a smaller number than \(1/2\) for the cutoff. This gives
\begin{eqnarray}
 { J}_{3a}^{\rm anomal}&=&2\frac{1}{2\pi}\int_{x}\left[  \frac{|{\bf 1}-\vec x|^{-4\tau}-1}{|\vec x|^{2T}}\right]\Theta(|\vec x|<1/2)\nn\\
 && \qquad\qquad \times   \Theta(\vec x\cdot {\bf 1}<0),
\end{eqnarray}
where we have made explicit that $\vec x$ is a vector, see Fig.\ \ref{f:sec}. Going to complex coordinates gives
\begin{align}
& { J}_{3a}^{\rm anomal}= \frac{2}{2\pi}\int_{x}\left[  \frac{(1-x)^{-2\tau} (1-\bar x)^{-2\tau}-1}{(x\bar x)^{T}}\right]\nn\\
 &\qquad\qquad \qquad \qquad \times \Theta(| x|<1/2)  \Theta(\Re(x)<0) \nn\\
&= \frac{2}{2\pi}\int_{x}\left[  \frac{2\tau(x+\bar x)+4\tau^2x \bar x+2\tau(2\tau+1)(x^2+\bar x^2)/2}{(x\bar x)^{T}}\right]\nn\\
 &\qquad\qquad  \quad \times\Theta(| x|<1/2)\  \Theta(\Re(x)<0) + \ldots\nn\\
&= 2\int_0^{\frac 12} {\rmd x}\, x^{4\tau-3} \frac{1}{2\pi}\int_{\pi/2}^{3\pi/2} \bigg[ 2\tau x(\rme^{i\phi}+\rme^{-i\phi})+4\tau^2x^2\nn\\
& \qquad\qquad\qquad\qquad\quad~~ +\tau(2\tau+1)x^2(\rme^{2i\phi}+\rme^{-2i\phi})+...\bigg] \nn\\
&= \mathcal{O}(\tau),
\end{align}
since the only (logarithmically at \(\tau=0\)) diverging terms are the term \(\sim 4\tau^2x^2\to \mathcal{O}(\tau)\) after integration, and the last one, which vanishes due to angular integration.

The  interesting term is
\begin{align}
& { J}_{3b}^{\rm anomal}\nn\\
&=-\frac{4 T}{2\pi}\int_{x} \frac{\vec x\cdot ({\bf 1}-\vec x)  }{|\vec x|^{2T} |{\bf 1}-\vec x|^{2} } \Theta(|\vec x|<1/2)  \nn\\
&\qquad\qquad\times [\max(|\vec x|,|{\bf 1}{-}\vec x|)^{-4\tau}-\max(1,|\vec x|,|{\bf 1}{-}\vec x|^{-4\tau})]
 \nn\\
\nn\\
&=- \frac{4 T}{2\pi}\int_{x} \frac{\vec x\cdot ({\bf 1}-\vec x) [|{\bf 1}-\vec x|^{-4\tau}-1] }{|\vec    x|^{2T} |{\bf 1}-\vec x|^{2} }
\nn\\
& \qquad\qquad \times \Theta(|\vec x|<1/2) \Theta(\vec x\cdot {\bf 1}>0)\nn\\
&=- \frac{4 T}{2\pi}\int_{x} \frac{\frac12(x+\bar x-2x \bar x)2 \tau(x+\bar x+...) }{(x\bar x)^T(1-x)(1-\bar x) }
\nn\\
& \qquad\qquad \times
\Theta(|x|<1/2)\Theta(\Re (x)>0) \nn\\
&= - \frac{4 T\tau}{2\pi}\int_{-\pi/2}^{\pi/2}\rmd\phi\,\int_{0}^{\frac12}\rmd x\, x^{4\tau-3}
\nn\\
& \qquad\qquad \times \frac{(x\rme^{i \phi}+x \rme^{-i \phi}-2x^2) (x\rme^{i \phi}+x\rme^{-i \phi}+...) }{(1-x\rme^{i \phi})(1-x \rme^{-i \phi}) } \nn\\
&= - \frac{4 T\tau}{2\pi}\int_{-\pi/2}^{\pi/2}\rmd\phi\,   (\rme^{i \phi}+\rme^{-i \phi})^2\int_{0}^{\frac12}\rmd x\, x^{4\tau-1}+...\nn\\
&=-T  +\mathcal{O}(\tau).
\label{4.17}\end{align}
The remaining anomalous terms only diverge logarithmically at small $x$, thus when expanding the factors of \(\max(...)^{-4\tau}\), this yields an additional factor of \(\tau |x|\), ensuring convergence; thus these terms can be neglected.
Therefore,
calculating (\ref{a22}) analytically, and using (\ref{4.10}) we conclude that {}
\begin{eqnarray}\label{a24}
{ I}_3 &=& -  { I}_{3}^{\mathrm{c, RG}}  +\frac1{4\tau} { J}_{3b}^{\rm anomal}\nn\\
&=&
\frac{(1-\tau )^2}{2} \frac{L^{4\tau}}{\tau^2}  -{2(1-\tau)}\frac{L^{4\tau}}{4\tau}+ \mathcal{O}(\tau^{0}) \ .\qquad
\end{eqnarray}

\subsection{Integral ${ I}_4$}\label{s:I4}
\begin{equation}\label{I4dbis}
{ I}_{4} =  \frac{1}{( 2\pi )^{2}}
\int_{x}\int_{z}\left(\frac{|x-z|}{|x|^{2} |z|} \right)^{T}\Theta(|x-z|,|x|,|z|<L).
\end{equation}
We follow the same strategy as for \({ I}_3\), first identifying the counter-terms.  The relevant counter-term to be added is
\begin{equation}\label{a27}
{ I}_{4}^{\mathrm{c,rel}} = -\frac{1}{(2\pi)^{2}}\int_{x}\int_{z} \frac{1}{|x|^{2T}} \,\Theta(|x-z|<L).
\end{equation}
We can again add the following (relevant) odd (i.e., vanishing) counter-term
\begin{equation}\label{a28}
{ I}_{4}^{\mathrm{c,odd}} = \frac{T}{(2\pi)^{2}} \int_{x}\int_{z}\frac{ x\cdot
z}{|x|^{2T} z^{2}}\,\Theta(|x|,|z|<L).
\end{equation}
This time, there are two marginal counter-terms.
The marginal counter-term for $x\to 0$ comes from the insertion of
(\ref{a7}):
\begin{eqnarray}\label{a29}
{ I}_{4}^{\mathrm{c,RG,1}}&=&- \frac{T}{(2\pi)^{2}} \int_{x}\int_{z}\frac{ (T-2) ( x\cdot
z)^{2} +x^{2} z^{2}}{2 |x|^{2T} |z|^{4}}
\nn\\
&& \qquad\qquad\qquad \times\Theta (|x|<|z|<L).\qquad
\end{eqnarray}
The marginal counter-term for $z\to 0$ comes from the sub-divergence
(\ref{a10}):
\begin{equation}\label{a29b}
{ I}_{4}^{\mathrm{c,RG,2}}=- \frac{1}{(2\pi)^{2}} \int_{x}\int_{z}\frac{1}{ |x|^{T} |z|^{T}}\,\Theta (|z|<|x|<L)
\end{equation}
Consider now the combination \begin{eqnarray}
{J}_4 &:=&L\frac{\partial }{\partial L} \left[ { I}_{4}  +{ I}_{4}^{\mathrm{c,rel}}+{ I}_{4}^{\mathrm{c,odd}}   +{ I}_{4}^{\mathrm{c,RG,1}}+{ I}_{4}^{\mathrm{c,RG,2}}\right]_{L=1}  \nn\\
&=& \frac{1}{2\pi}\int_x \bigg[ \left(\frac{1}{|x|^{2} |\boldsymbol1-x|} \right)^{\!T} \max(|x|,1,|\boldsymbol1-x|)^{-4\tau}  \nn\\
&& \qquad\qquad -\frac{1}{|x|^{2T}} \nn\\&&\qquad \qquad -T\frac{ x\cdot
({\bf 1}-x)}{|x|^{2T} |1-x|^{2}}\,  \max(|x|,|\boldsymbol1-x|)^{-4\tau}
\nn\\
&& \qquad\qquad-T\frac{ (T-2) [ x\cdot
({\bf 1}-x)]^{2} +x^{2} ({\bf 1}-x)^{2}}{2 |x|^{2T} |{\bf 1}-x|^{4}} \times
\nn\\
&& \qquad\qquad\qquad   \times\Theta (|x|<|{\bf 1}-x|) |\boldsymbol1-x|^{-4\tau}\nn\\
&&\qquad\qquad-\frac{1}{ |x|^{T} |\boldsymbol1-x|^{T}}\,\Theta (|\boldsymbol1-x|<|x|)|x|^{-4\tau}\bigg]\nn\\
&=& { J}_4^{\rm normal}+{ J}_4^{\rm anomal}\ . \end{eqnarray}
The normal contribution is\begin{align}
&{ J}_4^{\rm normal} =  \frac{1}{2\pi}\int_x \bigg[ \left(\frac{1}{|x|^{2} |\boldsymbol1-x|} \right)^{\!T} -\frac{1}{|x|^{2T}}
\nn\\
&   -T\frac{ x\cdot
({\bf 1}-x)}{|x|^{2T} |1-x|^{2}}\nn\\
&  -T\frac{ (T-2) [ x\cdot
({\bf 1}-x)]^{2} +x^{2} ({\bf 1}-x)^{2}}{2 |x|^{2T} |{\bf 1}-x|^{4}} \,\Theta (|x|<|{\bf 1}-x|)\nn\\
&-\frac{1}{ |x|^{T} |\boldsymbol1-x|^{T}}\,\Theta (|\boldsymbol1-x|<|x|)\bigg]\max(|x|,1,|\boldsymbol1-x|)^{-4\tau}\nn\\
&= \mathcal{O}(\tau)\ .
\end{align}
The integral is convergent,
since the counter-terms  where constructed in order to cancel {\em exactly} all sub-divergences. Taking the limit of $\tau \to 0$, one finds that the integrand identically vanishes, which shows that the expression is \(\mathcal{O}(\tau)\).

There are two potentially anomalous terms,
\begin{equation}
{ J}_4^{\rm anomal} = { J}_{4a}^{\rm anomal} +{ J}_{4b}^{\rm anomal}+ ...\ .
\end{equation}
For the first term, a calculation identical to (\ref{4.14}) shows that
\begin{eqnarray}
{ J}_{4a}^{\rm anomal} &=&   \frac{1}{2\pi}\int_x \frac{1}{|x|^{2T}}  \left[\max(|x|,1,|\boldsymbol1-x|)^{-4\tau}-1  \right] \nn\\
&=& \mathcal{O}(\tau)\ .
\end{eqnarray}
The second term is, up to a prefactor of \(1/4\)  identical  to (\ref{4.17}), and reads
\begin{align}
& { J}_{4b}^{\rm anomal}=\frac{ T}{2\pi}\int_{x} \frac{\vec x\cdot ({\bf 1}-\vec x)  }{|\vec x|^{2T} |{\bf 1}-\vec x|^{2} }
\Theta(|\vec x|<1/2) \times \nn\\
 & \qquad\quad \times \big[\max(1,|\vec x|,|{\bf 1}-\vec x|^{-4\tau})-\max(|\vec x|,|{\bf 1}-\vec x|)^{-4\tau}\big]\nn\\&=   \frac{ { J}_{3b}^{\rm anomal}}4=-\frac T4\ .
\end{align}
Calculating explicitly the integrals (\ref{a29}) and (\ref{a29b}) yields \begin{eqnarray}\label{I4rbis}
{ I}_4 &=&-{ I}_{4}^{\mathrm{c,RG,1}}-{ I}_{4}^{\mathrm{c,RG,2}}+\frac 1{4\tau }{J}_4^{\rm anomal}\nn\\
&=& \frac{(1-\tau )^2 L^{4 \tau }}{16 \tau ^2}
   +\frac{L^{4\tau}}{8\tau^2}-\frac{(1-\tau) L^{4\tau}}{8\tau} + \mathcal{O}(\tau^0)\ .\qquad
\end{eqnarray}We remark that the contributions proportional to $T^{2}\sim (1-\tau)^{2}$ from ${ I}_{3}$ and ${ I}_{4}$ cancel. We will see later that the only terms which appear in the RG-functions come from the anomalous terms.

\subsection{Integral ${ I}_5$}\label{s:I5}
The integral  ${ I}_5$ is
\begin{equation}\label{I5dbis}
{ I}_5=\frac{1}{(2\pi)^2}\int_x\int_z \frac{\Theta(|x|,|z|,|x-z|<L)}{|x|^{T}|z|^T}
\end{equation}
Clearly, the marginal counter-terms are subtracted by $(I_2)^2$:
\begin{align}
&{ I}_5-{ I}_2^2=\nn\\
&\frac{1}{(2\pi)^2}\int_x\int_z \frac{\Theta(|x|,|z|,|x-z|<L)-\Theta(|x|,|z|<L)}{|x|^{T}|z|^T}
\end{align}
Using the mapping prescription, we find
\begin{align}
&L\frac{ \rmd}{\rmd L} \left[{ I}_5-{ I}_2^2\right]_{L=1} \nn\\
&\  =\frac{1}{2\pi}\int_x \frac{\max(|x|,|{\bf 1}{-}x|,1)^{-4\tau}-\max(|x|,|{\bf 1}{-}x|)^{-4\tau}}{|x|^{T}|{\bf 1}-x|^T}\nn\\
&=\mathcal{O}(\tau)
\end{align}
Thus there is again no genuine contribution. This is not
astonishing for a bubble-chain.
The final result is \begin{equation}\label{I5rbis}
{ I}_5 = { I}_2^2+\mathcal{O}(\tau^0)
=\frac{L^{4\tau}}{4\tau^2}  + \mathcal{O}(\tau^0)\ .
\end{equation}

\subsection{Integral ${ I}_{6a}$}\label{s:I6a}
The integral ${ I}_{6a}$ is
\begin{equation}\label{I6adbis}
{ I}_{6a}= \frac{1}{(2\pi)^{2}}
\int_{x} \int_{z}
\frac{(x-z)^{2}\, \Theta(|x-y|,|y-z|,|z-x|<L)}{|x-y|^{T} |y-z|^{T} |z-x|^{T}}
\end{equation}
It has two subdivergences, due to 1-loop counter-terms for the coupling $g$: for $x-y\to 0$ and for $y-z \to 0$,
\begin{align}\label{lf5}
&{ I}_{6a}^{\mathrm{c,RG}} =  \frac{-1}{(2\pi)^{2}}
\int_{x} \int_{z}\\
& \qquad\qquad\bigg[ \frac{1}{|x-y|^{T}}
\frac{(y-z)^{2}}{|y-z|^{2T}}  \Theta (|x-y|<|y-z|<L)  \nn\\
&\qquad\qquad+\frac{1}{|z-y|^{T}}
\frac{(x-y)^{2}}{|x-y|^{2T}}  \Theta (|z-y|<|x-y|<L)\bigg] \nn
\end{align}
We have  explicitly written the subdivergence (first factor)
times the remaining term (second factor) times the restriction on the
sector (third factor, $\Theta$-function).  Note that we have used our
freedom to put the second factor at a point of our choice.

We now note that (i) when combining the integrands of ${
I}_{6a}+{ I}_{6a}^{\mathrm{c,RG}}$, there are no subdivergences,
and (ii)  the integrand vanishes in the limit of $\tau\to
0$. Therefore,
\begin{equation}\label{lf6}
{ I}_{6a} + { I}_{6a}^{\mathrm{c,RG}} = \mbox{finite}\ ,
\end{equation}
and
\begin{equation}\label{I6arbis}
{ I}_{6a} = -{ I}_{6a}^{\mathrm{c,RG}}+\mathcal{O}(\tau^0)= \frac{L^{6\tau}}{6\tau^2}+\mathcal{O}(\tau^0)\ .
\end{equation}

\subsection{Integral ${ I}_{6b}$}\label{s:I6b}
The integral ${ I}_{6b}$ is
\begin{eqnarray}\label{I6dbis}
{ I}_{6b}&=& \frac{1}{(2\pi)^{2}}
\int_{x} \int_{z}
\frac{(x-z)\cdot (y-z)}{|x-y|^{T} |y-z|^{T} |z-x|^{T}}\nn\\
&& \qquad\qquad\quad  \times \Theta(|x-y|,|y-z|,|z-x|<L).\qquad \ \
\end{eqnarray}
It has a sole subdivergence, when $x-y\to 0$. It is subtracted by
\begin{align}\label{lf7}
&{ I}_{6b}^{\mathrm{c,RG}}= - \frac{1}{(2\pi)^{2}}
\int_{x} \int_{z} \\
&\qquad\ \  \bigg[ \frac{1}{|x-y|^{T} }\times
\frac{(y-z)^{2}}{|y-z|^{2T}}\times  \Theta (|x-y|<|y-z|<L)\bigg] \nn
\end{align}
We have decided to use this specific form of the counter-term (for later convenience) but a symmetrized version would also be possible.
We claim that ${ I}_{6b}+{ I}_{6b}^{\mathrm{c,RG}}=\mathcal{O}(\tau^0)$. To prove this, we  set $y=0$,  vary $L$, and map onto $|x|=1$
\begin{eqnarray}\label{lf8}
{ J}_{6}&:=&\left.\frac{L \partial }{\partial L}\right|_{L=1} \left({
I}_{6b}+{ I}_{6b}^{\mathrm{c,RG}} \right) \nn\\
&=& \frac{1}{2\pi} \int_{z}
\bigg[ \frac{z\cdot({z-\bf 1}) \max(|z|,1,|{\bf 1}-z|)^{-6\tau}}{|z|^{T} |z-{\bf 1}|^{T}} \nn\\
&&\qquad\quad  -\frac{\Theta (|z|>1)\max(|z|,|{\bf 1}-z|)^{-6\tau}}{|z|^{2T-2}} \bigg],\qquad
\end{eqnarray}
The integral is finite, thus one can go to the critical dimension. This yields
\begin{eqnarray}\label{lf9}
{ J}_{6}&:=&\left.\frac{L \partial }{\partial L}\right|_{L=1} \left({
I}_{6b}+{ I}_{6b}^{\mathrm{c,RG}} \right) \nn\\
&=& \frac{1}{2\pi} \int_{z}
\left[ \frac{z\cdot({z-\bf 1})}{z^{2} (z-{\bf 1})^{2}}  -\frac{\Theta (|z|>1)}{z^{2}} \right]
\ .\end{eqnarray}
The integral can be split into two parts, ${ J}_{6}={ J}_{6}^{<}+{
J}_{6}^{>}$, calling ${ J}_{6}^{<}$  the part where $|z|<1$, and
the other term the part for $|z|>1$. We get
\begin{eqnarray}\nn
{ J}_{6}^{<} &=&\frac{1}{2\pi} \int \rmd^{2}z \,\Theta (|z|<1)
\frac{z\cdot({z-\bf 1})}{z^{2} |z-{\bf 1}|^{2}}  \\
&=&\frac{1}{4}\frac{1}{2\pi} \int \rmd^{2}z \,\Theta (|z|<1)  \frac{|2z-{\bf 1}|^{2} -1}{z^{2} |z-{\bf 1}|^{2}} \nn \\
&=&\frac{1}{4}\frac{1}{2\pi} \int \rmd^{2}z \,\Theta (|z|<1)
\frac{(2z-1) (2z^*-1)-1}{z z^{*}(z-1) (z^{*}-1)}\ ,\qquad \ \ \label{lf10}
\end{eqnarray}
where in the last line we have introduced complex coordinates. This
gives
\begin{equation}\label{lf11}
{ J}_{6}^{<} =-\frac{1}{2}\frac{1}{2\pi} \int \rmd^{2}z \,\Theta (|z|<1)
\left[\frac{1}{z (1-z^{*})} +\frac{1}{z^{*} (1-z)} \right]
\end{equation}
Since $|z|<1$, Taylor expansion can be used around zero. It shows that
there are only terms which vanish upon angular integration. Therefore
\begin{equation}\label{lf12}
{ J}_{6}^{<} =0\ .
\end{equation}
We now turn to ${ J}_{6}^{>}$:
\begin{eqnarray}\label{lf13}
{ J}_{6}^{>} &=&-\frac{1}{2}\frac{1}{2\pi} \int \rmd^{2}z \,\Theta
(|z|>1)  \nn\\
&& \qquad \times \left[\frac{1}{z (1-z^{*})} +\frac{1}{z^{*} (1-z)} +\frac{2}{zz^{*}} \right].\qquad
\end{eqnarray}
Using that the Taylor-expansion for $z\to \infty$ is
\begin{equation}\label{lf14}
\frac{1}{1-z} = -\frac{1}{z} \frac{1}{1-1/z} = -\frac{1}{z}\left(1+\frac{1}{z}+\frac{1}{z^{2}}+\dots  \right)\ ,
\end{equation}
we conclude that the terms of order $1/ (zz^{*})$ cancel, and the
remaining terms vanish upon angular integration. Therefore
\begin{equation}\label{lf15}
{ J}_{6}^{>} =0\ .
\end{equation}
Putting the pieces together, we conclude that
\begin{equation}\label{lf16}
{
I}_{6b}+{ I}_{6b}^{\mathrm{c,RG}} =\mbox{finite}\ .
\end{equation}
We checked that one can also map onto \(|x-z|=1\), yielding
\begin{eqnarray}
{ J}_{6b}&=&\left.\frac{L \partial }{\partial L}\right|_{L=1} \left({
I}_{6b}+{ I}_{6b}^{\mathrm{c,RG}} \right) \\
&=& \frac{1}{2\pi} \int_x \left[ \frac{\mathbf 1\cdot (\mathbf 1-x)}{|x|^2 |\mathbf 1-x|^2} -\frac{\Theta(|x|<1)}{|x|^2} \right] \nn\\
& =& \frac{1}{2\pi} \int_x \bigg[ \frac{1}{2xx^*} -\frac{1}{2(1-x)(1-x^*)}\nn\\
&& \qquad +\frac{1}{2x x^*(1-x)(1-x^*)}-\frac{\Theta(|x|<1)}{x x^*}\bigg] = 0\ ,\nn
\end{eqnarray}
using again the Taylor-expansion method, this time separating into $ |x|<1$
and $|x|>1$.
Thus, renaming the variables,
\begin{eqnarray}\label{I6brbis}
{ I}_{6b}  &=&-{ I}_{6b}^{\mathrm{c,RG}}+\mathcal{O}(\tau^0)\nn\\
&= & \frac{1}{(2\pi)^{2}}
\int_{x} \int_{z} \frac{1}{|x|^{T} }
\frac{(z)^{2}}{|z|^{2T}}  \Theta (|x|<|z|) +\mathcal{O}(\tau^0)\nn\\
&=& \frac{ L^{6 \tau }}{12 \tau ^2}+\mathcal{O}(\tau^0).
\end{eqnarray}
We note a consistency relation between integrals $I_{6a}$ and $I_{6b}$: Rewriting the numerator of (\ref{I6adbis}) as $(x-z)^2=(x-y)^2+(y-z)^2+2(x-y)\cdot(y-z)$, we deduce that $I_{6a}=2  I_{6b}$. This is indeed satisfied by our results (\ref{I6arbis}) and (\ref{I6brbis}).

\section{The conformal mapping theorem}
\label{confmap}

In this appendix we discuss a convenient method to extract the divergence of an integral, known as the conformal mapping theorem
\cite{WieseDavid1995,WieseDavid1997,WieseHabil}.
This method we use to extract the \(1/\tau\) contribution of the 2-loop integrals.

In general, we have to compute integrals over $N$ points, equivalent to $N(N-1)$ distances
$x,y,\ldots$, of the form
\be
{ I}^{ S}=\int_{\max(x,y,\ldots)\le L} \tilde f(x,y,\ldots)
\label{t:I integrale}
\ee
with a homogeneous function $f$ such that the integral has a conformal weight
(dimension in $L$) $\kappa$: $I(\E)\sim L^\kappa$.
For the integrals which appear in $N$-loop diagrams, this weight is simply
$\kappa = 2N\tau $ (renormalization of $g$), or $\kappa = 2(N+1)\,\tau $ (renormalization of $\sigma$).

The integral over the distances is defined by the integral over $N-1$ points, keeping one chosen point fixed. The residue is extracted from the dimensionless integral
\begin{eqnarray}\label{J2}
{ J}^{ S}&:=&\kappa\, L^{-\kappa} { I}^{ S}
=L^{-\kappa}\,
L{\partial\over\partial L}{ I}^{ S} \\\
&=&
L\,\int_{\max(x,y,\ldots)= L} \tilde f(x,y,\ldots)\,\max(x,y,\ldots)^{-\kappa}
\label{t:J integrale}
\ . \nn
\end{eqnarray}
Note that the last factor is nothing but $L^{-\kappa}$, but we have chosen this form as will become apparent shortly.
The domain of integration can be decomposed into ``sectors", for instance
\be
\{\ldots<y<x=L\}\ ,\ \{\ldots<x<y=L\}\ ,
\label{t:secteurs}
\ee
and we can map these different sectors onto each other by global conformal
transformations.
As we show below,  we can for instance rewrite the integral
\eq{t:J integrale} sector by sector as
\begin{eqnarray}
{ J}^{S}&\equiv&
L\,\int_{x= L;\,y,\ldots} \tilde f(x,y,\ldots)\,\max(x,y,\ldots)^{-\kappa}\nn\\
&\equiv&
L\,\int_{y= L;\,x,\ldots} \tilde f(x,y,\ldots)\,\max(x,y,\ldots)^{-\kappa}
\label{t:diff sect}
\ .\qquad
\end{eqnarray}
The constraint on the maximum of the distances is replaced by the constraint on an arbitrarily chosen distance. Due to our normalization $\int_z:=\frac1{2\pi}\int
\rmd^2 z$, this is
equivalent to fixing the both endpoints of this largest
distance.

To formalize the above, consider the integral over a
function $\tilde f$ at order $N-1$ loops.
Suppose $ \tilde f (z_{1}, \dots, z_{N})$ is a homogeneous function of
dimension
$-2 (N-1)+\kappa$. Define the  function
\begin{equation}
  f (z_{1}, \dots
,z_{N}, ):= \tilde
 f (z_{1}, \dots
,z_{N}, ) \, \left[\max_{i<j}\{|z_i-z_j|\}\right]^{-\kappa}
\end{equation}
(This is the combination which appears in (\ref{J2}).
Then the
integral  over $z_{1}, \dots , z_{N-1}$ (the relative coordinates
between points), cut off by ${ C}(z_1,\ldots,z_N):=\prod_{i<j}\Theta(|z_i-z_j|<L)$
\begin{equation}\label{lf33}
 I_{N} (a,L) := \int_{z_{1},\dots, z_{N-1}} f (z_{1}, \dots
,z_{N}) \, { C}(z_1,\ldots,z_N)
\end{equation}
has $L$-dimension 0. Consider a  sector $ S$ (ordering of the
distances).  Be $ x_{\alpha }:=|z_{i}-z_{j}|$, with $1\le \alpha \le m
:= N (N-1)/2$.  Then ${ S}:=\{z_{1}, \dots, \},\
\mbox{s.t.}\ x_{1}<x_{2}< \dots < x_{m}$. (Actually, we have chosen
the labeling of the distances $x_{\alpha }$ to account for the
ordering. This is not always the most practical thing to do.)  Also
define the characteristic function $\chi _{ S} (x_{1},\dots ,
x_{m})$ of a sector $ S$ as being 1 if all distances satisfy the
inequalities of the sector and 0 otherwise. The $L$-derivative of the
integral restricted to the sector $ S $ is
\begin{eqnarray}\nn
{ J}^{ S}&:=& L \frac{\partial }{\partial L} { I}_{N}^{
S} (a,L) \\  \label{lf34}
&=& \int f (z_{1}, \dots)\Big| _{x_{m}=L}  \chi _{ S} (x_{1}, \dots , x_{m})\ .
\end{eqnarray}
The conformal mapping theorem
\cite{WieseDavid1995,WieseDavid1997,WieseHabil}, whose proof we
reproduce below for completeness,  states that {\em if} the
integral (\ref{lf34}) is Riemann-integrable everywhere (or at least via finite-part prescription), then
\begin{equation} \label{mapped}
{ J}^{ S} \equiv \int f
(z_{1}, \dots)\Big| _{x_l=L}  \chi
_{S} (x_{1}, \dots ,  x_{m})\ .
\end{equation}
In words: The above integral can be evaluated by fixing any of the
distances to be $L$ (or 1 equivalently).  (\ref{t:diff sect})  is then a simple corollary of (\ref{mapped}).
\medskip

To prove the latter, we start from (\ref{lf34}). First of all, since $x_{m}=L$,
and introducing a $\delta$-function to enforce it, ${ J}^{ S}$
becomes
\begin{equation}\label{p1}
{ J}^{ S} = \int f (z_{1}, \dots) \delta (x_{m}- L)    \chi
_{ S}
(x_{1}, \dots ,  x_{m}) \ .
\end{equation}
We now aim at integrating over  the   distances $x_{1},\dots , x_{m}$  instead of the
coordinates. For an arbitrary function  $g$ of  the latter distances, this is\begin{align}\label{p2}
&\int \rmd^{2}  z_1 \dots \rmd^{2} z_{N-1}\, g (x_{1},\dots , x_{m}) \nn\\
&=\int
\rmd x_1 \dots  \rmd x_m \
  \mu (x_{1},\dots , x_{m}) g (x_{1},\dots , x_{m})\ .
\end{align}
The measure is easily constructed as
\begin{eqnarray}
 \mu (x_{1},\dots , x_{m}) &=&\int \rmd ^2
z_{1} \dots \rmd^{2} z_{N-1}\, \delta (x_{1}-|{z_{1}-z_{2}}|)\times \nn\\
&& \quad \times \dots
\delta (x_{m}- |z_{N-1}-z_{N}|) \ ,\
\end{eqnarray}
where the $\delta$-distributions enforce the $x_{i}$'s to be the
distances between the $z_{j}$'s.

We now want to map onto $x_{l}=L$. To achieve this, we can always do
the integration over $x_l$ last. This gives for ${ J}^{ S}$
\begin{eqnarray}\label{p3}
{ J}^{ S} &=&  \int \rmd x_{l}
\displaystyle \int \rmd x_1 \dots \rmd x_{l-1} \rmd x_{l+1} \dots
\rmd x_m\
 \mu  (x_{1},\dots , x_{m})\nn\\
 && \times  \delta (x_{m}-L )\,f (x_{1},\dots , x_{m}) \, \chi
_{ S} (x_{1}, \dots ,  x_{m})\ .\qquad
\end{eqnarray}
We now make a change of variables. For all $i$ but $l$, set
\begin{equation}\label{p4}
x_{i}:= \tilde x_{i} x_{l}/L\ .
\end{equation}
We also define $\tilde x_{l}:=L $, and introduce this into
(\ref{p3}) as $1=\int \rmd \tilde x_{l}\, \delta
(\tilde x_{l} -L)$:
\begin{eqnarray}\label{p5}
{ J}^{ S} = \int \rmd x_{l}
&\displaystyle \int& \rmd \tilde x_1 \dots
\rmd \tilde  x_m\
 \mu  (\tilde x_{1}, \dots  ,\tilde  x_{m}) \, \delta
(\tilde x_{l} -L)\nonumber \\
&& \times \,f (\tilde x_{1}, \dots , \tilde x_{m})  \chi _{ S}
(\tilde x_{1}, \dots , \tilde x_{m})\qquad \qquad\nonumber \\
&& \times\, \delta (\tilde x_{m}  x_{l}/L-L) \, \frac{L}{x_{l}}\ .
\end{eqnarray}
Note that the factor of $\frac L{x_{l}}$ consists of
$\left(\frac{x_{l}}{L} \right) ^{N (N-1)/2-1}$
from the terms $\rmd \tilde x_{i}$ but $\rmd \tilde x_{l}$; a factor
of $\left(\frac{x_{l}}{L} \right) ^{(N-1) (2-\frac{N}{2})}$ from the
measure; and a factor of $\left(\frac{x_{l}}{L} \right) ^{-2 ( N-1)}$
from $f$.  Using that
\begin{equation}\label{p6}
\int \rmd x_{l} \,  \delta (  \tilde x_{m}  x_{l}/L -L )
\frac{L}{x_{l}}  = 1\ ,
\end{equation}
we obtain
\begin{eqnarray}\label{p7}
{ J}^{ S} =
&\displaystyle \int& \rmd \tilde x_1 \dots
\rmd \tilde  x_m \
 \mu  (\tilde x_{1}, \dots  ,\tilde  x_{m} ) \, \delta
(\tilde x_{l} -L )\nonumber \\
&& \times \,f (\tilde x_{1}, \dots , \tilde x_{m}) \Theta (\tilde  x_{m}/\tilde x_{1}<L/a)
\nn\\
 && \times \chi _{ S}
(\tilde x_{1}, \dots , \tilde x_{m})\ .\qquad \qquad
\end{eqnarray}
Dropping the tildes, this is nothing but (\ref{p3}) with $x_m$
replaced by $x_{l}$ which completes the proof.


\begin{thebibliography}{65}%
\makeatletter
\providecommand \@ifxundefined [1]{%
 \@ifx{#1\undefined}
}%
\providecommand \@ifnum [1]{%
 \ifnum #1\expandafter \@firstoftwo
 \else \expandafter \@secondoftwo
 \fi
}%
\providecommand \@ifx [1]{%
 \ifx #1\expandafter \@firstoftwo
 \else \expandafter \@secondoftwo
 \fi
}%
\providecommand \natexlab [1]{#1}%
\providecommand \enquote  [1]{``#1''}%
\providecommand \bibnamefont  [1]{#1}%
\providecommand \bibfnamefont [1]{#1}%
\providecommand \citenamefont [1]{#1}%
\providecommand \href@noop [0]{\@secondoftwo}%
\providecommand \href [0]{\begingroup \@sanitize@url \@href}%
\providecommand \@href[1]{\@@startlink{#1}\@@href}%
\providecommand \@@href[1]{\endgroup#1\@@endlink}%
\providecommand \@sanitize@url [0]{\catcode `\\12\catcode `\$12\catcode
  `\&12\catcode `\#12\catcode `\^12\catcode `\_12\catcode `\%12\relax}%
\providecommand \@@startlink[1]{}%
\providecommand \@@endlink[0]{}%
\providecommand \url  [0]{\begingroup\@sanitize@url \@url }%
\providecommand \@url [1]{\endgroup\@href {#1}{\urlprefix }}%
\providecommand \urlprefix  [0]{URL }%
\providecommand \Eprint [0]{\href }%
\providecommand \doibase [0]{http://dx.doi.org/}%
\providecommand \selectlanguage [0]{\@gobble}%
\providecommand \bibinfo  [0]{\@secondoftwo}%
\providecommand \bibfield  [0]{\@secondoftwo}%
\providecommand \translation [1]{[#1]}%
\providecommand \BibitemOpen [0]{}%
\providecommand \bibitemStop [0]{}%
\providecommand \bibitemNoStop [0]{.\EOS\space}%
\providecommand \EOS [0]{\spacefactor3000\relax}%
\providecommand \BibitemShut  [1]{\csname bibitem#1\endcsname}%
\let\auto@bib@innerbib\@empty
\bibitem [{\citenamefont {Mermin}\ and\ \citenamefont
  {Wagner}(1966)}]{Mermin+66}%
  \BibitemOpen
  \bibfield  {author} {\bibinfo {author} {\bibfnamefont {N.~D.}\ \bibnamefont
  {Mermin}}\ and\ \bibinfo {author} {\bibfnamefont {H.}~\bibnamefont
  {Wagner}},\ }\href@noop {} {\bibfield  {journal} {\bibinfo  {journal} {Phys.
  Rev. Lett.}\ }\textbf {\bibinfo {volume} {17}},\ \bibinfo {pages} {1133}
  (\bibinfo {year} {1966})}\BibitemShut {NoStop}%
\bibitem [{\citenamefont {Berezinskii}(1972)}]{Berezinskii+72}%
  \BibitemOpen
  \bibfield  {author} {\bibinfo {author} {\bibfnamefont {V.~L.}\ \bibnamefont
  {Berezinskii}},\ }\href@noop {} {\bibfield  {journal} {\bibinfo  {journal}
  {Sov. Phys. JETP}\ }\textbf {\bibinfo {volume} {34}},\ \bibinfo {pages} {610}
  (\bibinfo {year} {1972})}\BibitemShut {NoStop}%
\bibitem [{\citenamefont {Kosterlitz}\ and\ \citenamefont
  {Thouless}(1973)}]{Kosterlitz+73}%
  \BibitemOpen
  \bibfield  {author} {\bibinfo {author} {\bibfnamefont {J.~M.}\ \bibnamefont
  {Kosterlitz}}\ and\ \bibinfo {author} {\bibfnamefont {D.~J.}\ \bibnamefont
  {Thouless}},\ }\href@noop {} {\bibfield  {journal} {\bibinfo  {journal} {J.
  Phys. C}\ }\textbf {\bibinfo {volume} {6}},\ \bibinfo {pages} {1181}
  (\bibinfo {year} {1973})}\BibitemShut {NoStop}%
\bibitem [{\citenamefont {Amit}\ \emph {et~al.}(1980)\citenamefont {Amit},
  \citenamefont {Goldschmidt},\ and\ \citenamefont {Grinstein}}]{Amit+80}%
  \BibitemOpen
  \bibfield  {author} {\bibinfo {author} {\bibfnamefont {D.~J.}\ \bibnamefont
  {Amit}}, \bibinfo {author} {\bibfnamefont {Y.~Y.}\ \bibnamefont
  {Goldschmidt}}, \ and\ \bibinfo {author} {\bibfnamefont {G.}~\bibnamefont
  {Grinstein}},\ }\href@noop {} {\bibfield  {journal} {\bibinfo  {journal} {J.
  Phys. A}\ }\textbf {\bibinfo {volume} {13}},\ \bibinfo {pages} {585}
  (\bibinfo {year} {1980})}\BibitemShut {NoStop}%
\bibitem [{\citenamefont {Korshunov}(2006)}]{Korshunov06}%
  \BibitemOpen
  \bibfield  {author} {\bibinfo {author} {\bibfnamefont {S.~E.}\ \bibnamefont
  {Korshunov}},\ }\href@noop {} {\bibfield  {journal} {\bibinfo  {journal}
  {Physics-Uspekhi}\ }\textbf {\bibinfo {volume} {49}},\ \bibinfo {pages} {225}
  (\bibinfo {year} {2006})}\BibitemShut {NoStop}%
\bibitem [{\citenamefont {Pokrovsky}\ and\ \citenamefont
  {Talapov}(1979)}]{Pokrovsky+79}%
  \BibitemOpen
  \bibfield  {author} {\bibinfo {author} {\bibfnamefont {V.~L.}\ \bibnamefont
  {Pokrovsky}}\ and\ \bibinfo {author} {\bibfnamefont {A.~L.}\ \bibnamefont
  {Talapov}},\ }\href@noop {} {\bibfield  {journal} {\bibinfo  {journal} {Phys.
  Rev. Lett.}\ }\textbf {\bibinfo {volume} {42}},\ \bibinfo {pages} {65}
  (\bibinfo {year} {1979})}\BibitemShut {NoStop}%
\bibitem [{\citenamefont {Chaikin}\ and\ \citenamefont
  {Lubensky}(1995)}]{Chaikin+}%
  \BibitemOpen
  \bibfield  {author} {\bibinfo {author} {\bibfnamefont {P.~M.}\ \bibnamefont
  {Chaikin}}\ and\ \bibinfo {author} {\bibfnamefont {T.~M.}\ \bibnamefont
  {Lubensky}},\ }\href@noop {} {\emph {\bibinfo {title} {Principles of
  condensed matter physics}}}\ (\bibinfo  {publisher} {Cambridge University
  Press},\ \bibinfo {year} {1995})\BibitemShut {NoStop}%
\bibitem{Haldane+83}{F.~D.~M.~Haldane, J. Phys. A
\textbf{15}, 507 (1982).}
\bibitem [{\citenamefont {Korepin}\ \emph {et~al.}(1993)\citenamefont
  {Korepin}, \citenamefont {Bogoliubov},\ and\ \citenamefont
  {Izergin}}]{Korepin+}%
  \BibitemOpen
  \bibfield  {author} {\bibinfo {author} {\bibfnamefont {V.~E.}\ \bibnamefont
  {Korepin}}, \bibinfo {author} {\bibfnamefont {N.~M.}\ \bibnamefont
  {Bogoliubov}}, \ and\ \bibinfo {author} {\bibfnamefont {A.~G.}\ \bibnamefont
  {Izergin}},\ }\href@noop {} {\emph {\bibinfo {title} {Quantum Inverse
  Scattering Method and Correlation Functions}}}\ (\bibinfo  {publisher}
  {Cambridge University Press},\ \bibinfo {year} {1993})\BibitemShut {NoStop}%
\bibitem [{\citenamefont {Giamarchi}(2003)}]{Giamarchi}%
  \BibitemOpen
  \bibfield  {author} {\bibinfo {author} {\bibfnamefont {T.}~\bibnamefont
  {Giamarchi}},\ }\href@noop {} {\emph {\bibinfo {title} {Quantum Physics in
  One Dimension}}}\ (\bibinfo  {publisher} {Clarendon press, Oxford},\ \bibinfo
  {year} {2003})\BibitemShut {NoStop}%
\bibitem [{\citenamefont {Toner}\ and\ \citenamefont
  {DiVincenzo}(1990)}]{Toner+90}%
  \BibitemOpen
  \bibfield  {author} {\bibinfo {author} {\bibfnamefont {J.}~\bibnamefont
  {Toner}}\ and\ \bibinfo {author} {\bibfnamefont {D.~P.}\ \bibnamefont
  {DiVincenzo}},\ }\href@noop {} {\bibfield  {journal} {\bibinfo  {journal}
  {Phys. Rev. B}\ }\textbf {\bibinfo {volume} {41}},\ \bibinfo {pages} {632}
  (\bibinfo {year} {1990})}\BibitemShut {NoStop}%
\bibitem [{\citenamefont {Giamarchi}\ and\ \citenamefont
  {Schulz}(1988)}]{Giamarchi+88}%
  \BibitemOpen
  \bibfield  {author} {\bibinfo {author} {\bibfnamefont {T.}~\bibnamefont
  {Giamarchi}}\ and\ \bibinfo {author} {\bibfnamefont {H.~J.}\ \bibnamefont
  {Schulz}},\ }\href@noop {} {\bibfield  {journal} {\bibinfo  {journal} {Phys.
  Rev. B}\ }\textbf {\bibinfo {volume} {37}},\ \bibinfo {pages} {325} (\bibinfo
  {year} {1988})}\BibitemShut {NoStop}%
\bibitem{Ristivojevic+12}{Z. Ristivojevic, A. Petkovi\'{c}, P. Le Doussal, and T. Giamarchi, Phys. Rev. Lett. \textbf{109}, 026402 (2012).}%
\bibitem [{\citenamefont {Cardy}\ and\ \citenamefont
  {Ostlund}(1982)}]{Cardy+82}%
  \BibitemOpen
  \bibfield  {author} {\bibinfo {author} {\bibfnamefont {J.~L.}\ \bibnamefont
  {Cardy}}\ and\ \bibinfo {author} {\bibfnamefont {S.}~\bibnamefont
  {Ostlund}},\ }\href@noop {} {\bibfield  {journal} {\bibinfo  {journal} {Phys.
  Rev. B}\ }\textbf {\bibinfo {volume} {25}},\ \bibinfo {pages} {6899}
  (\bibinfo {year} {1982})}\BibitemShut {NoStop}%
\bibitem [{\citenamefont {Hwa}\ and\ \citenamefont {Fisher}(1994)}]{Hwa+94}%
  \BibitemOpen
  \bibfield  {author} {\bibinfo {author} {\bibfnamefont {T.}~\bibnamefont
  {Hwa}}\ and\ \bibinfo {author} {\bibfnamefont {D.~S.}\ \bibnamefont
  {Fisher}},\ }\href@noop {} {\bibfield  {journal} {\bibinfo  {journal} {Phys.
  Rev. Lett.}\ }\textbf {\bibinfo {volume} {72}},\ \bibinfo {pages} {2466}
  (\bibinfo {year} {1994})}\BibitemShut {NoStop}%
\bibitem{vortex}{This can be done by giving them an infinite core energy or, on a lattice model, by constraining the phase differences on neighboring sites to be always less than some (small enough) threshold.}%
\bibitem [{\citenamefont {Carpentier}\ and\ \citenamefont
  {Le~Doussal}(1997)}]{Carpentier+97}%
  \BibitemOpen
  \bibfield  {author} {\bibinfo {author} {\bibfnamefont {D.}~\bibnamefont
  {Carpentier}}\ and\ \bibinfo {author} {\bibfnamefont {P.}~\bibnamefont
  {Le~Doussal}},\ }\href@noop {} {\bibfield  {journal} {\bibinfo  {journal}
  {Phys. Rev. B}\ }\textbf {\bibinfo {volume} {55}},\ \bibinfo {pages} {12128}
  (\bibinfo {year} {1997})}\BibitemShut {NoStop}%
\bibitem [{\citenamefont {Schehr}\ and\ \citenamefont
  {Le~Doussal}(2003)}]{SchehrLeDoussal2003}%
  \BibitemOpen
  \bibfield  {author} {\bibinfo {author} {\bibfnamefont {G.}~\bibnamefont
  {Schehr}}\ and\ \bibinfo {author} {\bibfnamefont {P.}~\bibnamefont
  {Le~Doussal}},\ }\href@noop {} {\bibfield  {journal} {\bibinfo  {journal}
  {Phys. Rev. E}\ }\textbf {\bibinfo {volume} {68}},\ \bibinfo {pages} {046101}
  (\bibinfo {year} {2003})}\BibitemShut {NoStop}%
\bibitem [{\citenamefont {Zeng}\ \emph {et~al.}(1996)\citenamefont {Zeng},
  \citenamefont {Middleton},\ and\ \citenamefont {Shapir}}]{Zeng+96}%
  \BibitemOpen
  \bibfield  {author} {\bibinfo {author} {\bibfnamefont {C.}~\bibnamefont
  {Zeng}}, \bibinfo {author} {\bibfnamefont {A.~A.}\ \bibnamefont {Middleton}},
  \ and\ \bibinfo {author} {\bibfnamefont {Y.}~\bibnamefont {Shapir}},\
  }\href@noop {} {\bibfield  {journal} {\bibinfo  {journal} {Phys. Rev. Lett.}\
  }\textbf {\bibinfo {volume} {77}},\ \bibinfo {pages} {3204} (\bibinfo {year}
  {1996})}\BibitemShut {NoStop}%
\bibitem [{\citenamefont {Rieger}\ and\ \citenamefont
  {Blasum}(1997)}]{Rieger+97}%
  \BibitemOpen
  \bibfield  {author} {\bibinfo {author} {\bibfnamefont {H.}~\bibnamefont
  {Rieger}}\ and\ \bibinfo {author} {\bibfnamefont {U.}~\bibnamefont
  {Blasum}},\ }\href@noop {} {\bibfield  {journal} {\bibinfo  {journal} {Phys.
  Rev. B}\ }\textbf {\bibinfo {volume} {55}},\ \bibinfo {pages} {R7394}
  (\bibinfo {year} {1997})}\BibitemShut {NoStop}%
\bibitem [{\citenamefont {Zeng}\ \emph {et~al.}(1999)\citenamefont {Zeng},
  \citenamefont {Leath},\ and\ \citenamefont {Hwa}}]{ZengLeathHwa1999}%
  \BibitemOpen
  \bibfield  {author} {\bibinfo {author} {\bibfnamefont {C.}~\bibnamefont
  {Zeng}}, \bibinfo {author} {\bibfnamefont {P.~L.}\ \bibnamefont {Leath}}, \
  and\ \bibinfo {author} {\bibfnamefont {T.}~\bibnamefont {Hwa}},\ }\href@noop
  {} {\bibfield  {journal} {\bibinfo  {journal} {Phys. Rev. Lett.}\ }\textbf
  {\bibinfo {volume} {83}},\ \bibinfo {pages} {4860} (\bibinfo {year}
  {1999})}\BibitemShut {NoStop}%
\bibitem [{\citenamefont {Perret}\ and\ \citenamefont
  {Schehr}()}]{PerretSchehr2011}%
  \BibitemOpen
  \bibfield  {author} {\bibinfo {author} {\bibfnamefont {A.}~\bibnamefont
  {Perret}}\ and\ \bibinfo {author} {\bibfnamefont {G.}~\bibnamefont
  {Schehr}},\ }\href@noop {} {\bibinfo  {journal} {(in preparation)}\
  }\BibitemShut {NoStop}%
\bibitem [{\citenamefont {Henley}(1997)}]{henley}%
  \BibitemOpen
\bibfield  {journal} {  }\bibfield  {author} {\bibinfo {author} {\bibfnamefont
  {C.~L.}\ \bibnamefont {Henley}},\ }\href@noop {} {\bibfield  {journal}
  {\bibinfo  {journal} {J. Stat. Phys.}\ }\textbf {\bibinfo {volume} {89}},\
  \bibinfo {pages} {483} (\bibinfo {year} {1997})}\BibitemShut {NoStop}%
\bibitem [{\citenamefont {Kenyon}(2009)}]{Kenyon}%
  \BibitemOpen
  \bibfield  {author} {\bibinfo {author} {\bibfnamefont {R.}~\bibnamefont
  {Kenyon}},\ }\href@noop {} {\bibfield  {journal} {\bibinfo  {journal}
  {arxiv:0910.3129}\ } (\bibinfo {year} {2009})}\BibitemShut {NoStop}%
\bibitem [{\citenamefont {Bogner}\ \emph {et~al.}(2004)\citenamefont {Bogner},
  \citenamefont {Emig}, \citenamefont {Taha},\ and\ \citenamefont
  {Zeng}}]{Bogner2004}%
  \BibitemOpen
  \bibfield  {author} {\bibinfo {author} {\bibfnamefont {S.}~\bibnamefont
  {Bogner}}, \bibinfo {author} {\bibfnamefont {T.}~\bibnamefont {Emig}},
  \bibinfo {author} {\bibfnamefont {A.}~\bibnamefont {Taha}}, \ and\ \bibinfo
  {author} {\bibfnamefont {C.}~\bibnamefont {Zeng}},\ }\href@noop {} {\bibfield
   {journal} {\bibinfo  {journal} {Phys. Rev. B}\ }\textbf {\bibinfo {volume}
  {69}},\ \bibinfo {pages} {104420} (\bibinfo {year} {2004})}\BibitemShut
  {NoStop}%
\bibitem [{\citenamefont {Propp}(2003)}]{Propp}%
  \BibitemOpen
  \bibfield  {author} {\bibinfo {author} {\bibfnamefont {J.}~\bibnamefont
  {Propp}},\ }\href@noop {} {\bibfield  {journal} {\bibinfo  {journal} {Theor.
  Comp. Sci.}\ }\textbf {\bibinfo {volume} {303}},\ \bibinfo {pages} {267}
  (\bibinfo {year} {2003})}\BibitemShut {NoStop}%
\bibitem [{\citenamefont {Perret}\ \emph {et~al.}(2012)\citenamefont {Perret},
  \citenamefont {Ristivojevic}, \citenamefont {Le~Doussal}, \citenamefont
  {Schehr},\ and\ \citenamefont {Wiese}}]{Perret+12}%
  \BibitemOpen
  \bibfield  {author} {\bibinfo {author} {\bibfnamefont {A.}~\bibnamefont
  {Perret}}, \bibinfo {author} {\bibfnamefont {Z.}~\bibnamefont
  {Ristivojevic}}, \bibinfo {author} {\bibfnamefont {P.}~\bibnamefont
  {Le~Doussal}}, \bibinfo {author} {\bibfnamefont {G.}~\bibnamefont {Schehr}},
  \ and\ \bibinfo {author} {\bibfnamefont {K.~J.}\ \bibnamefont {Wiese}},\
  }\href@noop {} {\bibfield  {journal} {\bibinfo  {journal} {arxiv:1204.5685}\
  ,\ \bibinfo {pages} {(submitted)}} (\bibinfo {year} {2012})}\BibitemShut
  {NoStop}%
\bibitem [{\citenamefont {Guruswamy}\ \emph {et~al.}(2000)\citenamefont
  {Guruswamy}, \citenamefont {LeClair},\ and\ \citenamefont
  {Ludwig}}]{Guruswamy+00}%
  \BibitemOpen
  \bibfield  {author} {\bibinfo {author} {\bibfnamefont {S.}~\bibnamefont
  {Guruswamy}}, \bibinfo {author} {\bibfnamefont {A.}~\bibnamefont {LeClair}},
  \ and\ \bibinfo {author} {\bibfnamefont {A.~W.~W.}\ \bibnamefont {Ludwig}},\
  }\href@noop {} {\bibfield  {journal} {\bibinfo  {journal} {Nucl. Phys. B}\
  }\textbf {\bibinfo {volume} {583}},\ \bibinfo {pages} {475} (\bibinfo {year}
  {2000})}\BibitemShut {NoStop}%
\bibitem [{\citenamefont {Le~Doussal}\ and\ \citenamefont
  {Schehr}(2007)}]{LeDoussal+07}%
  \BibitemOpen
  \bibfield  {author} {\bibinfo {author} {\bibfnamefont {P.}~\bibnamefont
  {Le~Doussal}}\ and\ \bibinfo {author} {\bibfnamefont {G.}~\bibnamefont
  {Schehr}},\ }\href@noop {} {\bibfield  {journal} {\bibinfo  {journal} {Phys.
  Rev. B}\ }\textbf {\bibinfo {volume} {75}},\ \bibinfo {pages} {184401}
  (\bibinfo {year} {2007})}\BibitemShut {NoStop}%
\bibitem [{\citenamefont {Emig}\ and\ \citenamefont {Kardar}(2001)}]{Emig+01}%
  \BibitemOpen
  \bibfield  {author} {\bibinfo {author} {\bibfnamefont {T.}~\bibnamefont
  {Emig}}\ and\ \bibinfo {author} {\bibfnamefont {M.}~\bibnamefont {Kardar}},\
  }\href@noop {} {\bibfield  {journal} {\bibinfo  {journal} {Nucl. Rev. B}\
  }\textbf {\bibinfo {volume} {604}},\ \bibinfo {pages} {479} (\bibinfo {year}
  {2001})}\BibitemShut {NoStop}%
\bibitem [{\citenamefont {Le~Doussal}\ and\ \citenamefont
  {Giamarchi}(1995)}]{rsb}%
  \BibitemOpen
  \bibfield  {author} {\bibinfo {author} {\bibfnamefont {P.}~\bibnamefont
  {Le~Doussal}}\ and\ \bibinfo {author} {\bibfnamefont {T.}~\bibnamefont
  {Giamarchi}},\ }\href@noop {} {\bibfield  {journal} {\bibinfo  {journal}
  {Phys. Rev. Lett.}\ }\textbf {\bibinfo {volume} {74}},\ \bibinfo {pages}
  {606} (\bibinfo {year} {1995})}\BibitemShut {NoStop}%
\bibitem [{\citenamefont {Tsai}\ and\ \citenamefont
  {Shapir}(1992)}]{TsaiShapir1992}%
  \BibitemOpen
  \bibfield  {author} {\bibinfo {author} {\bibfnamefont {Y.-C.}\ \bibnamefont
  {Tsai}}\ and\ \bibinfo {author} {\bibfnamefont {Y.}~\bibnamefont {Shapir}},\
  }\href@noop {} {\bibfield  {journal} {\bibinfo  {journal} {Phys. Rev. Lett.}\
  }\textbf {\bibinfo {volume} {69}},\ \bibinfo {pages} {1773} (\bibinfo {year}
  {1992})}\BibitemShut {NoStop}%
\bibitem [{\citenamefont {Tsai}\ and\ \citenamefont
  {Shapir}(1994{\natexlab{a}})}]{TsaiShapir1994a}%
  \BibitemOpen
  \bibfield  {author} {\bibinfo {author} {\bibfnamefont {Y.-C.}\ \bibnamefont
  {Tsai}}\ and\ \bibinfo {author} {\bibfnamefont {Y.}~\bibnamefont {Shapir}},\
  }\href@noop {} {\bibfield  {journal} {\bibinfo  {journal} {Phys. Rev. E}\
  }\textbf {\bibinfo {volume} {50}},\ \bibinfo {pages} {3546} (\bibinfo {year}
  {1994}{\natexlab{a}})}\BibitemShut {NoStop}%
\bibitem [{\citenamefont {Tsai}\ and\ \citenamefont
  {Shapir}(1994{\natexlab{b}})}]{TsaiShapir1994b}%
  \BibitemOpen
  \bibfield  {author} {\bibinfo {author} {\bibfnamefont {Y.-C.}\ \bibnamefont
  {Tsai}}\ and\ \bibinfo {author} {\bibfnamefont {Y.}~\bibnamefont {Shapir}},\
  }\href@noop {} {\bibfield  {journal} {\bibinfo  {journal} {Phys. Rev. E}\
  }\textbf {\bibinfo {volume} {50}},\ \bibinfo {pages} {4445} (\bibinfo {year}
  {1994}{\natexlab{b}})}\BibitemShut {NoStop}%
\bibitem [{\citenamefont {Cule}\ and\ \citenamefont
  {Shapir}(1995)}]{culeShapir1995}%
  \BibitemOpen
  \bibfield  {author} {\bibinfo {author} {\bibfnamefont {D.}~\bibnamefont
  {Cule}}\ and\ \bibinfo {author} {\bibfnamefont {Y.}~\bibnamefont {Shapir}},\
  }\href@noop {} {\bibfield  {journal} {\bibinfo  {journal} {Phys. Rev. B}\
  }\textbf {\bibinfo {volume} {51}},\ \bibinfo {pages} {3305(R)} (\bibinfo
  {year} {1995})}\BibitemShut {NoStop}%
\bibitem [{\citenamefont {Schehr}\ and\ \citenamefont
  {Le~Doussal}(2005)}]{SchehrLeDoussal2005}%
  \BibitemOpen
  \bibfield  {author} {\bibinfo {author} {\bibfnamefont {G.}~\bibnamefont
  {Schehr}}\ and\ \bibinfo {author} {\bibfnamefont {P.}~\bibnamefont
  {Le~Doussal}},\ }\href@noop {} {\bibfield  {journal} {\bibinfo  {journal}
  {Europhys. Lett.}\ }\textbf {\bibinfo {volume} {71}},\ \bibinfo {pages} {290}
  (\bibinfo {year} {2005})}\BibitemShut {NoStop}%
\bibitem [{\citenamefont {Schehr}\ and\ \citenamefont
  {Rieger}(2005{\natexlab{a}})}]{SchehrRieger2005}%
  \BibitemOpen
  \bibfield  {author} {\bibinfo {author} {\bibfnamefont {G.}~\bibnamefont
  {Schehr}}\ and\ \bibinfo {author} {\bibfnamefont {H.}~\bibnamefont
  {Rieger}},\ }\href@noop {} {\bibfield  {journal} {\bibinfo  {journal} {Phys.
  Rev. B}\ }\textbf {\bibinfo {volume} {71}},\ \bibinfo {pages} {184202}
  (\bibinfo {year} {2005}{\natexlab{a}})}\BibitemShut {NoStop}%
\bibitem [{\citenamefont {Schehr}\ and\ \citenamefont
  {Le~Doussal}(2004)}]{SchehrLeDoussal2004}%
  \BibitemOpen
  \bibfield  {author} {\bibinfo {author} {\bibfnamefont {G.}~\bibnamefont
  {Schehr}}\ and\ \bibinfo {author} {\bibfnamefont {P.}~\bibnamefont
  {Le~Doussal}},\ }\href@noop {} {\bibfield  {journal} {\bibinfo  {journal}
  {Phys. Rev. Lett.}\ }\textbf {\bibinfo {volume} {93}},\ \bibinfo {pages}
  {217201} (\bibinfo {year} {2004})}\BibitemShut {NoStop}%
\bibitem [{\citenamefont {Schehr}\ and\ \citenamefont
  {Rieger}(2005{\natexlab{b}})}]{Schehr+05}%
  \BibitemOpen
  \bibfield  {author} {\bibinfo {author} {\bibfnamefont {G.}~\bibnamefont
  {Schehr}}\ and\ \bibinfo {author} {\bibfnamefont {H.}~\bibnamefont
  {Rieger}},\ }\href@noop {} {\bibfield  {journal} {\bibinfo  {journal} {Phys.
  Rev. B}\ }\textbf {\bibinfo {volume} {71}},\ \bibinfo {pages} {184202}
  (\bibinfo {year} {2005}{\natexlab{b}})}\BibitemShut {NoStop}%
\bibitem [{\citenamefont {Abramowitz}\ and\ \citenamefont
  {Stegun}(1972)}]{Abramowitz}%
  \BibitemOpen
  \bibfield  {author} {\bibinfo {author} {\bibfnamefont {M.}~\bibnamefont
  {Abramowitz}}\ and\ \bibinfo {author} {\bibfnamefont {I.~A.}\ \bibnamefont
  {Stegun}},\ }\href@noop {} {\emph {\bibinfo {title} {Handbook of Mathematical
  Functions}}}\ (\bibinfo  {publisher} {Dover, New York},\ \bibinfo {year}
  {1972})\BibitemShut {NoStop}%
\bibitem [{Note1()}]{Note1}%
  \BibitemOpen
  \bibinfo {note} {One should notice that the Fourier transform of (\ref {G})
  is positive for all values of $a$ and $m$, as it must be. See Eq.~(2.3) of
  Ref.~\cite {Amit+80}.}\BibitemShut {Stop}%
\bibitem [{\citenamefont {Schulz}\ \emph {et~al.}(1988)\citenamefont {Schulz},
  \citenamefont {Villain}, \citenamefont {Br\'{e}zin},\ and\ \citenamefont
  {Orland}}]{Schulz+88}%
  \BibitemOpen
  \bibfield  {author} {\bibinfo {author} {\bibfnamefont {U.}~\bibnamefont
  {Schulz}}, \bibinfo {author} {\bibfnamefont {J.}~\bibnamefont {Villain}},
  \bibinfo {author} {\bibfnamefont {E.}~\bibnamefont {Br\'{e}zin}}, \ and\
  \bibinfo {author} {\bibfnamefont {H.}~\bibnamefont {Orland}},\ }\href@noop {}
  {\bibfield  {journal} {\bibinfo  {journal} {J. Stat. Phys}\ }\textbf
  {\bibinfo {volume} {51}},\ \bibinfo {pages} {1} (\bibinfo {year}
  {1988})}\BibitemShut {NoStop}%
\bibitem [{\citenamefont {Le~Doussal}(2010)}]{LeDoussal2008}%
  \BibitemOpen
  \bibfield  {author} {\bibinfo {author} {\bibfnamefont {P.}~\bibnamefont
  {Le~Doussal}},\ }\href@noop {} {\bibfield  {journal} {\bibinfo  {journal}
  {Ann. Phys.}\ }\textbf {\bibinfo {volume} {325}},\ \bibinfo {pages} {49}
  (\bibinfo {year} {2010})}\BibitemShut {NoStop}%
\bibitem [{Note2()}]{Note2}%
  \BibitemOpen
  \bibinfo {note} {If one uses cutoff procedures which do not respect STS, the
  renormalized $\tau $, measured from ${\protect \cal G}(x)$ may differ from
  the bare one from $G(x)$. Everywhere here $\tau $ means the renormalized
  one.}\BibitemShut {Stop}%
\bibitem [{\citenamefont {Zinn-Justin}(2002)}]{Zinn-Justin}%
  \BibitemOpen
  \bibfield  {author} {\bibinfo {author} {\bibfnamefont {J.}~\bibnamefont
  {Zinn-Justin}},\ }\href@noop {} {\emph {\bibinfo {title} {Quantum Field
  Theory and Critical Phenomena}}}\ (\bibinfo  {publisher} {Clarendon Press,
  Oxford},\ \bibinfo {year} {2002})\BibitemShut {NoStop}%
\bibitem [{\citenamefont {Kogut}(1979)}]{Kogut79}%
  \BibitemOpen
  \bibfield  {author} {\bibinfo {author} {\bibfnamefont {J.~B.}\ \bibnamefont
  {Kogut}},\ }\href@noop {} {\bibfield  {journal} {\bibinfo  {journal} {Rev.
  Mod. Phys.}\ }\textbf {\bibinfo {volume} {51}},\ \bibinfo {pages} {659}
  (\bibinfo {year} {1979})}\BibitemShut {NoStop}%
\bibitem [{Note3()}]{Note3}%
  \BibitemOpen
  \bibinfo {note} {To be more precise, for the choice (\ref {G}) we have $- m
  \partial _m \protect \qopname \relax o{ln}\protect \mathaccentV {tilde}07Eg =
  2 \tau _a$ with $\tau _a = \tau - \protect \frac {1}{4} (1-\tau ) m^2 a^2
  [\protect \qopname \relax o{ln}(m^2 c^2 a^2)-1] + \protect \mathcal {O}(m^4
  a^4)$. One easily checks that this amount to replace $\tau $ with $\tau _a$
  in the above equations, hence in the limit $m a \to 0$ these terms have no
  effect on the beta functions.}\BibitemShut {Stop}%
\bibitem [{\citenamefont {Itzykson}\ and\ \citenamefont
  {Drouffe}(1989)}]{ItzyksonDrouffe}%
  \BibitemOpen
  \bibfield  {author} {\bibinfo {author} {\bibfnamefont {C.}~\bibnamefont
  {Itzykson}}\ and\ \bibinfo {author} {\bibfnamefont {J.-M.}\ \bibnamefont
  {Drouffe}},\ }\href@noop {} {\emph {\bibinfo {title} {Statistical field
  theory}}}\ (\bibinfo  {publisher} {Cambridge University Press, Cambridge},\
  \bibinfo {year} {1989})\BibitemShut {NoStop}%
\bibitem [{Note4()}]{Note4}%
  \BibitemOpen
  \bibinfo {note} {For the matrix $M_{\alpha \beta }=M_1\delta _{\alpha \beta
  }+M_2$ we have used the following inversion formula, valid for $n\to 0$,
  $(M_{\alpha \beta })^{-1}=\delta _{\alpha \beta
  }/M_1-M_2/M_1^2$.}\BibitemShut {Stop}%
\bibitem [{\citenamefont {Bogoliubov}\ and\ \citenamefont
  {Parasiuk}(1957)}]{BogoliubovParasiuk1957}%
  \BibitemOpen
  \bibfield  {author} {\bibinfo {author} {\bibfnamefont {N.}~\bibnamefont
  {Bogoliubov}}\ and\ \bibinfo {author} {\bibfnamefont {O.}~\bibnamefont
  {Parasiuk}},\ }\href@noop {} {\bibfield  {journal} {\bibinfo  {journal} {Acta
  Math.}\ }\textbf {\bibinfo {volume} {97}},\ \bibinfo {pages} {227} (\bibinfo
  {year} {1957})}\BibitemShut {NoStop}%
\bibitem [{\citenamefont {Hepp}(1966)}]{Hepp1966}%
  \BibitemOpen
  \bibfield  {author} {\bibinfo {author} {\bibfnamefont {K.}~\bibnamefont
  {Hepp}},\ }\href@noop {} {\bibfield  {journal} {\bibinfo  {journal} {Comm.
  Math. Phys.}\ }\textbf {\bibinfo {volume} {2}},\ \bibinfo {pages} {301}
  (\bibinfo {year} {1966})}\BibitemShut {NoStop}%
\bibitem [{\citenamefont {Zimmermann}(1969)}]{Zimmermann1969}%
  \BibitemOpen
  \bibfield  {author} {\bibinfo {author} {\bibfnamefont {W.}~\bibnamefont
  {Zimmermann}},\ }\href@noop {} {\bibfield  {journal} {\bibinfo  {journal}
  {Comm. Math. Phys.}\ }\textbf {\bibinfo {volume} {15}},\ \bibinfo {pages}
  {208} (\bibinfo {year} {1969})}\BibitemShut {NoStop}%
\bibitem [{\citenamefont {Bergere}\ and\ \citenamefont
  {Lam}(1976)}]{BergereLam1975}%
  \BibitemOpen
  \bibfield  {author} {\bibinfo {author} {\bibfnamefont {M.}~\bibnamefont
  {Bergere}}\ and\ \bibinfo {author} {\bibfnamefont {Y.-M.}\ \bibnamefont
  {Lam}},\ }\href@noop {} {\bibfield  {journal} {\bibinfo  {journal} {J. Math.
  Phys.}\ }\textbf {\bibinfo {volume} {17}},\ \bibinfo {pages} {1546} (\bibinfo
  {year} {1976})}\BibitemShut {NoStop}%
\bibitem [{\citenamefont {David}\ \emph
  {et~al.}(1993{\natexlab{a}})\citenamefont {David}, \citenamefont
  {Duplantier},\ and\ \citenamefont {Guitter}}]{DDG1}%
  \BibitemOpen
  \bibfield  {author} {\bibinfo {author} {\bibfnamefont {F.}~\bibnamefont
  {David}}, \bibinfo {author} {\bibfnamefont {B.}~\bibnamefont {Duplantier}}, \
  and\ \bibinfo {author} {\bibfnamefont {E.}~\bibnamefont {Guitter}},\
  }\href@noop {} {\bibfield  {journal} {\bibinfo  {journal} {Phys. Rev. Lett.}\
  }\textbf {\bibinfo {volume} {70}},\ \bibinfo {pages} {2205} (\bibinfo {year}
  {1993}{\natexlab{a}})}\BibitemShut {NoStop}%
\bibitem [{\citenamefont {David}\ \emph
  {et~al.}(1993{\natexlab{b}})\citenamefont {David}, \citenamefont
  {Duplantier},\ and\ \citenamefont {Guitter}}]{DDG2}%
  \BibitemOpen
  \bibfield  {author} {\bibinfo {author} {\bibfnamefont {F.}~\bibnamefont
  {David}}, \bibinfo {author} {\bibfnamefont {B.}~\bibnamefont {Duplantier}}, \
  and\ \bibinfo {author} {\bibfnamefont {E.}~\bibnamefont {Guitter}},\
  }\href@noop {} {\bibfield  {journal} {\bibinfo  {journal} {Nucl. Phys. B}\
  }\textbf {\bibinfo {volume} {394}},\ \bibinfo {pages} {555} (\bibinfo {year}
  {1993}{\natexlab{b}})}\BibitemShut {NoStop}%
\bibitem [{\citenamefont {David}\ \emph {et~al.}(1994)\citenamefont {David},
  \citenamefont {Duplantier},\ and\ \citenamefont {Guitter}}]{DDG3}%
  \BibitemOpen
  \bibfield  {author} {\bibinfo {author} {\bibfnamefont {F.}~\bibnamefont
  {David}}, \bibinfo {author} {\bibfnamefont {B.}~\bibnamefont {Duplantier}}, \
  and\ \bibinfo {author} {\bibfnamefont {E.}~\bibnamefont {Guitter}},\
  }\href@noop {} {\bibfield  {journal} {\bibinfo  {journal} {Phys. Rev. Lett.}\
  }\textbf {\bibinfo {volume} {72}},\ \bibinfo {pages} {311} (\bibinfo {year}
  {1994})}\BibitemShut {NoStop}%
\bibitem [{\citenamefont {David}\ \emph {et~al.}(1997)\citenamefont {David},
  \citenamefont {Duplantier},\ and\ \citenamefont {Guitter}}]{DDG4}%
  \BibitemOpen
  \bibfield  {author} {\bibinfo {author} {\bibfnamefont {F.}~\bibnamefont
  {David}}, \bibinfo {author} {\bibfnamefont {B.}~\bibnamefont {Duplantier}}, \
  and\ \bibinfo {author} {\bibfnamefont {E.}~\bibnamefont {Guitter}},\
  }\href@noop {} {\bibfield  {journal} {\bibinfo  {journal}
  {arxiv:cond-mat/9702136}\ } (\bibinfo {year} {1997})}\BibitemShut {NoStop}%
\bibitem [{\citenamefont {Wiese}(1999)}]{WieseHabil}%
  \BibitemOpen
  \bibfield  {author} {\bibinfo {author} {\bibfnamefont {K.}~\bibnamefont
  {Wiese}},\ }in\ \href@noop {} {\emph {\bibinfo {booktitle} {Phase Transitions
  and Critical Phenomena, Vol. 19}}},\ \bibinfo {editor} {edited by\ \bibinfo
  {editor} {\bibfnamefont {C.}~\bibnamefont {Domb}}\ and\ \bibinfo {editor}
  {\bibfnamefont {J.}~\bibnamefont {Lebowitz}}}\ (\bibinfo  {publisher}
  {Acadamic Press, London},\ \bibinfo {year} {1999})\BibitemShut {NoStop}%
\bibitem [{\citenamefont {Wiese}\ and\ \citenamefont
  {David}(1997)}]{WieseDavid1997}%
  \BibitemOpen
  \bibfield  {author} {\bibinfo {author} {\bibfnamefont {K.}~\bibnamefont
  {Wiese}}\ and\ \bibinfo {author} {\bibfnamefont {F.}~\bibnamefont {David}},\
  }\href@noop {} {\bibfield  {journal} {\bibinfo  {journal} {Nucl. Phys. B}\
  }\textbf {\bibinfo {volume} {487}},\ \bibinfo {pages} {529} (\bibinfo {year}
  {1997})}\BibitemShut {NoStop}%
\bibitem [{\citenamefont {David}\ and\ \citenamefont
  {Wiese}(1996)}]{DavidWiese1996}%
  \BibitemOpen
  \bibfield  {author} {\bibinfo {author} {\bibfnamefont {F.}~\bibnamefont
  {David}}\ and\ \bibinfo {author} {\bibfnamefont {K.}~\bibnamefont {Wiese}},\
  }\href@noop {} {\bibfield  {journal} {\bibinfo  {journal} {Phys. Rev. Lett.}\
  }\textbf {\bibinfo {volume} {76}},\ \bibinfo {pages} {4564} (\bibinfo {year}
  {1996})}\BibitemShut {NoStop}%
\bibitem [{\citenamefont {Wiese}\ and\ \citenamefont
  {David}(1995)}]{WieseDavid1995}%
  \BibitemOpen
  \bibfield  {author} {\bibinfo {author} {\bibfnamefont {K.}~\bibnamefont
  {Wiese}}\ and\ \bibinfo {author} {\bibfnamefont {F.}~\bibnamefont {David}},\
  }\href@noop {} {\bibfield  {journal} {\bibinfo  {journal} {Nucl. Phys. B}\
  }\textbf {\bibinfo {volume} {450}},\ \bibinfo {pages} {495} (\bibinfo {year}
  {1995})}\BibitemShut {NoStop}%
\bibitem [{\citenamefont {Pinnow}\ and\ \citenamefont
  {Wiese}(2002)}]{PinnowWiese2001}%
  \BibitemOpen
  \bibfield  {author} {\bibinfo {author} {\bibfnamefont {H.}~\bibnamefont
  {Pinnow}}\ and\ \bibinfo {author} {\bibfnamefont {K.}~\bibnamefont {Wiese}},\
  }\href@noop {} {\bibfield  {journal} {\bibinfo  {journal} {J. Phys. A}\
  }\textbf {\bibinfo {volume} {35}},\ \bibinfo {pages} {1195} (\bibinfo {year}
  {2002})}\BibitemShut {NoStop}%
\bibitem [{\citenamefont {Ludwig}\ and\ \citenamefont
  {Wiese}(2003)}]{LudwigWiese2002}%
  \BibitemOpen
  \bibfield  {author} {\bibinfo {author} {\bibfnamefont {A.}~\bibnamefont
  {Ludwig}}\ and\ \bibinfo {author} {\bibfnamefont {K.}~\bibnamefont {Wiese}},\
  }\href@noop {} {\bibfield  {journal} {\bibinfo  {journal} {Nucl. Phys. B}\
  }\textbf {\bibinfo {volume} {661}},\ \bibinfo {pages} {577} (\bibinfo {year}
  {2003})}\BibitemShut {NoStop}%
\bibitem [{\citenamefont {Neudecker}(1982)}]{Neudecker82}%
  \BibitemOpen
  \bibfield  {author} {\bibinfo {author} {\bibfnamefont {B.}~\bibnamefont
  {Neudecker}},\ }\href@noop {} {\bibfield  {journal} {\bibinfo  {journal} {Z.
  Phys. B}\ }\textbf {\bibinfo {volume} {49}},\ \bibinfo {pages} {57} (\bibinfo
  {year} {1982})}\BibitemShut {NoStop}%
\bibitem [{\citenamefont {Ma}(1993)}]{Ma}%
  \BibitemOpen
  \bibfield  {author} {\bibinfo {author} {\bibfnamefont {S.-K.}\ \bibnamefont
  {Ma}},\ }\href@noop {} {\emph {\bibinfo {title} {Statistical Mechanics}}}\
  (\bibinfo  {publisher} {World Scientific},\ \bibinfo {year}
  {1993})\BibitemShut {NoStop}%
\bibitem [{Note5()}]{Note5}%
  \BibitemOpen
  \bibinfo {note} {Note that the pole at $w=1,c=-1$ is suppressed by the factor
  $\Theta (w+ 2 c>0)$.}\BibitemShut {Stop}%
\end{thebibliography}

%

\end{document}